\documentclass[]{fjq_tex/fjq}

\abstract{As quantum computing progresses from proof-of-principle demonstrations toward practical utility, a significant impediment is the need to augment algorithmic feasibility with system-level optimization across heterogeneous hardware and software stacks. 
Quantum resource estimation (QRE) plays a central role in this transition, yet existing approaches remain largely compilation-heavy or domain-knowledge-guided symbolic annotations, and tightly coupled to long-term fault-tolerant assumptions, limiting their topical applicability.

In this work, we introduce AutoQuREO, an Automated framework for full-stack Quantum Resource Estimation and Optimization. 
AutoQuREO is built around four core novelties: (i) a flexible, user-defined abstraction of the quantum computing stack; (ii) a modular library of reusable stack components enabling rapid full-stack prototyping; (iii) surrogate modeling of layer-wise resources via algorithmic profiling and neuro-symbolic learning; and (iv) integrated multi-objective optimization that embeds QRE directly into deployment pipelines. 
Together, these design choices enable AutoQuREO to serve as a digital twin for quantum computing stacks, supporting the tractable exploration of complex design spaces.

We demonstrate the capabilities of AutoQuREO through representative co-design case studies, including early-fault-tolerant quantum algorithms, small error correction codes, gate decomposition and variational training of parametric quantum circuits. 
These examples illustrate how AutoQuREO enables systematic discovery of unexploited resource trade-offs that are computationally intractable or abstruse using existing QRE tools.
AutoQuREO is positioned as a general-purpose platform for advancing quantum technology readiness. 
}

\newif\ifreviseshow
\reviseshowfalse
\definecolor{darkred}{rgb}{0.55,0,0}

\newcommand{\addch}[1]{\ifreviseshow\textcolor{darkred}{#1}\else#1\fi}

\begin{document}

\title{AutoQuREO: A Framework for Automated Quantum Resource Estimation and Optimization}

\author[\star]{Harshkumar Oza
\orcidlink{0000-0003-1239-5727}}

\author[\star]{Aritra Sarkar
\orcidlink{0000-0002-3026-6892}}

\author[]{Syed Naqi Abbas
\orcidlink{0009-0000-5348-2806}}

\author[]{Rahul Bhowmick
\orcidlink{0000-0002-5893-0847}}

\author[]{\authorcr \brandFont{Aryan Prakash}
\orcidlink{0009-0003-0835-0961}}

\author[]{Prateek P Kulkarni
\orcidlink{0009-0006-0716-0391}}

\author[]{Krishna Kumar Sabapathy
\orcidlink{0000-0003-3107-6844}}

\affiliation[]{Quantum Lab, Fujitsu Research of India}

\contribution[\star]{Equal contribution.}

\date{\today}
\correspondence{Aritra Sarkar at \url{aritra.sarkar@fujitsu.com}}

\maketitle
\tableofcontents
\addtocontents{toc}{\vspace{-1.3em}}
\section{Introduction}
\label{sec:intro}

Quantum computing has witnessed rapid progress over the past decade, marked by improvements in hardware scale, fidelity, and system integration. 
On the algorithmic side, the complexity-theoretic quantum advantage~\cite{bernstein1993quantum} through problems in the bounded-error quantum polynomial time (BQP) class continues to motivate provable long-term speedups.
This is exemplified by algorithms for factoring, simulating quantum many-body systems, and performing amplitude amplification. 
In parallel, a growing focus of heuristic and application-driven algorithms targets near-term and early fault-tolerant (EFT)~\cite{katabarwa2024early} regimes.
These developments have been accompanied by experimental demonstrations~\cite{arute2019quantum} of quantum computational supremacy in restricted sampling tasks, while a complementary line of work~\cite{google2025quantum} focuses on scalable quantum error correction (QEC) as a principled path toward fault-tolerant quantum computation (FTQC).

To manage the complexity of system design, the quantum computing community has increasingly adopted a layered abstraction of the quantum computing stack~\cite{bertels2020quantum}, enabling a separation of concerns across algorithms, compilation, error correction, and hardware execution. 
The stack has proven invaluable for structuring research and engineering workflows, allowing specialists to focus on individual layers without being overwhelmed by system-wide details. 
However, as an inherent drawback of the approach, design choices made at one layer can substantially affect the feasibility of other layers, and thereby of the full system.

Hardware-software co-design is therefore increasingly gaining traction for achieving early quantum advantage~\cite{shi2020resource,campbell2023superstaq}. 
Co-design considers inter-layer trade-offs that snowball into system-level efficiency~\cite{gratsea2025achieving}, particularly in the EFT regime. 
As a result, optimization objectives shift from studying asymptotic resource measures to multi-dimensional resource trade-offs involving qubits, depth, fidelity, and classical control cost, etc.
This underscores the need for systematic tools that can reason across stack boundaries.

Exploring such trade-offs analytically is non-trivial and requires system-wide expertise.
This motivates the development of quantum resource estimation (QRE) tools within a quantum software development kit (QSDK). 
QRE is intended to operate as a digital twin of a quantum system, estimating relevant resources without explicitly simulating the computation. 
This distinction is crucial as simulation quickly becomes intractable beyond modest system sizes, while resource estimation can remain scalable by disregarding state-level dynamics in favor of cost models. 
The utility of QRE extends beyond offline analysis and benchmarking. 
In quantum processing unit (QPU) execution pipelines, architectural insights from optimized small-scale configurations can inform design choices for scalable deployments. 
QRE can thus be embedded within iterative development workflows, becoming a tool for knowledge transfer across system scaling.
Several QRE frameworks~\cite{babbush2025grand,quetschlich2024utilizing,beverland2022assessing} have already been proposed and adopted by the community, recognizing QRE as a critical component of quantum software ecosystems.
Prominent examples include Microsoft's Azure Quantum Resource Estimator~\cite{beverland2022assessing,van2023using}, Google's Qualtran~\cite{harrigan2024expressing}, PsiQuantum's Bartiq~\cite{PsiQ}, Rigetti's Resource Estimator (RRE)~\cite{saadatmand2024superconducting}, QRE within Xanadu's PennyLane compiler~\cite{bergholm2018pennylane}, Infleqtion's resource-superstaq~\cite{campbell2026resource}, QRE within Qrisp compiler~\cite{krzyszkowski2025analysis} jointly developed with multiple European academic and industry partners, and the TopQAD platform~\cite{mohseni2024build} developed jointly by 1QBit, NVIDIA, and Synopsys, among others. 
These tools enable large-scale analyses of fault-tolerant quantum algorithms to set expectations for long-term quantum advantage. 
Consequently, the design choices of these tools largely reflect a fault-tolerant, algorithm-centric perspective.
Resource estimation is anchored at the quantum algorithm layer, while downstream layers, such as gate decomposition, routing, error correction, and hardware configurations, are implicitly assumed based on an FTQC roadmap. 
As a result, these platforms provide limited support for optimizing downstream layers or for benchmarking a new method.
Moreover, the QRE approaches rely heavily on either compilation-based resource extraction, which, while accurate, is computationally intractable for large-scale design space exploration (DSE), or on user-provided symbolic annotations that require a high level of cross-domain expertise. 
These limitations motivate the need for a more flexible and scalable QRE approach that remains applicable across NISQ, EFT, and FTQC regimes.

\begin{figure}[tbh]
\centering
\includegraphics[page=1, clip, trim=0cm 3cm 1.7cm 0cm,width=0.95\textwidth]{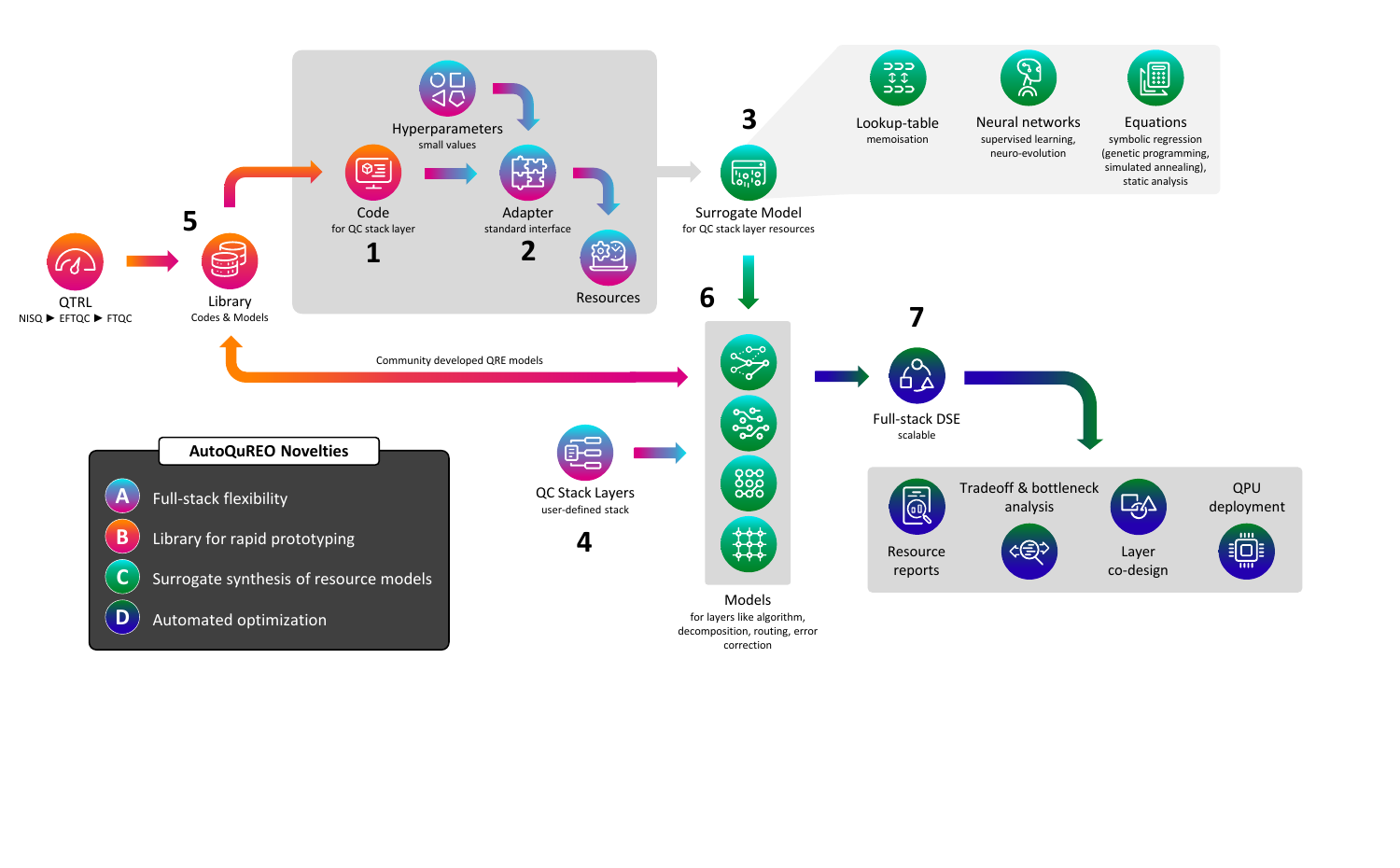}
\caption{Overview of a typical AutoQuREO workflow. (1) Layer-specific code is either imported from the rapid-prototyping library or provided by the user. The library supports NISQ, EFTQC, and FTQC primitives. (2) The adapter design pattern interfaces the code, mapping hyperparameters to resources. The code is compiled for small instances to generate data for modeling. (3) A surrogate resource model is created. AutoQuREO natively supports various models and modeling techniques shown in the inset. (4) Multiple layers can be flexibly added by the user from the library to create a full-stack scenario. (5) The models can be imported from the library, or newly created models can be added back to the library for reuse and community development. (6) The models of the QC stack layers of algorithm, decomposition, error correction, routing, etc., define the project. (7) Parameters and resources of the full-stack can be co-designed via scalable and interpretable DSE. The optimal full-stack can be analyzed to understand bottlenecks and deployed for the target backend.}
\label{fig:autoqureo}
\end{figure}

In this work, we introduce AutoQuREO, an automated quantum resource estimation and optimization framework designed to address these challenges. 
A visual overview of a typical AutoQuREO workflow is shown in Figure~\ref{fig:autoqureo}.
AutoQuREO is distinguished by four core contributions. 
First, it provides full-stack flexibility, allowing users to define arbitrary layers, corresponding hyperparameters, and resource metrics of interest, rather than restricting DSE to a fixed FTQC stack. 
Second, it offers an extensive library of reusable code components and resource models for exemplary stack layers, reducing the barrier to full-stack QRE research. 
Third, it introduces surrogate modeling of layer-wise resources via algorithmic profiling and neuro-symbolic learning, enabling scalable estimation that progressively replaces costly compilation-based ground-truth while remaining analytically explainable. 
Finally, AutoQuREO embeds automated multi-objective optimization directly into the deployment pipeline, supporting Pareto analysis and informed selection of optimal configurations under competing constraints.

We demonstrate the utility of AutoQuREO's salient features through three representative case studies: (i) co-design of Trotter-order and steps in Hamiltonian simulation with error-mitigated noisy hardware, (ii) co-design of a universal quantum error correction scheme, connectivity topology, gate decomposition accuracy, and early fault-tolerant algorithm for ground-state estimation on noisy hardware, and (iii) co-design of mixing ans\"atze, error mitigation, and routing for variational quantum optimization. 
Together, these examples illustrate how automated, full-stack quantum resource estimation can help navigate intricate design spaces to optimize system-level performance via surrogate models.

The remainder of this article is organized as follows.
Section~\ref{sec:background} reviews background concepts and related work on quantum computing stacks, resource estimation, and algorithmic profiling. 
In Section~\ref{sec:autoqureo}, we describe the design and architecture of the proposed AutoQuREO framework.
Section~\ref{sec:qureous} presents four detailed case studies demonstrating AutoQuREO's capabilities. 
Section~\ref{sec:conclusion} concludes the article with a discussion of broader implications, limitations, and future directions.

\addtocontents{toc}{\vspace{-1.0em}}
\section{Background and related works}
\label{sec:background}

This section reviews the conceptual and technical foundations underlying quantum resource estimation and algorithmic profiling, situating AutoQuREO within the broader landscape of quantum software tooling. 
We first summarize how resources are defined, propagated, and estimated across the quantum computing stack, and survey existing frameworks developed for this purpose. 
We then introduce algorithmic profiling as an empirical and interpretable approach to modeling resource costs, highlighting its relevance for scalable, full-stack quantum resource estimation.

\subsection{Quantum resource estimation}

Quantum advantage is ultimately a statement about resources rather than computability. 
The physical Church-Turing thesis (pCTT) provides the underlying foundation of computability in computer science, allowing any physical system to be simulated by a universal computer.
However, in view of pCTT, both classical and quantum computers (via their automata models, such as the universal Turing machine and the quantum Turing machine~\cite{deutsch1989quantum}) are equivalent in the Turing hierarchy, i.e., anything that can be computed by one can also be computed by the other.
The separation between classical and quantum computation arises from the complexity-theoretic version~\cite{bernstein1993quantum} of the thesis, which positions quantum computation (QC) as a superset of classical computation (CC) when resources are taken into account.
In view of that, all CC can be emulated with QC efficiency (i.e., with a worst-case polynomial overhead), for example, using gates like Toffoli that map to universal CC gates like NAND and FanOut.
On the other hand, while all QC can also be emulated with CC via unitary evolution, for the general case, the time and memory cost grow exponentially in the system size.
Thus, the quantum advantage boundary requires careful understanding of the inherent complexity in the problem formulation that is not amenable to efficient CC, as well as the exact resources of CC and QC for problem sizes of relevance in additional to asymptotics.
Even provably asymptotic quantum speedups may be rendered irrelevant by prohibitive constant factors or fault-tolerance overheads. 
Conversely, heuristic or approximate algorithms without asymptotic guarantees may offer practical advantages when resource trade-offs align favorably. 
Quantum computational resources~\cite{mishra2026eqisa,hoefler2023disentangling,giortamis2025qos} thus play a pivotal role in translating abstract algorithmic promise into engineered performance.

Quantum resource estimation (QRE) concerns the systematic quantification of the dependencies between the physical costs required to realize a quantum computation on a given execution stack and the various configurations in the design space. 
Typical resources include the number of logical and physical qubits (space), circuit depth and wall-clock runtime (time), structured gate counts (e.g., Clifford vs non-Clifford), composite measures (e.g., active volume), approximation (e.g., target precision, logical error rates), and hardware metrics (e.g., connectivity, gate latency, classical control bandwidth, noise). 
Importantly, these resources are often not analyzable independently.
For example, reductions in circuit depth may require additional qubits, while higher precision or fault tolerance can amplify space-time costs through error-correction overhead. 
Additionally, what constitutes a resource as input or output of the estimation or optimization is guided by the problem statement posed by the use case.
For example, a user may either optimize an algorithm within the qubit limits of a specific hardware platform or study the qubit-scaling behavior of the algorithm with respect to problem size.

Historically, large-scale QRE has been carried out through painstaking analytical derivations. 
Seminal works such as \cite{sanders2020compilation, bagherimehrab2022nearly,gratsea2025achieving,gidney2025factor} provide detailed end-to-end resource analyses, requiring substantial effort from domain experts. 
While these studies set important benchmarks, their methodology does not scale to the diversity of quantum algorithms, hardware platforms, or error-correction schemes under active investigation.
Moreover, such analyses are brittle in incorporating design choices; for instance, replacing surface codes with holographic codes~\cite{fan2024lego_hqec} can invalidate substantial portions of the analysis.
This limits their utility for rapid iteration, exploratory research, and cross-layer co-design.

The need for systematic tooling motivated some of the earliest QRE frameworks, including QuRE~\cite{suchara2013qure}, which formalized resource accounting at the architectural level. 
This direction has since been reinforced by large-scale initiatives such as DARPA's Quantum Benchmarking program that led to the development of tools such as pyLIQTR from MIT Lincoln Laboratory~\cite{pyLIQTR}, Rigetti's resource estimators~\cite{saadatmand2024fault}, Zapata AI's BenchQ~\cite{BenchQ}, and academic tools such as Jabalizer~\cite{bowen2025design} exemplify interest in software tools for resource analysis.
QRE tools under active development have matured significantly, with four platforms standing out for adoption and scope. 
Microsoft's Azure Quantum Resource Estimator (AQRE)~\cite{beverland2022assessing, van2023using} provides symbolic and compilation-based estimation tightly integrated with fault-tolerant assumptions.
Google's Qualtran~\cite{harrigan2024expressing} introduces a structured, compositional language for expressing quantum algorithms as resource-annotated blocks (Bloqs) and symbolic surface code QEC models. 
PsiQuantum's Bartiq and QREF focus on symbolic estimation of the resource requirements of algorithms aimed at the FTQC era. 
TopQAD~\cite{mohseni2024build} integrates algorithm-architecture co-design across partners, including 1QBit, NVIDIA, and Synopsys. 
It includes rigorous assessments of FTQC surface code superconducting processors.

QRE tools are increasingly used to support scientific and engineering studies. 
For example, \cite{van2023using} uses AQRE to assess the feasibility of fault-tolerant implementations of chemistry, optimization, and simulation workloads. 
Such studies demonstrate the value of automated QRE for grounding algorithmic proposals in realistic resource budgets~\cite{toshio2025practical,kanasugi2026enabling} and for exploring alternative design pathways.

Despite significant progress, existing QRE frameworks share several topical limitations. 
Most are anchored to a fault-tolerant quantum computing (FTQC) roadmap~\cite{su2026resource}, offering limited support for near-term or EFT research, characterized by experimenting with design choices across stack layers.
By weakly parameterizing the lower stack layers to the FTQC roadmap, resource estimation remains confined to the algorithmic layer.
This restricts the adoption of QRE tools by researchers of other layers.
Another consequence of the existing QRE design philosophy is that full-stack optimization across layers remains largely unsupported. 

Existing QRE tools infer resource estimates using either the compilation or symbolic annotation method.
Quantum software stacks, such as Qiskit, include sequential compilation passes~\cite{javadiabhari2015scaffcc,cross2022openqasm}, including high-level synthesis, gate-set decomposition, layout selection, routing, scheduling, etc. 
Each pass transforms the circuit structure, thereby updating the estimates of various resource costs.  
QRE propagates through the stack and tracks the resources induced by the passes.
Compilation provides accurate resource counts; however, it becomes computationally prohibitive for large design spaces.
To mitigate this, QRE tools often use compilation at the algorithm layer to be flexible to user input, while the lower layers are based on symbolic models hardcoded for the FTQC stack.
Symbolic annotation, though trivially scalable and interpretable, demands deep cross-domain expertise and is difficult to adapt as assumptions evolve. 
These gaps prompted us to redesign how QRE can be made flexible across regimes and stack, tractable at scale, yet requiring minimal domain expertise.

\subsection{Quantum algorithmic profiling}
\label{sec:algo_prof}

Complexity theory characterizes algorithmic scaling in terms of input size, while deliberately abstracting away constant factors, lower-order terms, hardware effects, and system-level interactions. 
In practice, however, these effects often dominate real-world performance in resource-constrained environments. 
In classical computing, software profiling has long served as a practical complement to theoretical complexity analysis. 
Profilers such as gprof~\cite{graham1982gprof} empirically measure execution time, memory consumption, cache behavior, and call frequencies, enabling developers to identify performance bottlenecks that are typically out of scope in asymptotic analysis. 
This empirical perspective is particularly valuable when performance depends on implementation details, compiler decisions, or architectural features that are difficult to model analytically. 
Recent works~\cite{suau2022qprof,zilk2025breaking} have adapted this paradigm to quantum settings.

Complementary to profiling, static analysis techniques such as automatic amortized resource analysis (AARA) provide sound upper bounds on resource usage (typically runtime) through type systems and abstract interpretation. 
Recent work has explored hybrid approaches that augment static guarantees with data-driven Bayesian inference to improve robustness and tightness~\cite{pham2024robust,pham2025integrating}.
While these methods demonstrate promising results for time and memory analysis in classical programs, their adoption in quantum computation~\cite{javadiabhari2015scaffcc,colledan2025flexible,meuli2020enabling} remains limited to verifiability. 
In particular, frameworks for jointly modeling multiple interacting resources, such as space, time, and fidelity, are still lacking.

Algorithmic profiling~\cite{zaparanuks2012algorithmic} extends classical software profiling by explicitly linking empirical measurements to inferred cost models. 
It acts as a bridge between computational complexity theory and runtime profiling, aiming to recover interpretable cost functions from observed execution data. 
Given a set of tuples mapping input characteristics to measured resource costs, an algorithmic profiler infers a functional relationship that approximates the underlying resource asymptote. 
The broader field of empirical algorithmics frames algorithmic profiling as the automatic inference of cost functions or asymptotic bounds from experimental data~\cite{mcgeoch2002using}. 
Superficially, this formulation appears to conflict with computability limits such as Rice's theorem~\cite{rice1953classes}, which states that all non-trivial semantic properties of programs are undecidable. 
Resource usage is precisely such a property. 
However, this theoretical limitation does not preclude practical profiling for several reasons. 
First, profiling operates on finite executions and bounded input regimes, rather than on all possible program behaviors. 
Second, the goal is not exact characterization, but approximate prediction within a domain of interest.
Third, profiling exploits regularities induced by algorithmic structure, compiler conventions, and hardware constraints, which significantly reduce the effective hypothesis space. 
Thus, algorithmic profiling adopts a pragmatic stance, aiming for actionable guidance rather than formal completeness.

Analogously, in the quantum circuit model, non-trivial resource properties, such as circuit depth after compilation, or logical error rates under noise, depend on the semantic behavior of quantum programs over transpiler transformations. 
Exact analytical characterization of such properties across arbitrary circuits and compilation stacks is therefore intractable and undecidable in general. 
However, as in classical empirical algorithmics, this limitation does not preclude practical estimation. 
AutoQuREO leverages quantum algorithmic profiling to empirically approximate these properties using sampled configurations and small-instance compilations and infer scalable cost models. 
\addch{Thus, quantum algorithmic profiling constitutes the data-generation stage of the surrogate modeling pipeline. The small-instance compilations produce the labeled tuples that map hyperparameter configurations to measured resources and serve as training data, and the empirical regularities these tuples expose justify modeling quantum circuit resources as learnable functions of circuit features.}
This explicit framing of QRE through empirical algorithmics implemented as a surrogate modeling pipeline, is novel in quantum computing.

Algorithmic profiling, however, is in itself a non-trivial modeling problem, trading off interpretability, generalization, and computational efficiency. 
For example, lookup-based models offer high fidelity but poor scalability, neural networks improve predictive performance at the cost of transparency, while symbolic regression methods provide interpretable closed-form expressions but may struggle with high-dimensional parameter spaces. 
AutoQuREO addresses this trade-off by treating surrogate model selection and configuration as a meta-optimization problem, enabling systematic exploration of model families and hyperparameters. 
This ensures the quantum resource estimates remain scientifically informative across diverse QRE scenarios without the user manually refining the model.
The model selection and configuration is further discussed in Sections~\ref{sec:autoqureo_model}-\ref{sec:autoqureo_lifelong}.

\addtocontents{toc}{\vspace{-1.0em}}
\section{The AutoQuREO framework}
\label{sec:autoqureo}

This section introduces the design and architecture of AutoQuREO, the proposed framework for automated, full-stack quantum resource estimation and optimization. 
We describe the core design principles underpinning the system, including flexible stack abstractions, modular interfaces, scalable surrogate modeling via algorithmic profiling, and an integrated optimization pipeline. 
The section further details the software design and architecture of our implementation.

\subsection{Design principles and novelty}
\label{sec:autoqureo_novelty}

The accelerating pace of quantum computing research has shifted the focus from isolated innovation of hardware and algorithms to system-level solutions.
Achieving a higher quantum technology readiness level (QTRL)~\cite{purohit2024building} increasingly depends on the ability to rapidly perform DSE and deploy these design choices across the full QC stack. 
Currently, this process remains largely manual, fragmented across tools, and biased toward long-term FTQC assumptions. 
AutoQuREO is designed to address this gap by treating quantum resource estimation as an integral component of quantum software development.
By automating the process and incorporating optimization as a post hoc step, we augment the QRE workflow beyond a static accounting exercise.

The framework is built around four core novelties \textbf{\textcolor{black}{$\mathcal{N}$}}, each addressing a fundamental limitation of existing QRE approaches. 
\begin{itemize}
    \item \textcolor{black}{$\mathcal{N}$1} - \textbf{Full-stack flexibility:} 
    AutoQuREO imposes no fixed notion of layers, architectures, or fault-tolerance assumptions. 
    Instead, users define arbitrary stacks composed of modular layers. 
    Layer interfaces are explicitly defined using the adapter design pattern, enabling compatibility across the stack and reuse across use cases.
    This allows our tool to be of relevance to quantum researchers across the stack, beyond quantum algorithms.
    \item \textcolor{black}{$\mathcal{N}$2} - \textbf{Library for rapid prototyping:}  
    The framework provides a library of composable code implementations (called modules) and models that allow users to instantiate end-to-end QRE pipelines with minimal boilerplate code, lowering the barrier to systematic co-design studies.
    \item \textcolor{black}{$\mathcal{N}$3} - \textbf{Surrogate synthesis of resource models:} 
    To overcome the scalability limits of compilation-based estimation and the domain-expertise requirement of symbolic annotation, AutoQuREO introduces surrogate synthesis for resource models.
    Models vary in estimation cost and accuracy, including neural networks, symbolic regression, and lookup-tables.
    Advanced AI methods, such as symbolic distillation of neuro-evolved networks, are used to autonomously synthesize and fine-tune scalable models.
    \item \textcolor{black}{$\mathcal{N}$4} - \textbf{Automated optimization:} 
    Resource estimation is embedded within an optimization loop that supports multi-objective user-guided Pareto front analysis and downstream deployment.
    This enables targeted analysis of performance bottlenecks.
    Eventually, the optimized configurations can be compiled and deployed on target backends.
    \addch{AutoQuREO turns the stack configuration from a passive object of analysis into an active decision variable, and integrates estimation, optimization, and deployment within a single interface, so that researchers need not switch to a separate quantum SDK for circuit generation and cloud execution. This tight coupling of QRE with deployment is atypical of existing QRE tools.}
\end{itemize}
Together, these principles position AutoQuREO as a system-level digital twin for quantum computing stacks, designed to support improvement across NISQ, EFTQC, and FTQC regimes.

\subsection{Layer definition}
\label{sec:autoqureo_layer}

AutoQuREO represents a quantum computing stack as an ordered collection of layers, each encapsulating a functional stage of running an end-to-end quantum solution.
Typical layers include input data preprocessing, quantum algorithms, quantum logic gate decomposition, quantum circuit routing, quantum error handling (e.g., error mitigation or error correction), quantum hardware execution, and output data postprocessing. 
Importantly, the framework does not enforce a fixed taxonomy: layers may be added, removed, merged, or reordered within the constraints of architectural dependencies.
The resource estimation is organized sequentially according to these layers. 

Each layer can have multiple implementations, either as a module or a model.
\begin{itemize}
    \item \textbf{Module}: is an implementation of a layer that can be compiled, and a quantum circuit is available as an output, which can further be used for QRE. 
    For example, for a quantum logic gate decomposition layer, this can be implemented as Python code that takes in Qiskit circuits and decomposes them by transpiling them into an equivalent circuit composed only of quantum logic gates supported by the lower layers, say, using the quantum Shannon decomposition and Solovay-Kitaev decomposition. 
    Resources estimated from modules are considered as ground-truths. 
    However, it is important to note that transpiling circuits does not scale well for large sizes. 
    Since we need to perform resource estimation across many DSE configurations, solely using modules becomes computationally intractable beyond toy use cases.
    \item \textbf{Model}: To circumvent the intractability of using modules for QRE, AutoQuREO introduces layer-specific resource models. 
    Users can define their own model type, e.g., neural networks, symbolic regression, and lookup-tables. 
    Models can be thought of as surrogates that take in the resources of the input to the current layer, a specific hyperparameter configuration for the current layer, and output an estimate of the updated resources expected upon adding the layer to the stack.
\end{itemize}
The general design philosophy is to start with modules for each layer, model them individually, and then perform DSE across all layers as models.
Once the optimal configuration is analyzed, the modules can be tuned to it and deployed.

In the context of the framework, it is imperative to clarify four additional terms that we have used loosely so far:
\begin{itemize}
    \item \textbf{Resources}: broadly refer to the set of metrics that influence a user's preference to prototype a quantum solution in a certain way. 
    For example, it may be the number of qubits required to deploy a quantum algorithm for a specific application, which may or may not be available in the selected quantum hardware. 
    Typically, we consider the number of qubits, the number and types of quantum logic gates, the quantum runtime, and the fidelity of the quantum circuit. 
    The AutoQuREO framework allows users to seamlessly add other resources of interest (e.g., energy consumption or active volume) as long as a specification for measuring or estimating them for a candidate configuration is provided.    
    \item \textbf{Hyperparameter}: Each layer (strictly need not, but typically) has some tunable hyperparameters. These can be tuned to holistically optimize a full-stack quantum prototype.     
    \item \textbf{Hyperparameter bounds}: Each hyperparameter of a layer has some typical range (or list of options, without any particular ordering) within which the QRE would be conducted. 
    These can be specified as a list, e.g., (A,B,C), or as a range and a step size, e.g., (1 to 100 in steps of 20), or as a continuous range bound, e.g., (0.900 to 0.999).    
    \item \textbf{Configurations}: When there are two or more independent hyperparameters of a layer, the permutation of their choices leads to different configurations of the layer. 
    For example, say Layer X has 2 hyperparameters with H$_1$ having options (1,2,3) and H$_2$ having options (A,B); then the $3 \times 2 = 6$ configurations of Layer X are (1A,1B,2A,2B,3A,3B).
    Configurations of subsequent layers are combinatorially expanded from those of previous layers.
\end{itemize}

\subsection{Adapters}
\label{sec:autoqureo_adapter}

AutoQuREO implements a plug-and-play design for modules and models via the adapter design pattern~\cite{gamma1995design}.
Adapters for each module/model maintain a specific Python-based template to define hyperparameters and log the resources. 
The modules and models might even be third-party tools that use different coding styles or languages (such as t$|$ket$\rangle$~\cite{sivarajah2021t}). 
\addch{Multiple adapters can also be composed into a composite adapter, which is useful when their corresponding hyperparameters are coupled, so that only the valid joint configurations are explored.}

\begin{figure}[htb]
\includegraphics[page=3, clip, trim=0cm 0cm 1cm 0.38cm,width=\textwidth]{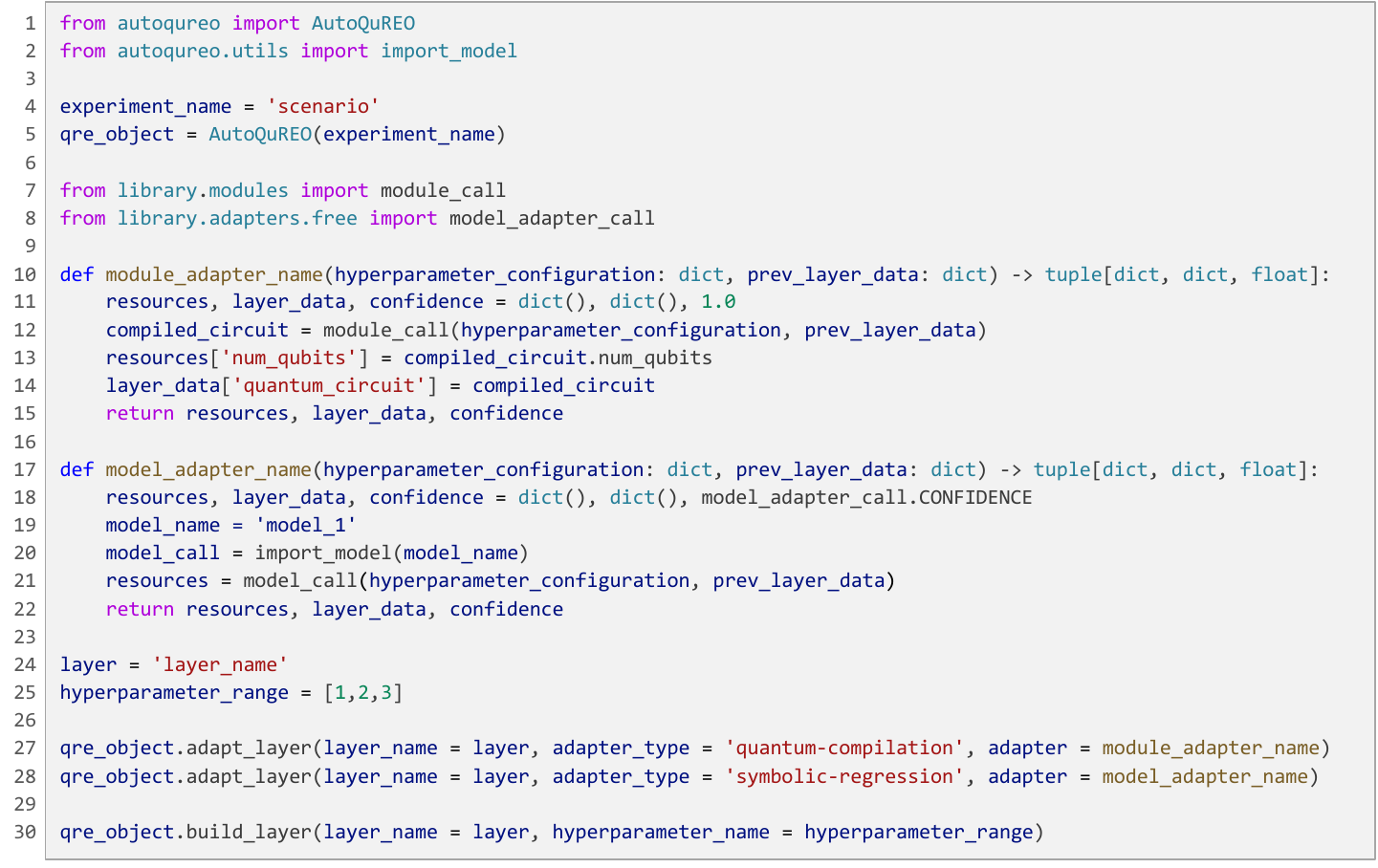}
\caption{API template for AutoQuREO project creation, adapter definition, and layer building.}
\label{fig:adapter_api}
\end{figure}

Figure~\ref{fig:adapter_api} presents an example use of the AutoQuREO API.
After installing the Python software package, a user starts by importing the package and creating an object with an experiment name (lines 1-5).
The next step is to define the adapters for each layer (lines 7-22). In this example, we define two adapters: (1) \verb|module_adapter_name| a quantum-compilation type, and (2) \verb|model_adapter_name| a symbolic-regression type.
Note that these might already be available in the library, so the user can import them directly from the adapter library.
The user might also define their own adapters for their own modules.
The module (or model) might be present in the AutoQuREO library (line 7) or accessed from a user-defined location.
The adapter is identified by the function name; the input interface of a dictionary of hyperparameter configuration and any data from the previous layer; and the output interface of a dictionary of resources, data from this layer to the next, and the confidence estimation of this adapter (lines 10-11 and 17-18).
The encapsulated module (or model) is accessed inside the adapter with the input parameters (lines 12 and 21).
For modules, a quantum circuit is returned and then inspected for various resources (line 13), whereas for models, the resources are returned directly (line 21).
Though we show only the number of qubits as an example, multiple such resources of interest might be added to the dictionary.
The layer data for modules might include the compiled quantum circuit when required for downstream layers, whereas for models, this is not available.
Finally, the confidence score is typically $1.0$ for resources assessed from direct circuit inspection, whereas for models it is fetched from the model adapter from the library (line 18), and updated during training, as explained in Section~\ref{sec:autoqureo_lifelong}.

Once the adapters are defined (or imported from the adapter library), they are added to a specific layer of the project, along with the adapter type metadata (lines 27-28).
After all the adapters are added, the layer is built, with the parameters specifying the layer name, the hyperparameter names, and their corresponding ranges (lines 24-25 and 30).
The layer building process includes (i) expanding the hyperparameters to configurations, (ii) choosing a particular adapter, and (iii) adding the resource estimates for each configuration to each existing configuration up to the previous layer to form the new layer of nodes in the resource tree.
These steps will be explained in further detail in the following sections.
The layer building progress can be tracked, as it might take significantly longer for large DSE experiments.

Adapter type is metadata used to select a specific adapter for performing the QRE on a layer. 
Typically, it is one of the following types: quantum-compilation, lookup-table, neural-network, symbolic-regression, preprocessing, postprocessing, or executor.
In addition to the type, each adapter has two additional metadata fields. 
Cost refers to the classical computational time to estimate the quantum resources. 
For example, as discussed earlier, estimating the resource via quantum compilation (or module) is not scalable and thus incurs a high cost. 
In contrast, a trained neural network at inference can quickly estimate resources. 
Confidence refers to a measure of the closeness of the resource estimate to the actual resource, reflecting uncertainty due to limited data or extrapolation. 
Obviously, since compilation is the ground-truth, the confidence of modules is 100\% or $1.0$. 
For models, the confidence is typically less than $1.0$. 
The significance of these metadata and the process of estimating and updating these will be clarified in Section~\ref{sec:autoqureo_agent} and \ref{sec:autoqureo_surrogate}, respectively.

\begin{figure}[htb]
\centering
\includegraphics[page=2, clip, trim=0cm 0cm 6.6cm 9.4cm,width=\textwidth]{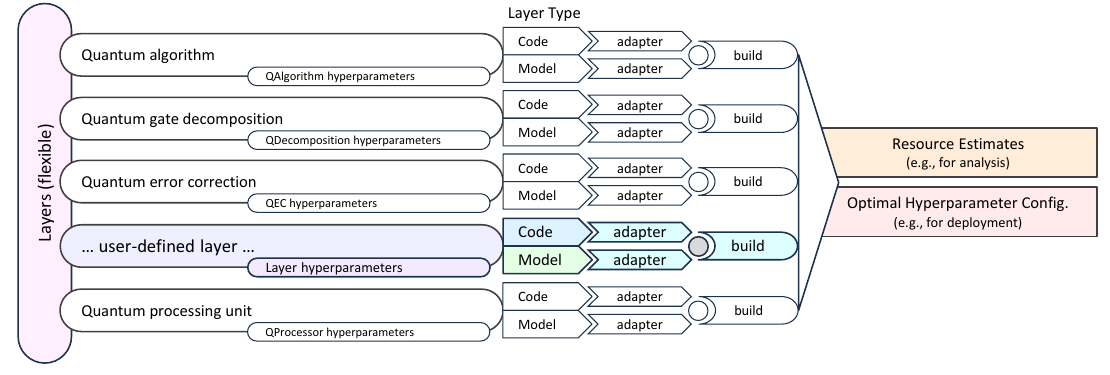}
\caption{
Internal workflow of AutoQuREO. A quantum computing stack is represented as a sequence of user-defined layers, each associated with hyperparameters and one or more adapters. For a given configuration, each layer produces resource estimates that are cumulatively aggregated across the stack. Surrogate models replace compilation where applicable, enabling scalable estimation. The aggregated resources are used for downstream multi-objective optimization and hardware deployment.
}
\label{fig:workflow_comparison}
\end{figure}

Figure~\ref{fig:workflow_comparison} presents the internal workflow of AutoQuREO, to aid the comparison with existing QRE tools.
The layers are not predefined in AutoQuREO.
Users can flexibly add or remove layers based on their QC stack of interest.
The specific layers shown are just an example.
Standard layer definitions are available in the library for rapid prototyping of full-stack use cases.
In existing tools, the quantum algorithm layer is typically the only user-defined component, leaving the lower stack monolithic.
Each layer has its corresponding set of tunable hyperparameters used for DSE and eventual optimization by AutoQuREO.
Since no layer is treated differently, these hyperparameters can thus be used to optimize any/all chosen layer/s, enabling co-design.
Each layer is defined by either a code implementation that can be compiled into a quantum circuit or by one or more surrogate models that map hyperparameter configurations to resources.
Autonomously synthesizing these models forms the core novelty of AutoQuREO.
The code and model across layers follow a standard adapter interface, allowing easy plug-and-play with the tool.
During the build process, one of the adapters is selected to estimate the resources corresponding to the configurations.
These resources are cumulated across the layers to produce the full-stack resource estimate.
This estimate is valuable for bottleneck and sensitivity analysis.
AutoQuREO also allows Pareto optimality analysis to select an optimal configuration for deployment on the actual quantum computing platform.

\subsection{Resource models}
\label{sec:autoqureo_model}

To achieve scalability, AutoQuREO employs various surrogate modeling~\cite{balaprakash2013active} techniques for resource estimation.
Some archetypal ones are discussed herein.
Lookup-table~\cite{magalhaes2018optimization} is employed in regimes where precomputed values can effectively be substituted at inference-time to accelerate the QRE without forgoing accuracy.
Symbolic regression can be used when interpretability and extrapolation are critical. 
Various library models use PySR\cite{cranmer2023interpretable} to perform symbolic regression on a dataset using genetic programming for the equational structure and simulated annealing for the numerical regression.
Neuro-evolution using augmented topologies (NEAT)~\cite{stanley2002evolving} is another technique employed in some models, especially as a flexible approximator for highly non-linear resource behavior.
As a neuro-symbolic (NeSy) technique, these two can be combined by using PySR for symbolic distillation on a trained NEAT model.
\addch{This denotes a two-stage procedure. A NEAT network is first evolved to fit the profiled resource data as a flexible sub-symbolic approximator, providing accuracy and robustness over irregular data. PySR then distills this trained network into a compact closed-form expression. The composite estimator combines the predictive quality of the network with the transparency and low inference cost of a formula.}
NEAT captures intricate empirical patterns, while PySR extracts structured representations that are formalizable and compositional.
NeSy aligns with the so-called third wave of artificial intelligence (AI) that seeks to combine sub-symbolic efficiency with symbolic interpretability. 
To the best of our knowledge, AutoQuREO is the first to systematically apply NeSy-AI to quantum computation in general and resource estimation in particular. 
NeSy-QRE enables the framework to reconcile competing demands of scalability across large design spaces, robustness to architectural variation, and transparency necessary for scientific insight and trust. 

\subsection{Adapter metadata and arbitration}
\label{sec:autoqureo_agent}

An adapter includes various metadata to guide the QRE process.
When a model is encapsulated by an adapter, an initial estimate of the cost and confidence metadata is assigned based on the model type.
This estimate is based on model training conducted prior to the QRE process, known as warm-start.
The cost is based on the model's inference time on the test set, while the confidence reflects the test-set performance (e.g., via RMS error or F1 score).

During the QRE process, an agent arbitrates among available module and model adapters for each layer.
Agents can be configured with a persona/policy as a scoring function, which encodes preferences such as accuracy versus speed, interpretability, and extrapolation compatibility for generalization.
For example, an agent may prefer a quick-and-dirty estimate for fast QRE, while another agent may focus on getting a better estimate, trading off longer time to reach the optimal prototype. 
We use the reinforcement learning terminology of agent and policy to bespeak the incorporation of more advanced arbitration in the future.
The adapter with the highest score is selected to estimate the QRE for the layer.
When an adapter is chosen by the agent, based on the actual time the QRE for the particular configuration took, the cost metadata is updated.
The confidence parameter is updated only during model retraining via lifelong learning, as explained in the following section.

\subsection{Surrogate synthesis}
\label{sec:autoqureo_surrogate}

\begin{figure}[p]
    \centering
    \begin{subfigure}[b]{\textwidth}
        \includegraphics[page=6, clip, trim=0.5cm 1.6cm 0.6cm 3cm,width=\textwidth]{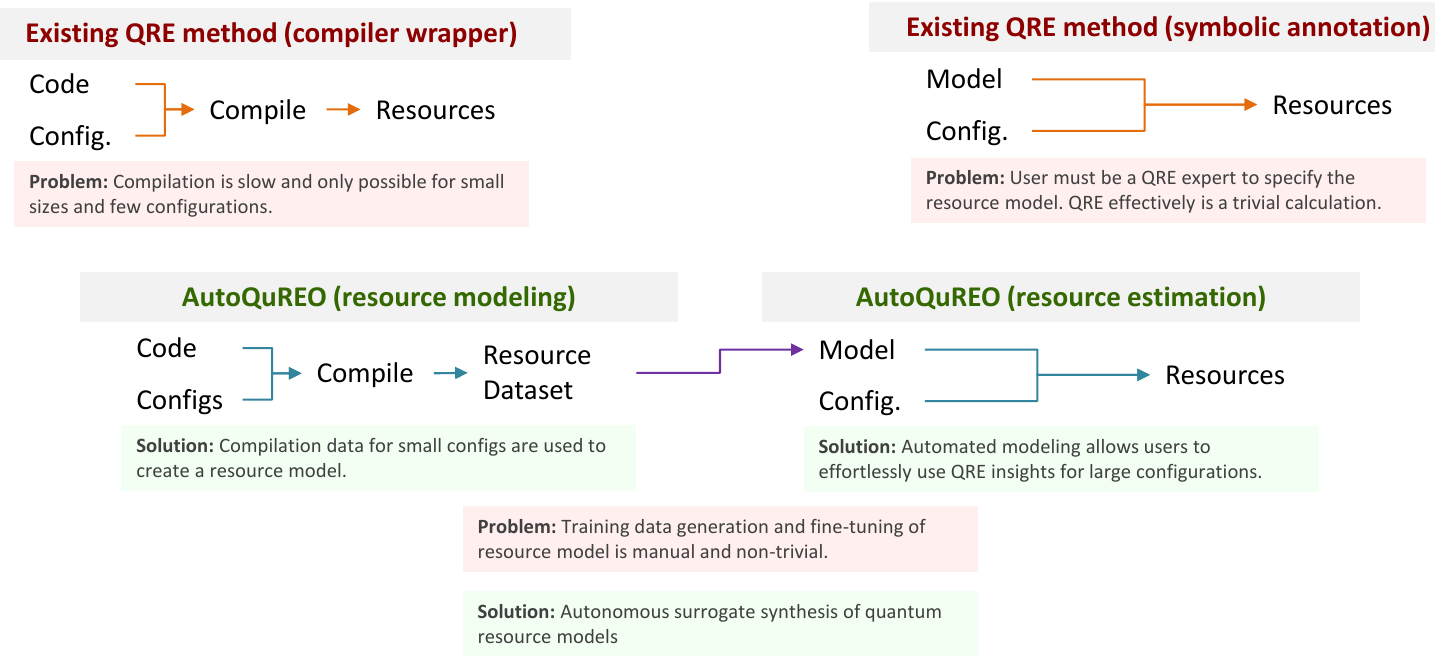}
        \caption{}
        \label{fig:qre_techniques_flow}
    \end{subfigure}
    ~
    \begin{subfigure}[b]{\textwidth}
        \centering
        \includegraphics[page=7, clip, trim=4.7cm 5cm 4.7cm 4.5cm,width=0.9\textwidth]{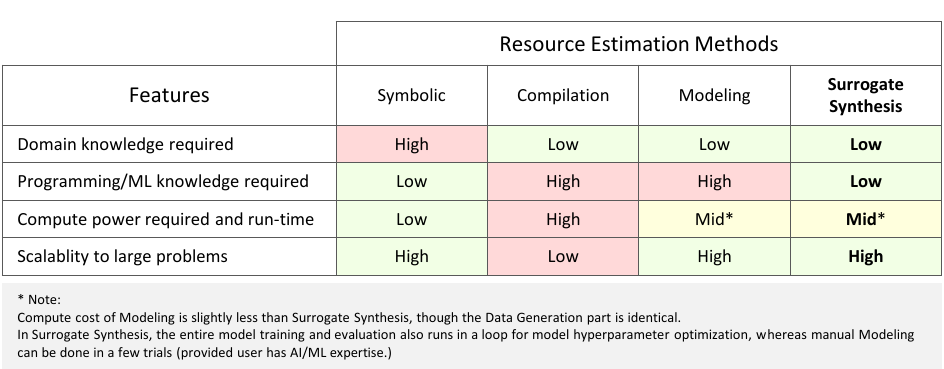}
        \caption{}
        \label{fig:qre_techniques_comparison}
    \end{subfigure}
    ~
    \begin{subfigure}[b]{\textwidth}
        \centering
        \includegraphics[page=8, clip, trim=0cm 13.3cm 5.6cm -0.3cm,width=\textwidth]
        {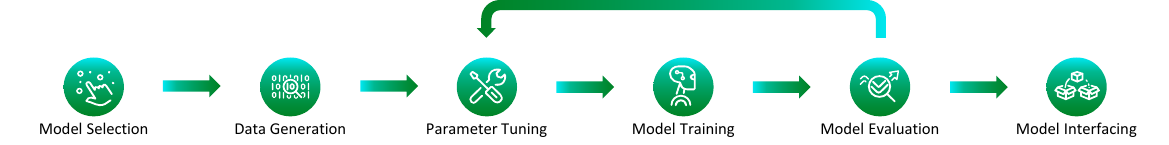}
        \caption{}
        \label{fig:surrogate_workflow}
    \end{subfigure}
    \caption{Surrogate synthesis workflow and comparison with existing QRE methods. \textbf{(a)} Comparison of QRE workflow of compilation-based, symbolic-annotation-based and the proposed modeling-based approach, highlighting trade-offs in scalability and automation. \textbf{(b)} Comparison of resource estimation features across criteria such as required expertise, computational cost, and scalability. \textbf{(c)} Surrogate synthesis pipeline including model selection, data generation, model tuning, training and evaluation, and adapter sketching. This enables AutoQuREO to perform scalable resource estimation.}
\end{figure}

Resource models, as introduced in the previous section, alleviate the scalability issue with QRE while being amenable to non-experts in quantum complexity analysis.
However, generating the training data, training the models, and evaluating them is fairly involved, often requiring multiple iterations and prior working knowledge of AI.
As illustrated in Figures~\ref{fig:qre_techniques_flow} and \ref{fig:qre_techniques_comparison}, surrogate synthesis automates and optimizes the entire QRE workflow with minimal user intervention.

The surrogate synthesis process proceeds in several stages, as shown in Figure~\ref{fig:surrogate_workflow}.
To initiate the process, the user can inspect available adapters of a layer and their corresponding cost and confidence metadata.
Based on the use case, the total QRE time for an adapter can be estimated and cumulated across the full stack.
If this total time is prohibitive, the user can decide to synthesize a new surrogate model to accelerate the QRE DSE.
The bottleneck adapter is chosen as the input adapter, and the user specifies a model class for the surrogate model.
Each adapter also stores additional metadata of the typical hyperparameter ranges it is expected to work with.
For modules, this corresponds to sizes that can be compiled tractably, while for models, it reflects the input data on which they have been trained.
This metadata of the input adapter is accessed to generate the hyperparameter configurations.
Then, the input adapter is invoked with these configurations to generate training and test data for the surrogate.
Note that in this process, the input adapter need not be a module type; for example, a symbolic model can be derived from a neural network model, or vice versa.
The surrogate model is then trained using specialized routines for each type.
It is then evaluated on the test data to update the cost and confidence metadata.
If these metrics are not within the user-set acceptable bounds, the model's hyperparameters are further tuned via Optuna~\cite{akiba2019optuna}, then retrained and reevaluated (for a set maximum number of trials).
The final surrogate model is saved in the library, and the corresponding adapter code is synthesized, maintaining the same interface as the input adapter.
This output adapter is also saved in the library for seamless addition in the QRE use cases.

\subsection{Lifelong learning}
\label{sec:autoqureo_lifelong}

AutoQuREO supports a lifelong learning mechanism~\cite{ring1997child} in which resource estimation data generated for each evaluated configuration are persistently retained and reused to improve surrogate models across the stack. 
When an adapter of module type is selected by the agent, it is treated as a source of ground-truth supervision. 
This is propagated to retrain other models of the layer if the corresponding adapter metadata indicates that lifelong learning is supported and enabled for them.
The internal process of lifelong learning is similar to surrogate synthesis, and also updates the cost and confidence metadata.

This continual model update enables AutoQuREO to progressively replace expensive compilation-based estimation with efficient approximations.
In a typical use case, if we start with a module and a model corresponding to a layer, after reaching a threshold for arbitrating the module for the QRE during lifelong learning and improving the model, the model's metadata would be favorable to the agent's scoring function.
Thereafter, not only for subsequent configuration's QRE, but also for further use of the layer (in the same project or another use case), the model can serve as a trusted and efficient resource estimator.

\subsection{Resource analysis and optimization}
\label{sec:autoqureo5}

\begin{figure}[!bht]
\centering
\includegraphics[trim=0cm 0cm 0cm 0cm,width=0.99\textwidth]{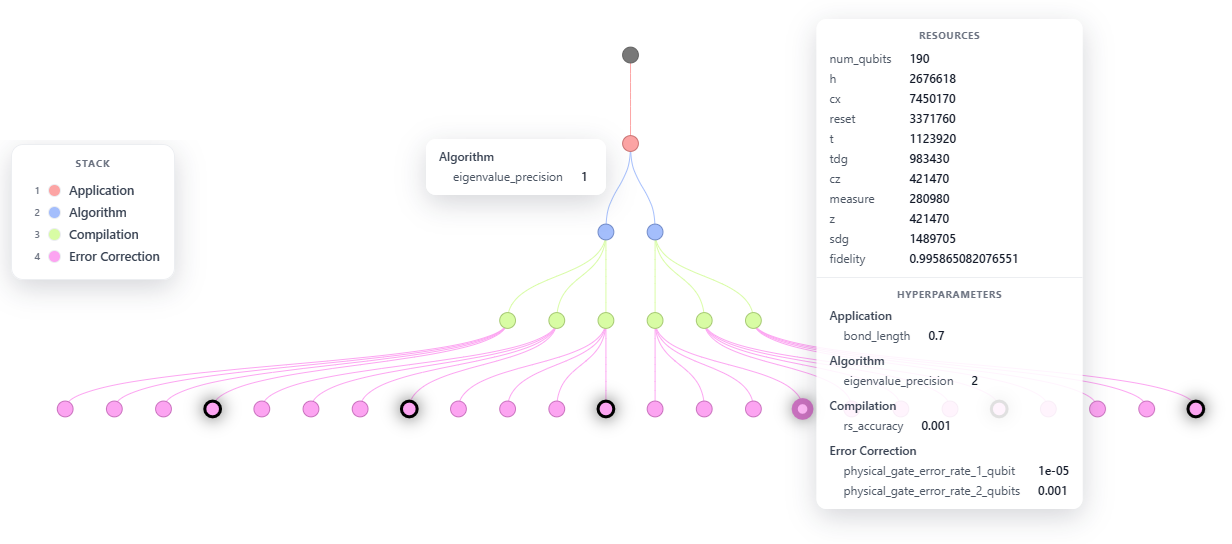}
\caption{An example resource tree representation of the design space. Each level corresponds to a stack layer, and edges represent hyperparameter choices. Nodes encode cumulative configurations and aggregated resource estimates. Leaf nodes correspond to complete system configurations, among which Pareto-optimal solutions (highlighted nodes) are identified based on user-defined objectives.}
\label{fig:resource_tree}
\end{figure}

The QRE iteratively estimates the cumulative resources for each layer based on the modules/models.
This is referred to as building a layer for the project, as explained previously in Section~\ref{sec:autoqureo_adapter}.
AutoQuREO's build process internally updates a tree data structure called the resource tree.
Each cumulative design space up to a layer of the stack corresponds to a level of the tree. 
Each edge in the resource tree corresponds to a specific hyperparameter configuration for a particular layer.
Each node of the resource tree stores the hyperparameter configurations up to that node from the root, along with the cumulative resources at the end of processing for that layer.
For example, if layer 1 has 10 configurations and layer 2 has 6 configurations, the first layer of the tree will have 10 nodes, while the second layer will have $6*10$ nodes, for each possible configuration of the entire compute stack. 
Thus, each progressive layer represents the cumulative configuration of all the layers up to that point. 
An example tree is shown in Figure~\ref{fig:resource_tree} for $4$ layers.
The layer names are annotated with colors on the left.
The optimal leaf nodes after the optimization process, as discussed shortly, are highlighted with a black border.
Each node can be inspected to view the corresponding resources and hyperparameter configurations.
The HTML version of the tree can be interactively explored for the node/edge data via a pointing device.
However, visualizing the tree becomes unruly for realistic DSE due to the combinatorial explosion. 
\addch{To keep the exploration tractable, AutoQuREO samples configurations from the specified hyperparameter ranges under user control rather than exhaustively enumerating every combination. The resources at each node are estimated by surrogate models rather than by compiling every configuration, so the design space is traversed over inexpensive model evaluations. Besides a configurable sampling control, more advanced search strategies, such as grid search or differential evolution, would be considered in future releases.}

To further analyze the configurations, AutoQuREO allows plotting any 2 or 3 quantities of interest from the full set of hyperparameters and resources of all layers.
This can be plotted for all leaf nodes or for a subset.
For example, the space (number of qubits) vs time (circuit depth) vs error (observable fidelity) plot for the leaf nodes.

\begin{figure}[htb]
\includegraphics[page=5, clip, trim=0cm 0cm 7.8cm 8.5cm,width=0.8\textwidth]{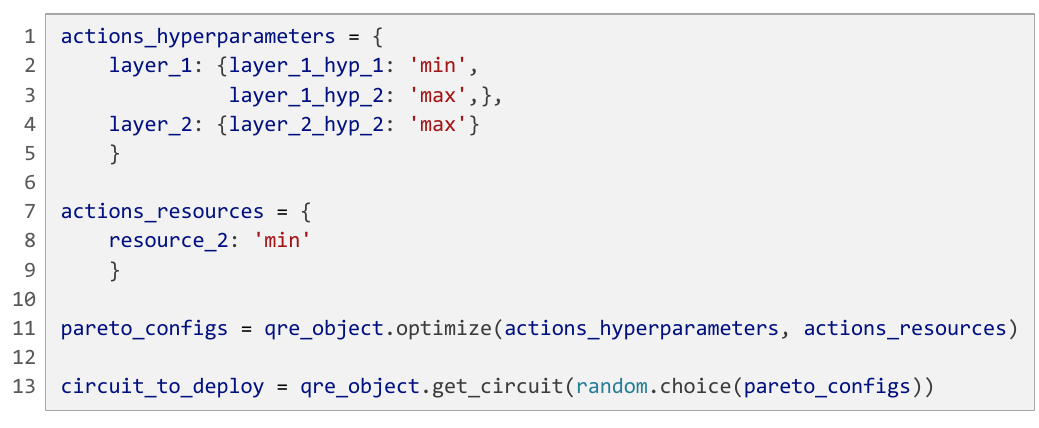}
\caption{API template for AutoQuREO resource optimization and deployment. Users define optimization objectives over hyperparameters and resource metrics. The framework performs multi-objective optimization to identify Pareto-optimal configurations, which can then be compiled into executable circuits. Estimation, optimization, and deployment are integrated within a unified interface.}
\label{fig:optimization}
\end{figure}

Resources and hyperparameters jointly define a multi-objective optimization problem. 
\addch{We emphasize that this optimization is distinct from the surrogate modeling step. Models first estimate the resources across the design space. The optimization then operates on these estimates by filtering Pareto-optimal configurations. Because the estimates come from models rather than full compilation, the design space is searched without compiling every configuration, and only the finally selected configuration is compiled for deployment.}
The optimization is guided by the user's goal in the use case.
These define the actions for each resource or hyperparameter.
Actions can be set to either minimize or maximize, as shown in Figure~\ref{fig:optimization}.
These actions guide the optimization, while the other resources and hyperparameters (e.g., layer 1 hyperparameter 1 and resource 1 are unspecified in the example) are outputs that the user wants to study.
Note that the min/max actions might not be sufficient to single out a configuration.
There might be trade-offs that form a Pareto frontier.
For example, if the goal is to minimize the number of qubits while maximizing the input problem size, the Pareto front typically filters out configurations that do not dominate at least one of the goals, keeping the other fixed.

Once optimal configurations are identified, the layers can be compiled and deployed on target backends, closing the loop between design and execution.
The framework is designed for integration with existing platforms, including support for Fujitsu's quantum SDK~\cite{kakuko2025practical} and backend, enabling seamless transition from resource estimation to hardware execution.

\subsection{Workflow of a QRE scenario}
\label{sec:autoqureo3}

To effectively use AutoQuREO, a user would typically start by defining a blueprint of the layers to add to the QC stack and the trade-offs they are interested in analyzing and optimizing. 
We refer to such a non-trivial use case of the AutoQuREO tool as a scenario.

After installing AutoQuREO, users may interact with the framework through either a Python-based application programming interface (API) or a graphical user interface (GUI). 
The scenario is used to add the corresponding layers to the QRE stack, either directly from the library if available or manually by the user.
The QRE is performed iteratively per added layer, and the user can track the progress (especially for large-scale DSE).
The final estimates can be visualized and analyzed with custom tools within AutoQuREO.
These aid the user in analyzing the design bottlenecks and addressing them in the layer definitions.
Once the user is satisfied with the optimization, the optimal configuration might (i) be the end-goal for exploratory research on projected configurations and resources (e.g., how many qubits are needed to run a problem size of interest), or, (ii) be deployed on an available hardware backend (e.g., what is the largest problem size that can be run on a hardware of interest).
Note that, though the surrogate synthesis of resource models is the core novelty, the user is effectively agnostic to the details of this involved process.

The API enables scripting of large-scale design space exploration and integration into existing SDK workflows, while the GUI supports interactive inspection of resource trade-offs, Pareto fronts, and profiling results. Internally, execution is orchestrated by a controller that manages data flow between layers, invokes appropriate estimators, and records provenance for reproducibility.

\subsection{Software architecture and execution flow}
\label{sec:autoqureo_flow}

\begin{figure}[!htb]
    \centering
    \includegraphics[width=1.0\linewidth]{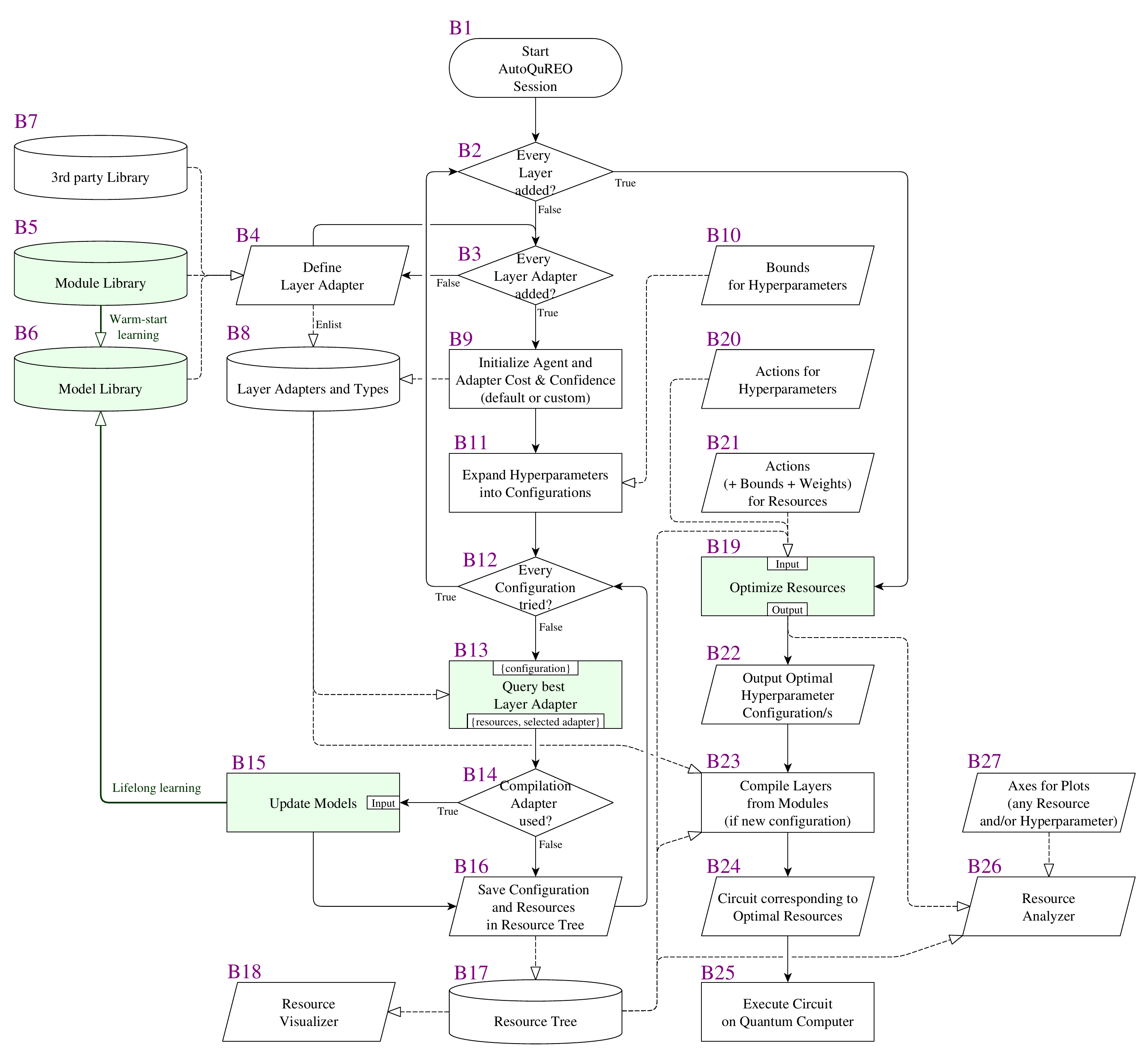}
    \caption{Software flowchart of AutoQuREO with the core novelties highlighted in green. The workflow includes layer definition  and interfacing via adapters, resource estimation for hyperparameter configuration, surrogate model training and utilization, and multi-objective optimization for Pareto-optimal deployment.}
    \label{fig:fc2}
\end{figure}

The detailed software design of AutoQuREO is depicted via the flowchart in Figure~\ref{fig:fc2}. 
Each block \textcolor{violet}{B} is numbered for ease of following the discussion. 
Figures \ref{fig:fc3}, \ref{fig:fc4}, and \ref{fig:fc5} are insets further detailing the blocks \textcolor{violet}{B13}, \textcolor{violet}{B15}, and \textcolor{violet}{B19} respectively.

\subsubsection*{Project setup}

\begin{tcolorbox}[colback=blue!10]
Here's the specification of a toy Scenario we will use to explain the software flow.
\begin{itemize}[nolistsep,noitemsep]
    \item We will consider a scenario with 2 layers, L$_1$, followed by L$_2$. 
    \item Let's assume L$_1$ has 2 tunable hyperparameters with the following values: L$_1$H$_1$ := (A,B,C) and L$_1$H$_2$ := (D,E). To estimate the resource, let's assume L$_1$ has 3 choices of adapters:
    \begin{itemize}[nolistsep,noitemsep]
        \item L$_1$A$_1$ of type quantum-compilation (available in the module library)
        \item L$_1$A$_2$ of type lookup-table (new empty model user will create and use), and 
        \item L$_1$A$_3$ of type symbolic-regression (new model user will create, warm-start, and use). 
    \end{itemize}
    \item L$_2$ has 1 tunable hyperparameter L$_2$H$_1$ := (F,G). L$_2$ has only 1 adapter.
    \begin{itemize}[nolistsep,noitemsep]
        \item L$_2$A$_1$, of type neural-network (imported from a 3rd-party library). 
    \end{itemize}
    \item Let's also assume that we are interested in 3 quantum resources at each stage, namely (R$_1$,R$_2$,R$_3$).
    \item Our interest is to specifically study the dependence of R$_2$ and L$_1$H$_2$, and what are the associated optimal values of L$_1$H$_1$, L$_2$H$_1$, R$_1$, and R$_3$ for minimizing R$_2$ while maximizing L$_1$H$_2$.
\end{itemize}
\end{tcolorbox}

The AutoQuREO session starts in \textcolor{violet}{B1} by creating a new project with a user-specified project name. 
This project can be saved and loaded later. 
The saved/loaded data includes layer information (name, hyperparameters, adapters), agents, the resource tree, and optimization actions with the corresponding optimized configurations. 
For the rest of this discussion, let's consider an empty project. 

In \textcolor{violet}{B2}, the user decides if the current project layers match the layers in the scenario blueprint. 
Currently, no/false, as we want to add the 2 layers, starting from an empty project. 
There's no point in jumping to \textcolor{violet}{B19} without adding any layer to optimize over, and it would result in an error flag. 
Let's start with adding layer L$_1$.

\subsubsection*{Adapter definition and library}

The next step is to add the layer adapters for this layer. 
As in the blueprint, we need to add 3 adapters; thus, the \textcolor{violet}{B3}-\textcolor{violet}{B4} loop would be executed 3 times. 
For the 1st time, i.e., for L$_1$A$_1$, the user selects a quantum-compilation module from the module library \textcolor{violet}{B5}. 
The adapter code (i.e., the Python code that interfaces with the library function) is also available in the library \textcolor{violet}{B5}, so the user can use it directly. 
After successful interfacing, adapter L$_1$A$_1$ gets linked to L$_1$. 
This adapter-layer map is logged in the database \textcolor{violet}{B8}. 

The flow returns to \textcolor{violet}{B4} via \textcolor{violet}{B3} now for the 2nd adapter L$_1$A$_2$. 
This is a lookup-table type estimator that the user wants in this scenario, so that whenever the same configuration needs to be estimated, it is faster to retrieve the saved results than to recompile it with L$_1$A$_1$. 
Since this is a new model, the user defines and adds it to the AutoQuREO model library \textcolor{violet}{B6}. 
The adapter L$_1$A$_2$ for the lookup-table is then added via \textcolor{violet}{B4} to \textcolor{violet}{B8} for the layer L$_1$.

We move on to the last adapter of this layer, L$_1$A$_3$. 
This time, say, the user wants to compress the resource estimate into a set of equations that map the hyperparameters to the resources. 
The user creates the new symbolic model, say, as a set of SymPy~\cite{meurer2017sympy} expressions, and defines a way to update/train the model, say, via PySR~\cite{cranmer2023interpretable} or similar tools~\cite{muthyala2025symantic,muthyala2024torchsisso}.
To initialize the model at this stage (i.e., before being used in the QRE process), the user might warm-start the model by training it on the compilation module of L$_1$A$_1$ for an input dataset of the user's choice (e.g., a set of Haar random quantum states/unitaries, or a set of benchmark quantum algorithms like MQT Bench~\cite{quetschlich2023mqt} or QARP~\cite{martinez2026pauli}.
Note, this warm-start training can also be done independently before running the AutoQuREO session, thus keeping the warm-started model already available in the model library while adding the corresponding adapter.
Warm-start learning works similarly to lifelong learning, as described later.
In either case, the adapter L$_1$A$_3$ of the symbolic model is listed in \textcolor{violet}{B8} against L$_1$.
From \textcolor{violet}{B4}, we return back to \textcolor{violet}{B3}. 
This time, we have added all three adapters as per the Scenario blueprint. 
Next, we will estimate the resources for the L$_1$ layer.

\subsubsection*{Agent-based adapter arbitration}

In \textcolor{violet}{B9}, we begin preparing for this step by initializing an agent that arbitrates between the 3 adapters in this layer. 
There are a few predefined agent policies/personas available within AutoQuREO, but users can create and fine-tune their own agents. 
The agent assigns initial cost and confidence estimates for the 3 adapters. 
Say, in our case, the cost is (0.9, 0.4, 0.3), while confidence is (0.9, 0, 0.2). 
The default agent's policy is simple: pick the adapter with the highest confidence-divided-by-cost score, i.e., (1, 0, 2/3) for this case. 
Thus, initially, the agent would pick L$_1$A$_1$ to estimate the resource for this layer. 
Note that the values the agents assigned reflect the general understanding that the compilation is costly but close to the true value. 
The user's confidence in the symbolic equations can be reflected in the initialization by manually overriding the agent's default assignment. 
The lookup-table is initially empty, thus its confidence is zero.

\subsubsection*{Configuration expansion}

After that, we move to \textcolor{violet}{B11}. 
In this step, we expand the hyperparameters to configurations. 
The hyperparameter bounds of L$_1$H$_1$:= (A,B,C) and L$_1$H$_2$:= (D,E) are invoked from the user via \textcolor{violet}{B10}. 
These are combined to create the configurations for layer L$_1$ as (AD, AE, BD, BE, CD, CE). 

\subsubsection*{Resource estimation}

\begin{figure}[p]
    \centering
    \begin{subfigure}[b]{0.5\textwidth}
        \centering
        \includegraphics[width=0.9\linewidth]{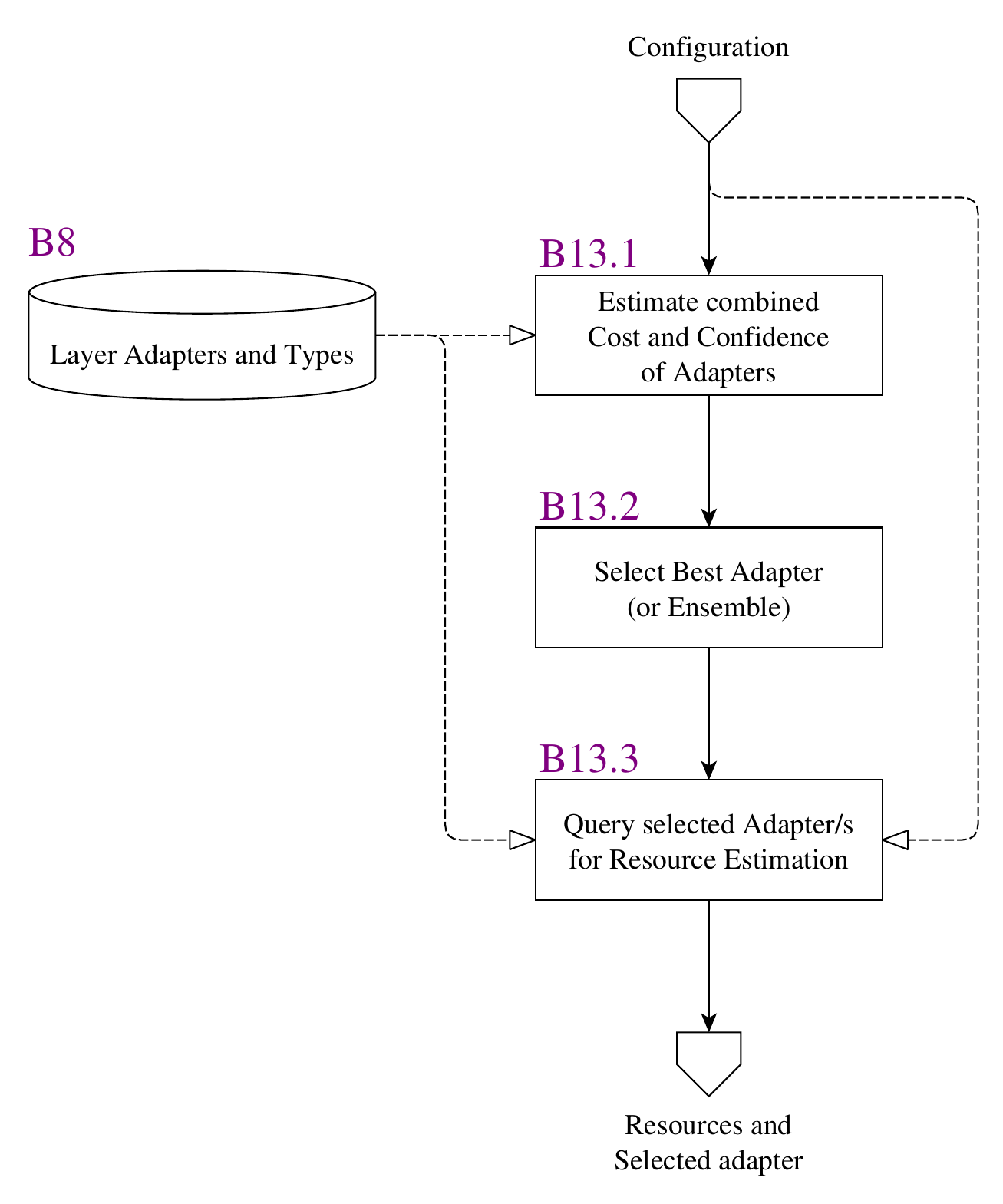}
        \caption{}
        \label{fig:fc3}
    \end{subfigure}%
    \begin{subfigure}[b]{0.5\textwidth}
        \centering
        \includegraphics[width=0.9\linewidth]{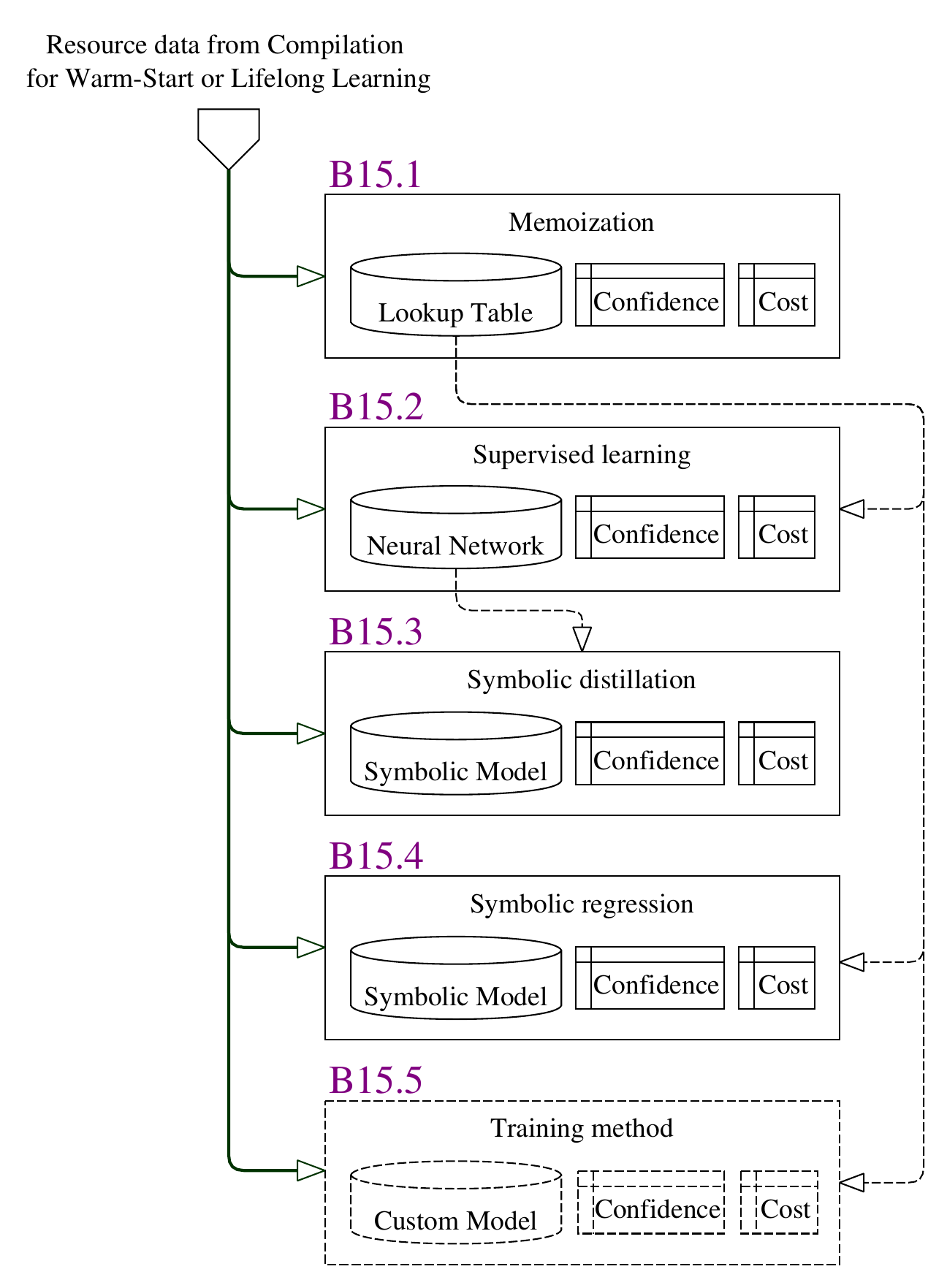}
        \caption{}
        \label{fig:fc4}
    \end{subfigure}
    ~
    \begin{subfigure}[b]{0.7\textwidth}
        \centering
        \includegraphics[width=0.9\linewidth]{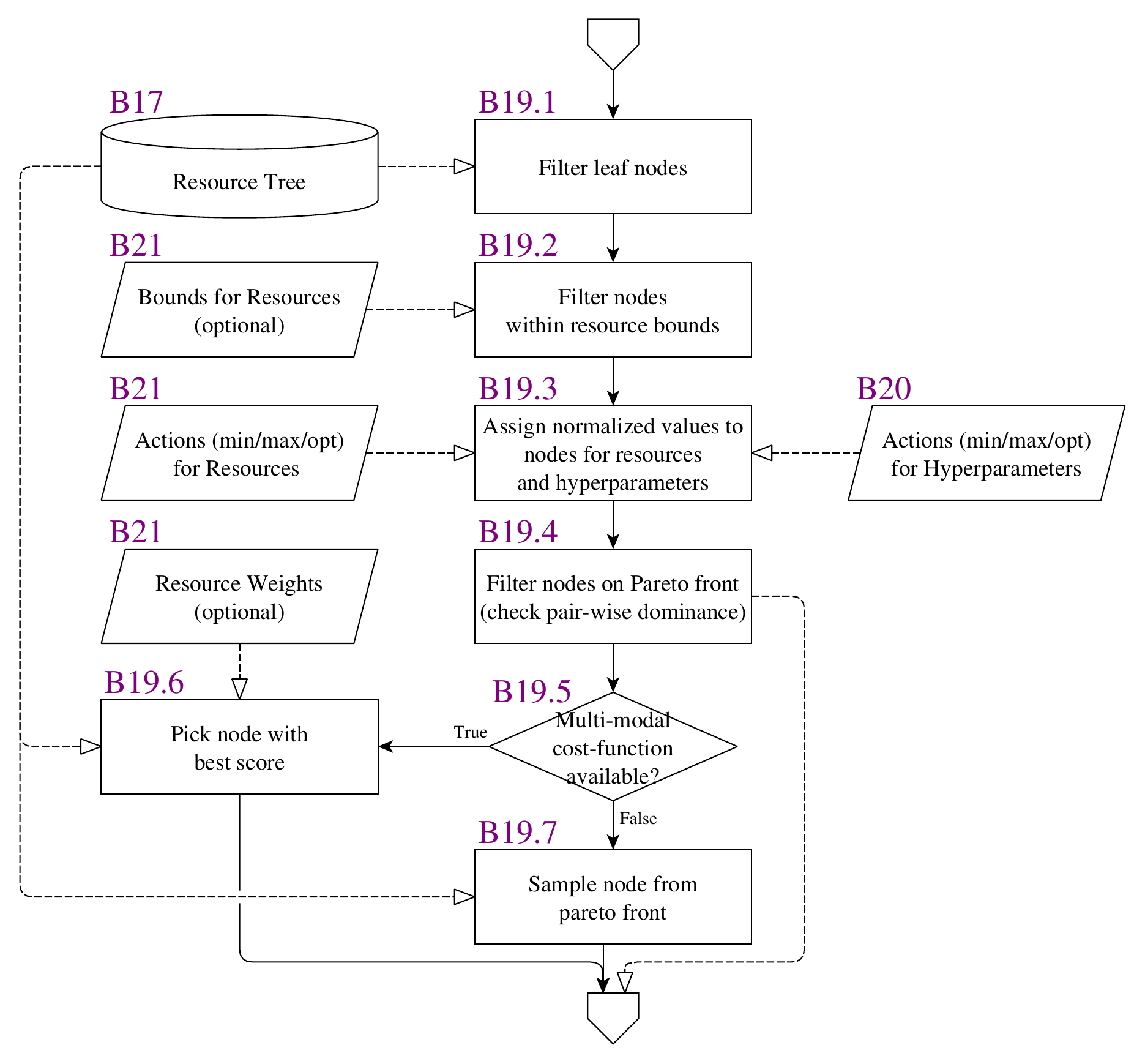}
        \caption{}
        \label{fig:fc5}
    \end{subfigure}
    \caption{Detailed views of key blocks in the flowchart. \textbf{(a)} Inset A for block \textcolor{violet}{B13} for adapter arbitration based on cost and confidence metrics. \textbf{(b)} Inset B for block \textcolor{violet}{B15}. The model creation and update follow the surrogate synthesis process described in Figure~\ref{fig:surrogate_workflow} incorporating warm-start and lifelong learning. \textbf{(c)} Inset C for block \textcolor{violet}{B19} for multi-objective Pareto filtering.}
\end{figure}

The loop from \textcolor{violet}{B12} to \textcolor{violet}{B16} estimates and logs the resources for each configuration. 
For each configuration, the agent arbitrates one of the adapters to generate a resource estimate. 
This operation of \textcolor{violet}{B13} is further expanded in inset A, in Figure~\ref{fig:fc3}. 
As discussed before, in \textcolor{violet}{B13.1}, the agent uses its policy to score the available adapters of the layer from \textcolor{violet}{B8} based on the current cost and confidence. 
The highest-scoring adapter is selected in \textcolor{violet}{B13.2} to estimate the resource. 
The score is based on the agent's policy function. 
Then, in \textcolor{violet}{B13.3}, this chosen adapter is fed with the configuration, and the resource estimate is obtained. 

\subsubsection*{Lifelong learning of surrogate models}

Back in the main flowchart of Figure~\ref{fig:fc2}, we move to \textcolor{violet}{B14}, where we check whether the adapter the agent chose is a compilation type or a model. 
In the event of the layer not having any compilation type adapter, or if it had not been chosen by the agent, we skip to block \textcolor{violet}{B16}, which stores the configuration and the corresponding resource estimate from the adapter into the resource tree in \textcolor{violet}{B17}. 

Alternatively, if the chosen adapter is of quantum-compilation type, we move to \textcolor{violet}{B15} to update any model adapters that allow lifelong learning. 
This is further explained in inset B, in Figure~\ref{fig:fc4}.
In contrast to traditional machine learning, lifelong learning refers to an AI system's ability to continuously learn and adapt to new data and tasks throughout its deployment, without needing to be retrained from scratch.
AutoQuREO implements lifelong learning based on compilation-based resource estimates for 4 typical models: memoization via lookup-table in \textcolor{violet}{B15.1}, supervised training of neural network in \textcolor{violet}{B15.2}, symbolic distillation of neural network \textcolor{violet}{B15.3}, and symbolic regression of lookup-table in \textcolor{violet}{B15.4}.
Each of these includes a specification of how the corresponding model is stored, and how it can be retrained, updating both the model and the confidence.
Note that users can build their custom models and lifelong learning routine in \textcolor{violet}{B15.5}.
Examples of such models include the Kolmogorov-Arnold network, transformer neural networks, diffusion neural networks, neuro-evolution, etc.
The model update process follows the surrogate synthesis steps described in Section~\ref{sec:autoqureo_surrogate} via Figure~\ref{fig:surrogate_workflow}.
When a model is updated, it does not affect the resource estimates of the current configuration. 
The updated, higher-confidence model metrics affect the QRE of subsequent configurations; for example, the agent might now choose a different adapter.

\subsubsection*{Resource tree and layer composition}

Back from the inset into the main flowchart of Figure~\ref{fig:fc2}, once the configuration is added in the resource tree in \textcolor{violet}{B17} via \textcolor{violet}{B16}, the flow returns to \textcolor{violet}{B12} for the next configuration.
Similarly, all configurations and their corresponding resources are stored in the resource tree.
When all the configurations have been explored, the flow returns to \textcolor{violet}{B2}.
The resource estimation of Layer L$_1$ is complete at this stage.
The resource tree can be visualized in \textcolor{violet}{B18}.

Following our scenario, we move on to layer L$_2$. 
There's a single adapter L$_2$A$_1$ of type neural-network.
The flow returns to \textcolor{violet}{B4} via \textcolor{violet}{B3}. 
This time, the user wants to use a trained neural network from a 3rd party, available as a TensorFlow model in \textcolor{violet}{B5} (alternatively, it can also be a cloud-based AI-powered transpilation~\cite{kremer2024practical} service). 
The user now has to write the adapter code corresponding to L$_2$A$_1$, that loads the neural network (or connects to the cloud via an access token-based API) and returns the resource estimates from the model. 
There's no scope to update this model on the fly. 
The adapter is enlisted again in \textcolor{violet}{B8}.
The flow returns to \textcolor{violet}{B3}, and since there are no more adapters, we move to \textcolor{violet}{B9} where the agent initializes the adapter with the default/expected cost and confidence of the neural network (which, of course, the user can override).
The single hyperparameter L$_2$H$_1$ maps to the configurations (F,G) of the layer in \textcolor{violet}{B11}.
However, note that this is the 2nd layer; thus, each configuration must be applied to all configurations already considered in the previous layer (6 in our case, for the 1 layer before it).
Thus, the full configuration becomes (ADF, ADG, AEF, AEG, BDF, BDG, BEF, BEG, CDF, CDG, CEF, CEG).
For each of these 12 configurations, the loop from \textcolor{violet}{B12} to \textcolor{violet}{B16} estimates and logs the resource in the tree, building the next layer.
Since there is only a single adapter, the agent always queries L$_2$A$_1$ in \textcolor{violet}{B13}.
Since there are no compilation-type adapters, the neural network cannot improve through lifelong learning, even if the 3rd party has that feature.
After the layer has been built, the flow returns to \textcolor{violet}{B2}.
We can again visualize the updated resource tree of \textcolor{violet}{B17} at this stage via \textcolor{violet}{B18}.
At this stage, our \textit{resource estimation} flow is complete. 

\subsubsection*{Multi-objective Pareto-optimization}

Next, we discuss the \textit{resource optimization} feature of AutoQuREO.
The optimization requires actions for each hyperparameter and resources for the current stack.
Recall our optimization goal from the scenario blueprint.
The action for the Hyperparameters in \textcolor{violet}{B20} would specify (L$_1$H$_1$:\texttt{opt}, L$_1$H$_2$:\texttt{max}) for L$_1$ and (L$_2$H$_1$:\texttt{opt}) for L$_2$, while for the Resources in \textcolor{violet}{B21}, it would be (R$_1$:\texttt{opt}, R$_2$:\texttt{min}, R$_3$:\texttt{opt}).
\textcolor{violet}{B19} optimizes as further explained in inset C, in Figure~\ref{fig:fc5}.
The leaf nodes in the resource tree of \textcolor{violet}{B17} are filtered out in \textcolor{violet}{B19.1}, as they correspond to a cumulative set of configurations and resources across all layers.
These nodes can optionally be filtered by a hard-bound on resources in \textcolor{violet}{B19.2}, specified by the user via \textcolor{violet}{B21}.
This is required as during the hyperparameter exploration, the user might not have a clear idea of how much of each resource the final prototype would need.
Many of these configurations might require resources beyond our current interest (e.g., a larger quantum computer than we currently have).
Filtering out such unreasonable settings helps declutter downstream resource analysis and optimization.
Thereafter, based on the supplied actions of hyperparameters in \textcolor{violet}{B20} and resources in \textcolor{violet}{B21}, the hyperparameters/resources with only \texttt{min} or \texttt{max} actions are filtered out in \textcolor{violet}{B19.3}.
Then, they are normalized such that the minimization action is inverted to maximization.
Recall, in our scenario, the optimization target of (R$_2$:\texttt{min}, L$_1$H$_2$:\texttt{max}) becomes (1/R$_2$:\texttt{max}, L$_1$H$_2$:\texttt{max}).
Thus, in \textcolor{violet}{B19.4}, we can jointly maximize over the selected hyperparameters/resources.
As described in the glossary, these can be multiple configurations that are Pareto-optimal for a set of actions.
The user can optionally supply an additional cost-function to weigh these actions in \textcolor{violet}{B21}. 
If that is the case, from \textcolor{violet}{B19.5} we choose the node with the best score, \textcolor{violet}{B19.6}.
Alternatively, a sample node from the Pareto front is returned in \textcolor{violet}{B19.7}.
Note that the full node information is returned, not just the optimal configuration based on the actions.
This is accessed from the resource tree in \textcolor{violet}{B17}, which contains information on all the associated hyperparameters and resources for the chosen leaf node.
The full list of Pareto nodes for the optimization scenario is also returned by \textcolor{violet}{B19.4} for analysis.
Back in the main flowchart of Figure~\ref{fig:fc2}, this optimal node is returned to the user in \textcolor{violet}{B22} as the output of the full-stack optimization.
The user can also analyze the Pareto front in \textcolor{violet}{B26} and plot subsets of the hyperparameters/resources as 2D/3D plots specified via \textcolor{violet}{B27}.

\subsubsection*{Compilation and deployment pipeline}

Once the optimal hyperparameter configuration that adheres to resource constraints and the scenario goal is obtained, we can compile the actual code for this configuration in \textcolor{violet}{B23}.
Note that if all the adapters used for the node expansion were of the quantum-compilation type, this full implementation is already available in the resource tree and only needs to be retrieved from \textcolor{violet}{B17}.
However, this is not the advocated approach, as fully compiling so many configurations to arrive at the optimal setting quickly becomes intractable.
Thus, we would likely have models that estimated resource requirements for at least some layers.
If we have used a model in any layer, we now have to compile the actual code using a quantum-compilation adapter for that layer, provided it is available (otherwise, we inform the user that a way to compile the code is missing).
The compiled quantum circuit for the optimal configuration is then output to the user in \textcolor{violet}{B24} and can be deployed on a quantum computer in \textcolor{violet}{B25}.
Blocks \textcolor{violet}{B23} to \textcolor{violet}{B25} embeds the AutoQuREO tool in a broader quantum computing platform-as-a-service.

\vspace{1em}

In summary, AutoQuREO introduces a flexible, scalable, and automated framework for full-stack quantum resource estimation and optimization. 
By combining modular stack abstractions, neuro-symbolic surrogate modeling, and integrated optimization workflows, the framework addresses key limitations of existing QRE tools. 
The following section demonstrates the practical impact of these design choices through a series of representative co-design case studies.

\addtocontents{toc}{\vspace{-1.0em}}
\section{Exemplary co-design use cases}
\label{sec:qureous}

In this section, we demonstrate how AutoQuREO can be used on representative full-stack co-design problems in quantum algorithms, quantum compilation, quantum error correction, and quantum machine learning.
For each use case, we first describe the problem motivation and related work, then define the scenario for resource estimation and optimization, including the stack layers, adapted modules and models, hyperparameters and ranges, and optimization actions.
Thereafter, we present the results obtained from our framework and interpret the non-trivial insights obtained from the experiment.
These examples are not intended as exhaustive benchmarks but rather as illustrative scenarios that highlight how full-stack, model-driven QRE can be performed systematically and scalably to gain novel, decisive insights.
These demonstrate co-design challenges across NISQ, EFTQC, and FTQC regimes, illustrating how the layer abstractions, surrogate modeling workflows, and optimization primitives introduced in Section~\ref{sec:autoqureo} enable exploration of otherwise computationally intractable design spaces.

\subsection{Scenario I: Trotterized Hamiltonian simulation and error mitigation}
\label{sec:scenario_hsim}


\begin{tcolorbox}[colback=pink!20, colframe=pink!75, boxrule=1pt, arc=2pt,
left=5pt, right=5pt, top=5pt, bottom=5pt,
title={\vspace{0.2em}\textcolor{pink!25!black}{\sffamily{{Summary}}}},
fonttitle=\bfseries, coltitle=black]

\begin{itemize}[leftmargin=*]
    \item \textbf{Novelties used:} Full-stack flexibility ($\mathcal{N}$1), library support ($\mathcal{N}$2), and automated optimization ($\mathcal{N}$4).
    
    \item \textbf{Co-design variables:} Trotterization order, Trotter step count, noise-scaling factors, noisy backend, and zero-noise extrapolation (ZNE).
    
    \item \textbf{Key insight:} Joint exploration of Trotterization and ZNE identifies regimes in which error mitigation provides benefit.
\end{itemize}

\end{tcolorbox}

\textbf{Background}

Hamiltonian simulation is a foundational quantum primitive for a wide range of applications in quantum physics and chemistry, enabling the time evolution of many-body systems that are intractable for classical methods. 
It is useful for computing dynamical properties, ground-state energies, and correlation functions, which are essential for applications in materials discovery, molecular design, and condensed-matter analysis. 
Efficient and accurate Hamiltonian simulation~\cite{daley2022practical} is widely regarded as one of the most promising pathways toward achieving practical quantum advantage, motivating this case study.
In this study, we focus on Trotterized simulation of the transverse-field Ising model (TFIM), using magnetization as the observable of interest.

Digital Hamiltonian simulation approximates the time-evolution unitary,
\begin{equation}
    U(t) = e^{-iHt},
\end{equation}
where $\hbar=1$, using a sequence of gates implementable on a quantum computer. 
A general decomposition of the time-evolution unitary cannot exploit the inherent structure of the Hamiltonian.
In most cases simulating physical system dynamics, the Hamiltonian can be written as,
\begin{equation}
    H = \sum_{\ell=1}^{L} H_\ell,
\end{equation}
where each term $H_\ell$ represents a basic interaction (such as a local coupling or a Pauli string) whose short-time evolution $e^{-iH_\ell  \Delta t}$ can be efficiently compiled. 
Hence, the total evolution time $t$ is split into $r$ equal short-time slices of duration $\Delta t = \frac{t}{r}$.
A circuit implementing the product formula $S_p(\Delta t)$ is then constructed to approximate the short-time evolution $e^{-iH\Delta t}$. 
Here, $S_p(\Delta t)$ denotes a $p$-th order product formula approximation to $e^{-iH  \Delta t}$~\cite{hatano2005finding}.
Repeating this circuit $r$ times yields an approximation to the full evolution as,
\begin{equation}
    U(t) = e^{-iHt} = (e^{-iH\Delta t})^r = (e^{-i\sum_{\ell} H_\ell\Delta t})^r \approx \big(S_p(\Delta t)\big)^r.
\end{equation}
This process is known as Trotterization, and in practice, it concerns two interdependent choices:
\begin{itemize}
    \item \textbf{Trotter steps ($r$)}: determine how many slices are used. Each slice of duration $\Delta t$, together with a single application of the short-time circuit, is called a Trotter step. Here, $r$ denotes the number of steps, and $\Delta t$ denotes the step size. Increasing the number of steps, i.e., smaller $\Delta t$, reduces the approximation error, but increases the circuit depth because the short-time circuit must be repeated more times.
    \item \textbf{Trotter order ($p$)}: determines how the single-step product formula $S_p(\Delta t)$ is constructed from the Hamiltonian terms $\{H_\ell\}$. The parameter $p$ denotes the order of the product formula and characterizes how rapidly the approximation improves as $\Delta t \to 0$. For the single-step approximation,
    \begin{equation}
        \big\|e^{-iH\Delta t} - S_p(\Delta t) \big\| = \mathcal{O}\!\left(\Delta t^{p+1}\right).
    \end{equation}
    For example, first-order Trotterization is expressed as,
    \begin{equation}
        S_1(\Delta t) = \prod_{l=1}^{L} e^{-iH_l \Delta t},
        \label{eqn:first_order_trotterization_step}
    \end{equation}
    whose single-step error scales as $\mathcal{O}(\Delta t^2)$ in Big-$\mathcal{O}$ complexity.
\end{itemize}

A higher-order means each step is a more accurate approximation to $e^{-iH\Delta t}$, so one can often use fewer steps (larger $\Delta t$) to reach a fixed target error, but at the cost of a more complicated circuit for each step, and hence more circuit depth.

The global error of this algorithm typically scales as,
\begin{equation}
    \big\| U(t) - \big(S_p(\Delta t)\big)^r \big\|
    = \mathcal{O}\!\left(r\,\Delta t^{p+1}\right)
    = \mathcal{O}\!\left(\frac{t^{p+1}}{r^{p}}\right),
\end{equation}
up to problem-dependent constants determined by the Hamiltonian terms $\{H_\ell\}$.
Hence, when running this algorithm on noisy quantum devices, a trade-off is observed.
Using a higher-order product formula or increasing the number of Trotter steps reduces the algorithmic error, but typically increases the circuit depth. 
The deeper circuit is susceptible to noise, thereby increasing the physical error. 
Let $\langle O\rangle$ denote the expectation value of the observable of interest. 
The algorithmic and physical errors contribute to the overall error in the estimated observable, which can be bounded as,
\begin{equation}
\Delta\langle O\rangle_{\mathrm{total}}
\lesssim
\Delta\langle O\rangle_{\mathrm{alg}}
+
\Delta\langle O\rangle_{\mathrm{phys}}.
\label{eq:total_error}
\end{equation}

This physical error can be reduced in postprocessing using quantum error mitigation (QEM) techniques, such as zero-noise extrapolation (ZNE)~\cite{giurgica2020digital}. 
ZNE estimates the zero-noise expectation value by executing the circuit at multiple noise levels and extrapolating the results to the zero-noise limit.
A common approach to generate noise-scaled circuits is unitary folding, which maps $U \rightarrow  U(U^{\dagger} U)^n$ for some integer $n$. Since  $U^{\dagger} U = I$, the ideal unitary is preserved while the noise is amplified.
This transformation can be applied globally or locally, i.e., $U$ may represent either the entire circuit or a subset of gates within the circuit.
The results obtained from these noise-scaled circuits are then extrapolated using a suitable technique.
Let $\lambda$ denote the noise-scaling factor, and infers the noise-scaled expectation value as $\langle O(\lambda)\rangle \approx f(\lambda)$, where $f(\lambda)$ is an extrapolation technique, such as linear, polynomial, or exponential function.
For example, a polynomial extrapolation technique can be written as:
\begin{equation}
    f(\lambda) = a_1 + a_2\lambda + a_3\lambda^2 + \dots +a_m\lambda^{m-1}.
\end{equation}
Here, $m$ denotes the number of coefficients in the polynomial function, corresponding to the polynomial of degree $m-1$.
The function is then fitted to the noise-scaled expectation values of the observable $O$, and the zero-noise estimate is obtained as $\langle O(0) \rangle  \approx \tilde{f}(0)$, where $\tilde{f}$ denotes the fitted function.
The ability to jointly optimize algorithmic and error mitigation hyperparameters while simultaneously monitoring performance metrics relies on the resource-action and Pareto analysis engine discussed in Section~\ref{sec:autoqureo5}.

Choosing the Trotter order and number of steps in noisy settings is non-trivial, as increasing either parameter simultaneously reduces discretization error while increasing circuit depth. 
In realistic quantum devices, this depth amplification exacerbates decoherence and gate noise, such that beyond a certain point, higher-order formulas or finer step sizes degrade overall fidelity rather than improve it. 
This implies an optimal regime in which the Trotterization error and the noise-induced error are balanced. 
Furthermore, ZNE can partially suppress noise effects in specific regimes, thereby shifting the optimal choice of order and steps.
This type of multi-layer co-design scenario directly motivates the flexible stack abstraction of AutoQuREO introduced in Section~\ref{sec:autoqureo_layer}, where platform constraints on Hamiltonian simulation can be composed within a unified resource estimation and optimization workflow.
Understanding the interplay among Trotter order, steps, the ZNE strategy, and hardware noise is a central objective of this study.

\textbf{Previous work}

Previous work on Trotterized Hamiltonian simulation has examined the balance between algorithmic Trotter error and physical noise using analytical arguments, direct simulation, and hardware benchmarks. 
In \cite{avtandilyan2024optimal}, an analytic study of TFIM and XY simulation under a gate-error model compared first, third, and higher-order Suzuki formulas and found that higher-order only becomes advantageous when gate errors are sufficiently small, with a critical scale in the range $10^{-4}$ and $10^{-3}$. 
Ref.~\cite{xu2025exponentially} modeled noisy Trotter evolution with local depolarizing noise and studied state trace distance for TFIM under second-order Trotterization, concluding that an optimal step count still exists but that a state-dependent treatment gives a less pessimistic estimate of the trade-off. 
In \cite{chatterjee2025comprehensive}, Trotterized Hamiltonian simulation was benchmarked on Qiskit Aer and BlueQubit simulators, as well as IBM hardware, using first-order Trotter circuits with $t=1$ and $5$ steps, and the measured quantity was a hardware and algorithmic fidelity comparison of the output distributions.
It concluded that the observed performance is governed by the trade-off between circuit depth and Trotter approximation error, with hardware noise generally being more important than the isolated Trotter error. 
Finally, \cite{siekierski2025software} used scalable benchmark circuits built from Hamiltonian simulation to separate algorithmic and hardware effects, evaluating process fidelity for first- and second-order Trotter circuits compiled and run on IBM systems and Qiskit Aer simulations.
They showed that deeper Trotterization reduces approximation error but increases noise sensitivity, with a second-order decomposition at three steps giving the best balance in their Heisenberg example. 

Building on these studies, we extend the analysis of the Trotter-noise trade-off in several practically relevant ways. 
Most importantly, we incorporate an additional error-mitigation layer into the optimization and study the trade-off directly through observable error, rather than treating fidelity-based quantities as the main figure of merit.
We also examine the optimal regime by varying both the Trotter order and the number of Trotter steps within a single framework, providing a mitigation-aware, observable-focused extension of previous analytical and benchmarking studies.

\textbf{Setup}

In this case study, we use AutoQuREO to generate both mitigated and unmitigated Hamiltonian simulation benchmarks to examine the trade-off between algorithmic error and physical noise, as discussed above, and to determine how the accuracy of Hamiltonian simulation can be maximized across different depolarizing error rates. 
We quantify this trade-off by estimating the error in a chosen observable, using the corresponding classically computed value obtained from QuTiP~\cite{johansson2012qutip} as the reference baseline.
The Hamiltonian considered in this analysis is a 1D TFIM chain with periodic boundary conditions (PBC), which is analytically solvable, defined as:
\begin{equation}
H = -J \sum_{i=1}^{N} Z_i Z_{i+1} - h\sum_{i=1}^{N}X_i
\hfill \rlap{\qquad (PBC: $Z_{N+1}=Z_1$)}
\end{equation}
and the observable of interest is longitudinal magnetization (i.e., along $Z$ axis):
\begin{equation}
    M_z = \sum_{i=1}^{N} Z_i
\end{equation}

We considered a $4$-qubit TFIM and, for fixed values of the Hamiltonian parameters, evolution time, and initial state, performed simulations over a range of Trotter orders and step counts at different depolarizing error rates, considering both unmitigated and mitigated cases. 
The noise model was defined on the $\{RZZ, RX, H, Measure\}$ gate set, with depolarizing noise applied to both single- and two-qubit gates using the same depolarizing error parameter. 
The simulation is performed on Qiskit Aer~\cite{qiskit2024}.
Table \ref{tab:hsim_stack} shows the stack along with hyperparameters, corresponding ranges, and actions. 

\begin{table}[thb]
\caption{Layers, hyperparameters, ranges, and resources for the quantum resource estimation scenario. The optimization minimizes the observable error while jointly analyzing the Trotter order, Trotter steps, depolarizing error rate, and error mitigation strategy.}
\centering

\begin{minipage}{0.73\textwidth}
\centering

\small

\begin{tabular}{|>{\centering\arraybackslash}m{3.5cm}|
                  >{\centering\arraybackslash}m{3.5cm}|
                  >{\centering\arraybackslash}m{4.0cm}|}

    \hline
    \rowcolor{lavender}
    \textbf{Layers} & \textbf{Hyperparameters} & \textbf{Ranges}\\
    \hline
    \multirow{4}{*}{\makecell{System definition}}
     & Transverse field (h) & [5.0] \\ \cline{2-3}
     & Coupling strength (J) & [10.0] \\ \cline{2-3}
     & Evolution time (t) & [10.0] \\ \cline{2-3}
     & Initial state & [$\ket{00...0}$]\\ \hline

    \multirow{2}{*}{\makecell{Algorithm compilation}}
     & Trotter order & [1, 2, 4] \\ \cline{2-3}
     & Trotter steps & [200, 250, 300] \\ \hline

     \multirow{2}{*}{\makecell{Circuit formation}}
     & Shots & [20000] \\  \cline{2-3}
     & Depolarizing error rate & [$10^{-5}$, $10^{-4}$, $10^{-3}$, $10^{-2}$]\\ \hline

    Error mitigation
     & Strategy & [None, ZNE Richardson] (noise scale factors: [1,3,5])\\ \hline
\end{tabular}

\end{minipage}%
\hfill
\begin{minipage}{0.27\textwidth}
\centering
\begin{tabular}{|>{\centering\arraybackslash}m{3.5cm}|}
\hline
\rowcolor{pink}
\textbf{Resources} \\
\hline
Number of qubits \\
\hline
Observable error  \\
\hline
Depth \\
\hline
\end{tabular}
\end{minipage}
\label{tab:hsim_stack}
\end{table}

\begin{figure}[htb]
    \centering
        \includegraphics[width=0.55\linewidth]{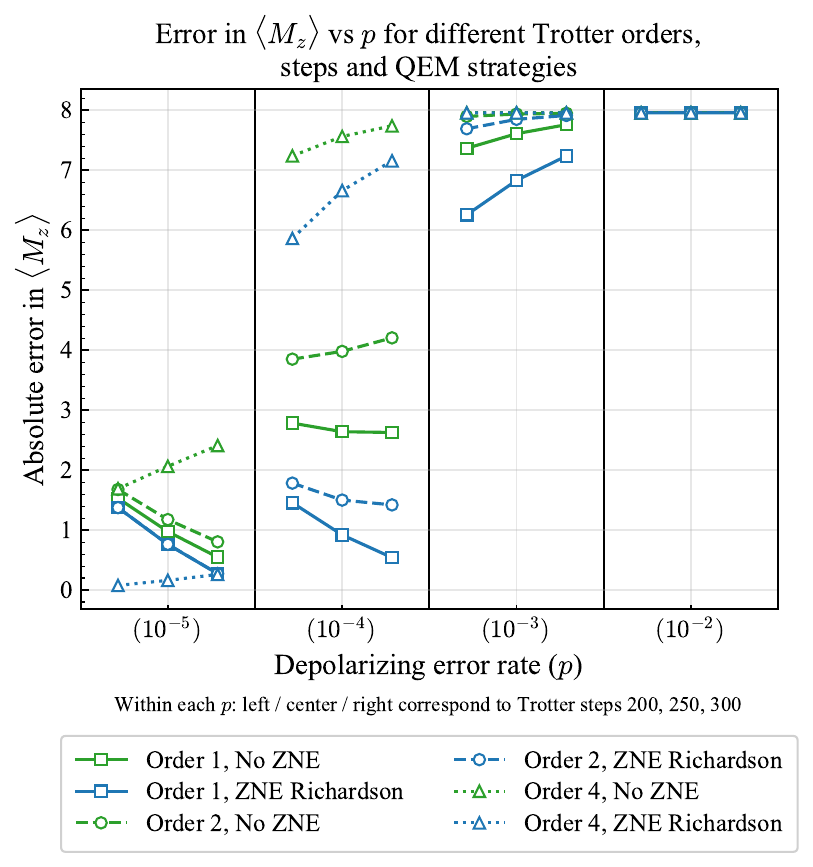}
        \caption{The absolute error in expectation value of magnetization $\langle M_z \rangle$ (for 20000 shots) for different settings observed for different depolarizing error rates in our noise model. The absolute error is calculated relative to the QuTiP result. For each point, the xticks signify Trotter steps, the marker shapes signify the Trotter order, and the color identifies the error mitigation scheme. Note: The x-axis entries (e.g., $10^{-5}$, $10^{-4}$, $10^{-3}$, $10^{-2}$) are categorical labels.}
        
        \label{fig:hsim_accuracy_vs_deperror}
\end{figure}

\begin{figure}[!hbt]
        \centering
        \includegraphics[width=0.5\linewidth]{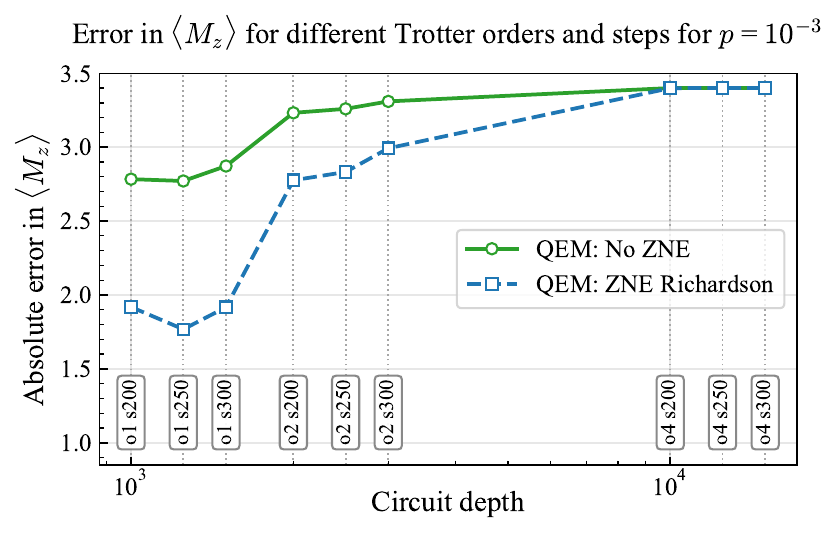}
        \caption{The variation of absolute error in expectation value of magnetization with the resulting circuit depth due to different Trotter settings for the mitigated and unmitigated cases for $10^{-3}$ depolarizing error rate. Note: `o$n$ s$m$' corresponds to Trotter order $n$ and steps $m$.}
        \label{fig:hsim_accuracy_vs_cirdepth}
\end{figure}

\textbf{Results}

From the analysis presented in Figure~\ref{fig:hsim_accuracy_vs_deperror}, we observe that even at very low depolarizing error rates (e.g., $1 \times 10^{-5}$), trivially increasing the Trotter order or the number of Trotter steps does not guarantee improved accuracy. 
This indicates that the effect of hardware noise needs to be considered as well. 
Further, we find that for every depolarizing error rate, there exists an optimal Trotter configuration that balances the algorithmic error $\Delta O_{alg}$ and physical error $\Delta O_{phys}$. 
Beyond this optimum, the deeper circuits required for higher-order or higher-step Trotterization accumulate noise faster than $\Delta O_{alg}$ is reduced, leading to an increase in the total error. 
Additionally, the use case demonstrates that ZNE's effectiveness depends strongly on the accumulated circuit error.
When the underlying noise becomes too high, scaling the noise further (a part of the ZNE procedure) deteriorates the extrapolated estimates rather than improving them. 
This trend is visible in Figure~\ref{fig:hsim_accuracy_vs_cirdepth}, where the benefit of ZNE diminishes as circuit depth grows. 
At higher depolarizing noise levels, as shown in Figure~\ref{fig:hsim_accuracy_vs_deperror}, essentially all Trotter settings produce circuits deep enough that accumulated noise dominates, causing both no-QEM and ZNE-corrected results to degrade. 
Together, these observations highlight the importance of co-optimizing algorithmic parameters and error-mitigation strategies, rather than treating them as independent implementation parameters.

In this use case, AutoQuREO enabled streamlining multi-objective exploration of hyperparameters for Trotter order, step count, error mitigation strategies, and error rates, as introduced in Section~\ref{sec:autoqureo5}.
The resulting Pareto fronts reveal regimes in which mitigation compensates for lower-order Trotterization, and hence can serve as guiding principles for the utility of QEM based on circuit depth. 
The layered adapter approach is modular and can thus be easily reused and extended to other use cases that share these layers. 

\textbf{Future work}

A natural extension of this co-design study is to move beyond closed-system Trotterized Hamiltonian simulation and apply it to a broader class of quantum simulations where these trade-offs can be further explored. 
Of immediate interest is open-system simulation governed by Lindblad dynamics~\cite{lindblad1976generators} via collision models~\cite{di2024efficient}. 
For both Hamiltonian and Lindbladian simulations, the study can be further extended to more sophisticated formulations, such as interpolated simulation schemes~\cite{rendon2024improved,mohammadipour2025reducing}, which offer additional flexibility in balancing algorithmic and physical errors and may better couple with QEM strategies. 
Beyond these, the framework can also be used to explore similar trade-offs in other simulation methods, such as qDRIFT~\cite{campbell2018random} and quantum-trajectory-based approaches~\cite{daley2014quantum}.
Furthermore, QEC can be combined with QEM~\cite{suzuki2022quantum,wahl2023zero} and with gate decomposition using the layers developed in the subsequent sections.

In this use case, we leveraged resource estimation via compilation to demonstrate AutoQuREO's features within an optimization workflow.
As discussed in Section~\ref{sec:autoqureo_model}, synthesizing and integrating surrogate resource models would circumvent expensive circuit-level evaluations by progressively replacing learned symbolic or neural estimators.
This would allow scalable predictive optimization of the use case.
The next scenario will elucidate this advantage.

\subsection{Scenario II}
\label{sec:scenario_dcmp}


\begin{tcolorbox}[colback=pink!20, colframe=pink!75, boxrule=1pt, arc=2pt,
left=5pt, right=5pt, top=5pt, bottom=5pt,
title={\vspace{0.2em}\textcolor{pink!25!black}{\sffamily{{Summary}}}},
fonttitle=\bfseries, coltitle=black]

\begin{itemize}[leftmargin=*]
    \item \textbf{Novelties used:} Full-stack flexibility ($\mathcal{N}$1), library support ($\mathcal{N}$2), surrogate synthesis ($\mathcal{N}$3), and automated optimization ($\mathcal{N}$4).
    
    \item \textbf{Co-design variables:} Molecular Hamiltonian size, eigenvalue precision of iterative quantum phase estimation (iQPE), Trotter steps, gate decomposition accuracy (GridSynth $\epsilon$), quantum error correction scheme (Steane with Reed-Muller T-teleportation, surface code), hardware noise.

    
    \item \textbf{Key insights:} 
    
    \begin{itemize}[leftmargin=*]
        \item Symbolic LER and PER models for the Steane code gates are characterized across all-to-all and square topologies, exposing gate-wise thresholds and a routing-induced threshold window.
        \item Optimization of single-qubit GridSynth decomposition fidelity is achieved by balancing synthesis accuracy against gate noise. We obtain a symbolic model matching the analytical derivation.
        \item Layer-wise neuro-symbolic surrogate models compose into an iQPE stack for the Hydrazine molecule with Trotterization and GridSynth, enabling scalable resource estimation for problem sizes beyond the reach of direct compilation.
        \item Propagating the optimal logical stack to quantum error correction codes yields compounding reduction in resources.
    \end{itemize}
    
\end{itemize}

\end{tcolorbox}

Given the inherent complexity of analyzing systems where algorithmic structure, gate decomposition, error correction, and hardware constraints interact, we present this case in three parts, each building on the outcomes of the previous, thereby highlighting the composability of surrogate resource models within the framework, as illustrated in Section~\ref{sec:autoqureo_flow}.

\subsubsection{Universal error correction and connectivity topology}
\label{sec:scenario_qec}

\textbf{Background}

Small quantum error correction (QEC) codes, such as the Steane code~\cite{steane1996multiple} and Reed-Muller constructions~\cite{daguerre2025code}, play a central role in early fault-tolerant architectures. 
The effective performance of QEC depends critically on the chosen gate set, assumed noise model, and resulting pseudo-threshold. 
While such considerations are often abstracted away in theoretical analyses, they become decisive in EFT regimes. 
Connectivity topology further complicates the picture. 
While many theoretical analyses assume all-to-all connectivity, realistic devices with thousands of qubits are constrained to sparse topologies. 
In EFT regimes, routing overheads can dominate error budgets and depth, necessitating explicit modeling.
AutoQuREO explicitly represents these dependencies within its stack abstraction.

\textbf{Setup}

In this study, we measure logical error rates (LER) and physical error rates (PER) under representative noise models and symbolically annotate their dependence on code parameters and topology.
This enables us to map any PER to LER and infer the pseudo-threshold analytically.
Routing is performed using SABRE \cite{li2019tackling,zou2024lightsabre}, allowing comparison between routed and non-routed circuits. 
The resulting rates are incorporated into surrogate models that capture the interaction between topology, routing, and error correction.

\begin{figure}[htb]
    \centering
    \includegraphics[clip, trim=0cm 0cm 4.7cm 1.9cm, width=1\linewidth]{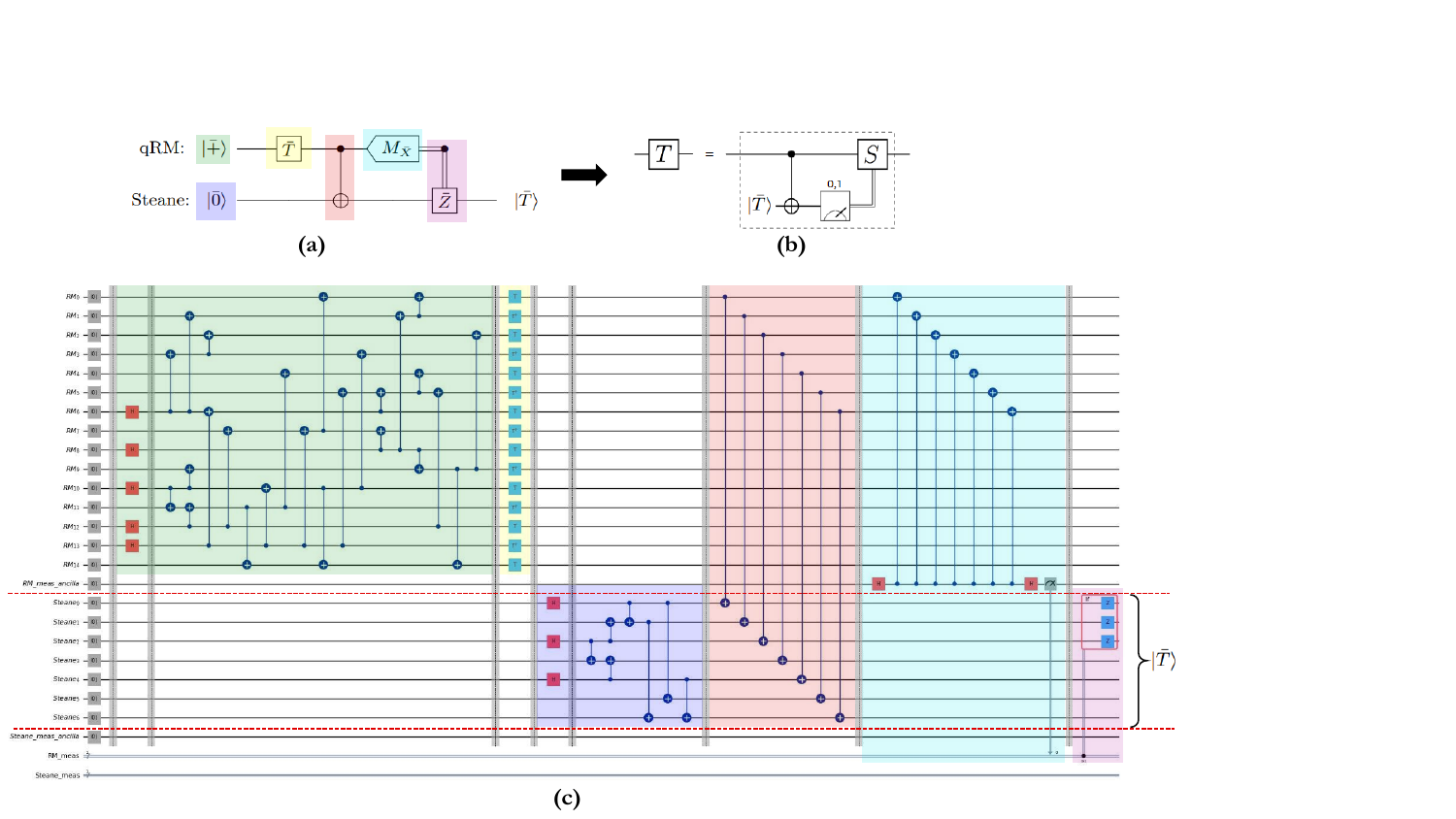}
    \caption{Universal error correction scheme using Steane code and T-teleportation from Reed-Muller~(qRM) code. (a) Logical circuit for $\ket{\overline{T}}$ preparation using Reed-Muller code. (b) Circuit to implement logical T by consuming the prepared state. (c) Physical circuit corresponding to logical circuit in (a).}    
    \label{fig:s1_t_gate_circuit}
\end{figure}

The $[[7,1,3]]$ Steane code encodes a single logical qubit into seven physical qubits and can correct any arbitrary single-qubit error (i.e., a code distance of $3$). 
Logical Clifford gates $\{H, S, CX\}$ are implemented transversely within this code. 
To achieve universality, the $T$ gate can be implemented using the technique described in \cite{daguerre2025code, daguerre2025experimental}. 
Figure~\ref{fig:s1_t_gate_circuit}a illustrates the logical circuit for magic state ($\ket{\overline{T}}$) preparation encoded in Steane code, while Figure~\ref{fig:s1_t_gate_circuit}c shows the corresponding physical-level implementation of the logical circuit. 
Once prepared, the magic state is teleported using the gadget shown in the Figure~\ref{fig:s1_t_gate_circuit}b.

\begin{table}
\caption{Stack table for characterization of LER and PER for Clifford gates in Steane code.}
\centering

\begin{minipage}{1.0\textwidth}
\centering

\small

\begin{tabular}{|>{\centering\arraybackslash}m{4.0cm}|
                  >{\centering\arraybackslash}m{4.5cm}|
                  >{\centering\arraybackslash}m{4.0cm}|}

    \hline
    \rowcolor{lavender}
    \textbf{Layers} & \textbf{Hyperparameters} & \textbf{Ranges}\\
    \hline
    Characterization Circuits & Gates & [H, S, CX] \\ 

     \hline

     \multirow{3}{*}{Simulation}
     & Depolarizing error rate ($p$) & $[10^{-5}, ... ,10^{0}]$ \\  \cline{2-3}
     & Connectivity & \text{[all-to-all, square]}\\  \cline{2-3}
     & Shots & [10000] \\
     \hline

    LER/PER calculation & - & -\\ \hline
\end{tabular}

\end{minipage}%
\label{tab:qec_stack_table}
\end{table}

\begin{figure}[htb]
    \centering
    \includegraphics[clip, trim=0cm 0cm 4.5cm 11.5cm, width=0.99\textwidth]{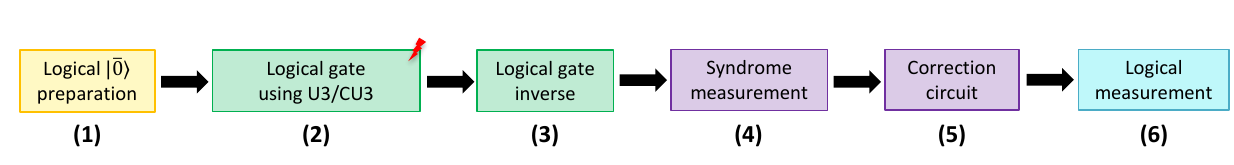}
    \caption{Flowchart for building a characterization circuit corresponding to the gates (H, S, CX, T). In step (1), Steane-encoded logical $\ket{\overline{0}}$ is prepared without any errors. In step (2), we apply a noisy version of the gate to be characterized using the corresponding $U3/CU3$ implementation to selectively apply the noise ($U3/CU3$ representations for target gates are given in Equations~\ref{eqn:u3_matrix} and \ref{eqn:cu3_matrix}). In step (3), a logical inverse gate is applied in the original operator form. For example, $H$ gate after applying $U3$ corresponding to $H$ in step (2). In step (4), the syndrome is extracted via the syndrome measurement circuit. In step (5), the error correction circuit inverts the detected error during syndrome measurement in step (4). In step (6), logical measurement is performed, and the data is stored for further postprocessing. Note that the noisy physical gates are applied only in steps (2), and everything else is assumed noiseless.} 
    \label{fig:s1_characterization_flowchart}
\end{figure}

\begin{figure}[htb]
    \centering
    \includegraphics[clip, trim=0cm 0cm 2cm 0cm, width=0.99\textwidth]{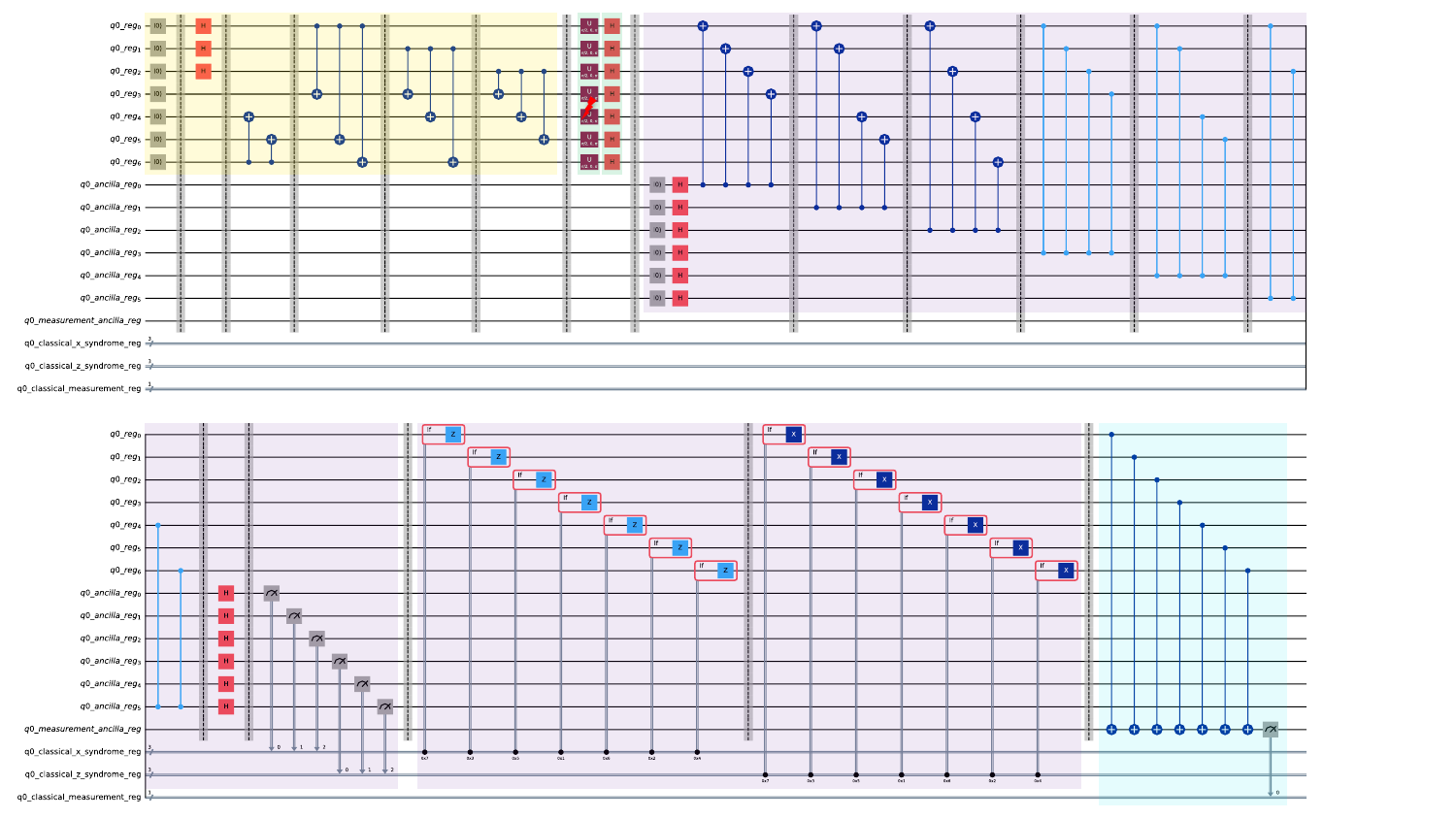}
    \caption{Physical-level circuit for characterizing the LER for $H$ gate in the Steane code, constructed via Figure~\ref{fig:s1_characterization_flowchart}: encoded $\ket{\overline{0}}$ preparation, a noisy $H$ (applied as $U3$) and its logical inverse, syndrome measurement, error correction, and logical measurement, with only the $H$ gate made noisy.}  
    \label{fig:s1_H_physical_circuit}
\end{figure}

Table \ref{tab:qec_stack_table} shows the layers and hyperparameters for this study. In the first layer, we prepare both logical and physical circuits for characterizing the PER and LER, respectively, for all the gates $\{H, S, CX\}$. Figure \ref{fig:s1_characterization_flowchart} illustrates the recipe for constructing the PER characterization circuits. The corresponding logical circuits are obtained by omitting steps (5) and (6) in Figure~\ref{fig:s1_characterization_flowchart}. An example of a PER characterization circuit for $H$ gate is shown in Figure~\ref{fig:s1_H_physical_circuit}. The prepared circuits are then simulated on a noisy simulator for various gate depolarizing error rates ($p$) and qubit connectivity topologies (all-to-all and square). In the last layer, the PER and LER are computed as the fraction of trials in which the expected output states are successfully measured.

\begin{align}
    U3(\theta,\phi,\lambda) =
    e^{i \frac{\phi + \lambda}{2}} R_Z(\phi) R_Y(\theta) R_Z(\lambda)  
    =
        \begin{bmatrix}
        \cos\left(\frac{\theta}{2}\right) &
        - e^{i\lambda}\sin\left(\frac{\theta}{2}\right) \\[6pt]
        e^{i\phi}\sin\left(\frac{\theta}{2}\right) &
        e^{i(\phi+\lambda)}\cos\left(\frac{\theta}{2}\right)
        \end{bmatrix}
    \label{eqn:u3_matrix}
\end{align}

\begin{align}
    CU3(\theta,\phi,\lambda) =
        \ket{0} \bra{0} \otimes I + \ket{1} \bra{1} \otimes U3(\theta,\phi,\lambda)
    \label{eqn:cu3_matrix}
\end{align}

\begin{align*}
    U3 \left( \frac{\pi}{2}, 0, \pi \right) = H, \quad U3 \left( 0, 0, \frac{\pi}{2} \right) = S, \quad U3 \left( 0, 0, \frac{\pi}{4} \right) = T, \quad CU3 \left( \pi, 0, \pi \right) = CX
\end{align*}

\textbf{Results}

\begin{figure}[htb]
    \centering
    
    \begin{subfigure}[t]{0.48\linewidth}
        \centering
        \includegraphics[clip, width=\linewidth]{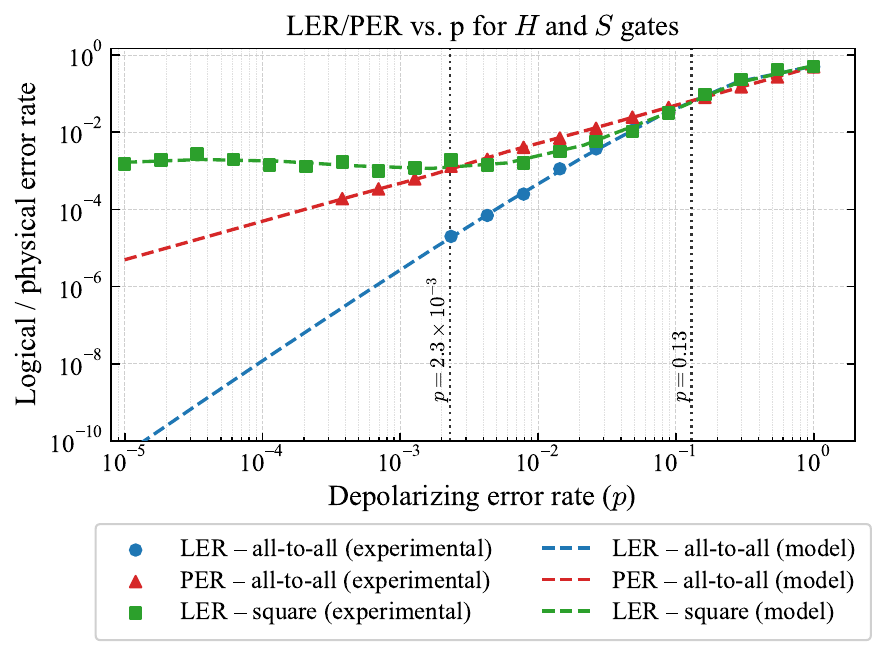}
        \caption{}
        \label{fig:s1_LER_PER_data_and_model_H_S_gates}
    \end{subfigure}
    \hfill
    \begin{subfigure}[t]{0.48\linewidth}
        \centering
        \includegraphics[clip, width=\linewidth]{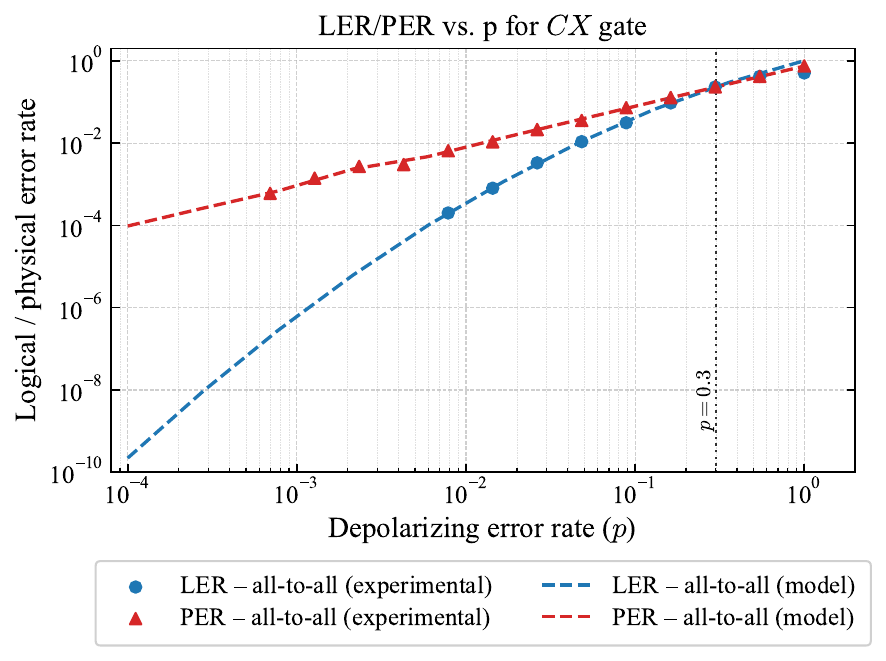}
        \caption{}
        \label{fig:s1_LER_PER_data_and_model_CX_gate}
    \end{subfigure}
    
    \caption{Experimental characterization data and the corresponding PySR models for $H, S$ and $CX$ gates. \textbf{(a)} Experimental data and model of $H$ and $S$ gates for all-to-all and square topology. For all-to-all connectivity, LER becomes lower than PER for $p = 1.3 \times 10^{-1}$. For square topology, \textit{SWAP} gates are injected, and the error rate for \textit{SWAP} gates was set to $10^{-4}$. The green curve corresponding to square topology shows a \textit{threshold window} from $2.3 \times 10^{-3}$ to $1.3 \times 10^{-1}$, i.e., for $p$ in this range, LER is lower than PER. \textbf{(b)} Experimental data and model of $CX$ gate for all-to-all topology. The threshold is estimated to be $3 \times 10^{-1}$.}
    \label{fig:s2_gler_gper}
\end{figure}

Since the applied noise is unbiased, the experimentally obtained PER and LER data yield similar values for the $H$ and $S$ gates, as shown in Figure~\ref{fig:s1_LER_PER_data_and_model_H_S_gates}. 
Consequently, a single model is sufficient for both gates, i.e., $\tilde{\varepsilon}_{H} = \tilde{\varepsilon}_{S}$. 
The red data points correspond to experimentally measured PER values, and the associated PySR model is shown by the red dashed curve. 
Similarly, the blue data points and curve represent the experimental PER data and the corresponding PySR model for the all-to-all topology, with a threshold value of $1.3 \times 10^{-1}$. 
The green data points and curve correspond to the square topology. We observe a range of $p$ values for which the LER is lower than the PER for the square topology. 
We refer to this as the \textit{threshold window}, which is identified as $[ 2.3 \times 10^{-3}, 1.3 \times 10^{-1}]$ for this experiment. 
Note that the \textit{SWAP} gate error rate is fixed at $10^{-4}$. 

Following the assumption given in \cite{gidney2024magic}, we construct an LER model for the $T$ gate by scaling the $S$ gate LER model by a factor of two, i.e., $\tilde{\varepsilon}_{T} = 2 \cdot \tilde{\varepsilon}_{S}$. 
For cases where $\tilde{\varepsilon}_{T} \geq 0.5$, the resulting value can exceed unity; therefore, $\tilde{\varepsilon}_{T}$ is clipped to a maximum value of $1.0$. 
Figure~\ref{fig:s1_LER_PER_data_and_model_CX_gate} presents the experimental data and the corresponding PySR model for the $CX$ gate, yielding a threshold value of $3 \times 10^{-1}$. 

The resulting analysis informs a symbolic resource model as introduced in Section~\ref{sec:autoqureo_model}, enabling efficient reuse without repeating the analysis.
These modeled error profiles are useful for applications such as estimating end-to-end circuit fidelities and integrating with other transpilation steps. 
In the next section, we leverage these results to study the co-design of gate decomposition and logical error rates, demonstrating how hardware-aware error modeling can guide the identification of optimal operating regimes.


\textbf{Future work}

Looking forward, the models can be utilized in a comparative study across quantum error detection (QED) codes (e.g., Iceberg-code~\cite{self2024protecting}), partial fault tolerance (e.g., STAR architecture~\cite{akahoshi2024partially}, dirty-qubits~\cite{bultrini2023battle}, CliNR~\cite{tham2025optimized}), and full fault tolerance (e.g., surface code~\cite{litinski2019game}, quantum low density parity code~\cite{breuckmann2021quantum}).
These approaches introduce distinct trade-offs between logical error rates, code size, circuit depth, and hardware constraints.
Symbolic annotations from existing QRE tools, including those developed in \cite{beverland2022assessing} and \cite{gidney2024magic}, can be integrated seamlessly, demonstrating interoperability with already developed QRE primitives.
Such experiments give insight into practical implementations of quantum error correction strategies.

\subsubsection{Gate decomposition accuracy and logical error rate}
\label{sec:scenario_qec_dcmp}

\textbf{Background}

Gate decomposition is a critical layer bridging algorithmic intent and hardware execution. 
While near-term devices tolerate relatively coarse approximations, fault-tolerant regimes impose stringent accuracy requirements that dramatically increase circuit depth. 
Understanding the optimal balance between decomposition accuracy and the accumulation of logical errors is therefore essential.
In noisy settings, higher decomposition accuracy does not monotonically improve overall fidelity. 
Instead, a noisy optimal zone emerges in which decomposition error and noise-induced error balance. 
This phenomenon exemplifies why AutoQuREO treats decomposition fidelity, logical error rate, and circuit depth as simultaneously co-optimized resources rather than isolated post-compilation metrics, as formalized in Section~\ref{sec:autoqureo5}.
A similar study was conducted in \cite{hao2026reducing} for only $T$ gate errors.
Using AutoQuREO, we model a universal gate set and identify optimal accuracy regimes as a function of noise strength and thereafter broaden this study to incorporate quantum error correction and algorithms.

Here, we study this trade-off for single-qubit unitaries using GridSynth \cite{ross2016optimal} for synthesis.
The GridSynth algorithm is designed to synthesize arbitrary $R_z$ rotations up to an approximation error of $\epsilon$ (defined in operator norm as the difference between the original and approximated unitary) using a sequence of $H$, $S$, and $T$ gates.
The $T$ gate count scales as $ 3 \log_2 \left({\frac{1}{\epsilon}} \right) + O \left( \log \left( \log \left({\frac{1}{\epsilon}} \right) \right) \right)$ with an expected runtime of $O \left( \text{polylog} \left( \frac{1}{\epsilon} \right) \right)$. 
To synthesize an arbitrary single-qubit unitary, one first computes its Euler-angle decomposition. 
Then, via conjugation with Clifford gates, an equivalent circuit consisting of three $R_z$ rotations is obtained. 
Finally, GridSynth approximates each $R_z$ individually. 
In this case, the $T$ gate count scales as $ 9 \log_2 \left({\frac{1}{\epsilon}} \right) + O \left( \log \left( \log \left({\frac{1}{\epsilon}} \right) \right)  \right)$~\cite{ross2016optimal}.

\textbf{Setup}

To characterize GridSynth, we begin by sampling 100 Haar-random single-qubit unitaries. 
Each unitary is then synthesized using GridSynth for various accuracies $\epsilon$. 
The resulting sequences of $H, S$, and $T$ gates are applied to the initial state $\ket{0}$, and run on a noisy density-matrix simulator. 
We use a logical-error model with depolarizing noise of varying strength $p$. 
From the final density matrices, $\rho$, the state fidelity is computed with the initial state $\rho_0 = \ket{0} \bra{0}$, as $F = \bra{0} \rho \ket{0}$. 
Table \ref{tab:decomposition_accuracy_LER_stack_table} summarizes the stack, hyperparameters,  corresponding ranges, and resources.

\begin{table}
\caption{The left panel presents the stack table for gate decomposition accuracy and LER co-design, whereas the right panel shows the corresponding resource table. For AutoQuREO-based optimization, the actions correspond to maximizing the average state fidelity for each depolarizing error rate.}
\centering

\begin{minipage}{0.73\textwidth}
\centering

\small

\begin{tabular}{|>{\centering\arraybackslash}m{4cm}|
                  >{\centering\arraybackslash}m{4cm}|
                  >{\centering\arraybackslash}m{2.5cm}|}

    \hline
    \rowcolor{lavender}
    \textbf{Layers} & \textbf{Hyperparameters} & \textbf{Ranges}\\
    \hline
    
    Unitary generation & Samples & [100] \\ 
    \hline

    GridSynth decomposition & Accuracy ($\epsilon$) & $[10^{-10}, ... ,10^{-1}]$ \\ 
    \hline

    Simulation & Depolarizing error rate ($p$) & $[10^{-8}, ... ,10^{-4}]$\\ 
    \hline
    
\end{tabular}

\end{minipage}%
\hfill
\begin{minipage}{0.27\textwidth}
\centering
\begin{tabular}{|>{\centering\arraybackslash}m{3.5cm}|}
\hline
\rowcolor{pink}
\textbf{Resources} \\
\hline
Number of qubits\\
\hline
Average state fidelity  \\
\hline
\end{tabular}
\end{minipage}
\label{tab:decomposition_accuracy_LER_stack_table}
\end{table}

\begin{figure}[htb]
    \centering
    
    \begin{subfigure}[t]{0.48\linewidth}
        \centering
        \includegraphics[clip, width=\linewidth]{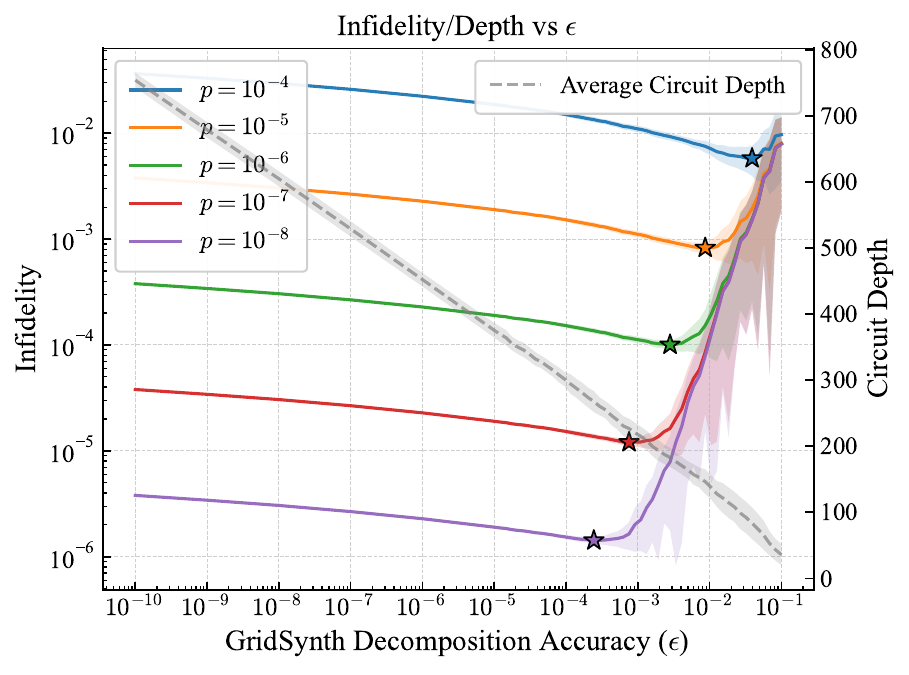}
        \caption{}
        \label{fig:s2_infidelity_depth_vs_accuracy}
    \end{subfigure}
    \hfill
    \begin{subfigure}[t]{0.48\linewidth}
        \centering
        \includegraphics[clip, width=\linewidth]{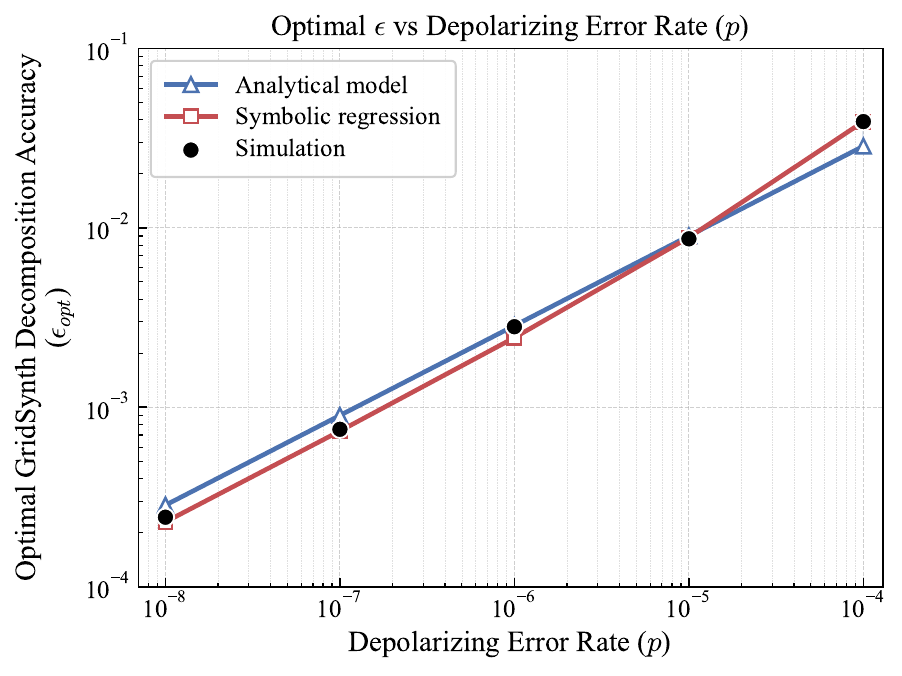}
        \caption{}
        \label{fig:s2_optimal_accuracy_vs_error_rate}
    \end{subfigure}
    
    \caption{Performance metrics of GridSynth decomposition under depolarizing noise. 
    \textbf{(a)} Infidelity ($1-F$) and circuit depth vs decomposition accuracy ($\epsilon$) for various depolarizing error rates ($p$). Star markers indicate the filtered configuration after AutoQuREO optimization.
    \textbf{(b)} Optimal decomposition accuracy ($\epsilon_{\mathrm{opt}}$) vs depolarizing error rate ($p$). Red curve shows the PySR model ($\epsilon_{\mathrm{opt}} \approx 2.28 \sqrt{p} + 162.43p$) generated using the simulation data. Blue curve shows the analytical model ($\epsilon_{\mathrm{opt}} \approx 2.85 \sqrt{p}$).}
    \label{fig:s2_infidelity_depth_vs_error_rate_and_optimal_accuracy_vs_error_rate}
\end{figure}

\textbf{Results}

Figure~\ref{fig:s2_infidelity_depth_vs_accuracy} shows the infidelity $1-F$ (left y-axis) and circuit depth (right y-axis) vs GridSynth accuracy ($\epsilon$) graphs for various depolarizing error rates ($p$). 
We clearly observe that the depth increases linearly with $-\log_{10}(\epsilon)$. 
A PySR symbolic regression model infers the average depth as $D \approx 75 \log_{10} \left(\frac{1}{\epsilon} \right)$. 
Since the noise model is unbiased, the average statistics capture random initial states instead of $\ket{0}$. 

AutoQuREO's optimization feature filters configurations based on specified hyperparameter and resource actions. For this study, we set the actions to maximizing the average state fidelity for each depolarizing error rate. 
These resulting filtered configurations are marked with a star in Figure~\ref{fig:s2_infidelity_depth_vs_accuracy} for all values of $p$. 
These minima indicate the existence of an optimal $\epsilon$ across all noise levels that minimizes infidelity (thereby maximizing fidelity). 
These filtered configurations are then used to model the optimal decomposition accuracy, $\epsilon_{\mathrm{opt}}$, as a function of $p$. 
\addch{This illustrates the iterative interplay between optimization and modeling in AutoQuREO. A full sweep over the decomposition accuracy is first optimized against noise to isolate the optimal Pareto points, these points are then modeled with PySR into a closed-form relation, and the resulting equation is reused as the resource model for the decomposition layer in the full-stack estimation. Optimization is therefore not a one-shot filtering, but a step that feeds reusable surrogate models back into the estimation loop for efficient full-stack DSE. Only the final selected configuration needs compilation.}
We use two methods to derive $\epsilon_{\mathrm{opt}}$: 
\begin{enumerate}
    \item Using PySR for symbolic regression (shown as the red curve in \ref{fig:s2_optimal_accuracy_vs_error_rate}), we obtain: 
    \begin{align}
    \epsilon_{\mathrm{opt}} \approx 2.28 \sqrt{p} + 162.43p
    \label{eqn:rs_eps_model}
    \end{align}
    \item By analytically deriving the model (shown as the blue curve in \ref{fig:s2_optimal_accuracy_vs_error_rate} and derived later in this section), we obtain:
    \begin{align}
    \epsilon_{\mathrm{opt}} \approx 2.85 \sqrt{p}
    \label{eqn:rs_eps_math}
    \end{align}
\end{enumerate}  

Comparing these two methods for model generation, we observe that the PySR model closely matches both the analytical model and the simulation results, while requiring significantly less time and effort. 
In contrast, the manual analytical method is non-trivial and requires expertise.

Since we assume a depolarizing logical error model and perform direct simulations, the resulting plots show exact infidelities. 
However, when a QEC code such as Steane is used, density-matrix simulation of Steane-encoded circuits becomes impractical due to the large number of $T$ gates and their implementation using a Reed-Muller code. 
Consequently, it is essential to develop a model that estimates fidelity as a function of $\epsilon$ and $p$. 
To this end, we empirically create a fidelity estimation model, denoted by $F_{\mathrm{est}}$, using exact fidelity data as follows:
\begin{align}
    F_{\mathrm{est}} &= \frac{1}{2}[1+K(1-2\epsilon^2)]
\label{eqn:fidelity_estimation}
\end{align}
where $K$ is the estimated success probability (ESP) \cite{escofet2025accurate}, defined as:
\begin{align}
    K = \prod_{i=1}^{m} (1-\tilde{\varepsilon}_{g_i})^{n_i}
\label{eqn:product_formula}
\end{align}

Here we assume that $G = \{ g_1, g_2, ... ,g_m \}$ is the gate set, and $N = \{ n_i |$ $n_i$ is counts of $g_i \}$ is the set of counts for each gate $g_i$ in a given circuit. $\tilde{\varepsilon}_{g_i}$ is the LER, and it is a function of $p$ in general. 

For this study, we are assuming $G = \{ H(g_1), S(g_2), T(g_3) \}$ and $\tilde{\varepsilon}_{H} = \tilde{\varepsilon}_{S} = \tilde{\varepsilon}_{T} = p$. Therefore,
\begin{align*}
    K &= \prod_{i=1}^{3} (1-\tilde{\varepsilon}_{g_i})^{n_i} = \prod_{i=1}^{3} (1-p)^{n_i} = (1-p)^{\sum n_i}.
\end{align*}
Substituting the values for the gate counts for the gate set $G$ as defined in Equation~\ref{eqn:GridSynth_PySR_HST_counts},
\begin{align*}
    K &\approx (1-p)^{75 \log_{10} \left( \frac{1}{\epsilon} \right)} = e^{-\frac{75}{\ln(10)} \ln(\epsilon) \ln(1-p)} = \epsilon^{A},
\end{align*}
where $A = -\frac{75}{\ln(10)} \ln(1-p)$. Since $0 < p < 1$, $\ln (1-p) < 0$, $\therefore A > 0$. Substituting in (\ref{eqn:fidelity_estimation}), we get $F_{\mathrm{est}} (\epsilon) = \frac{1}{2}[1 + \epsilon^A (1-2\epsilon^2)]$. 

\vspace{0.5em}
\textit{\emph{Analytical model derivation}}    
\vspace{0.2em}

To calculate $\epsilon_{\mathrm{opt}}$ theoretically, we maximize $F_{\mathrm{est}}$ w.r.t $\epsilon$. Therefore,
\begin{align*}
    F_{\mathrm{est}}^{'}(\epsilon) &= \frac{1}{2} \left[ A \epsilon^{A-1} - 2 (A+2) \epsilon^{A+1} \right]. \\
\end{align*}

From $F_{\mathrm{est}}^{'}(\epsilon) = 0$, we obtain,
\begin{align}
    \epsilon_{\mathrm{opt}} = \sqrt{\frac{A}{2(A+2)}}.
\label{eqn:epsilon_opt}
\end{align}

If $p \ll 1 $, then $ \ln (1-p) \approx -p$ and $A \approx 32.57p \ll 1 < 2$. 
Therefore, $\epsilon_{\mathrm{opt}} \approx \sqrt{\frac{A}{4}} = 2.85 \sqrt{p}$. 

Substituting $\epsilon_{\mathrm{opt}}$ in $F_{\mathrm{est}}(\epsilon)$, the analytical maximum fidelity is
\begin{align*}
F^{\mathrm{max}}_{\mathrm{est}}
&= \frac{1}{2}\left[1+\left(\frac{A}{2(A+2)}\right)^{A/2}\frac{2}{A+2}\right].
\end{align*}

For $A\ll 1$, using $\frac{A}{2(A+2)}\approx\frac{A}{4}$ and the further approximation
$\frac{2}{A+2}\approx 1$, we get
\begin{align*}
    F^{\mathrm{max}}_{\mathrm{est}} &\approx \frac{1}{2} \left[ 1 + \left( \frac{A}{4} \right)^{\frac{A}{2}} \right] \\
    &= \frac{1}{2} \left[ 1 + \exp \left(  \frac{A}{2} \ln \left( \frac{A}{4} \right) \right) \right] \\
    &\approx \frac{1}{2} \left[ 1 + 1 + \frac{A}{2} \ln \left( \frac{A}{4} \right) \right] (\because e^x \approx 1+x) \\
    &\approx 1 + \frac{A}{4} \ln \left( \frac{A}{4} \right) \\
    & \approx 1 + 8.14 p \ln(8.14p)
\label{eqn:fidelity_estimation_max}
\end{align*}


As $p$ becomes small, $\epsilon_{\mathrm{opt}}$ scales like $\sqrt{p}$, and the minimum infidelity $1 - F_{\mathrm{est}}^{\mathrm{opt}}$ scales like $p|\ln p|$ (up to constants).


Note that the above analysis assumes that all the gates in $G$ have identical LERs. 
However, the $T$ gate generally exhibits a higher LER because it is implemented using techniques such as magic-state distillation \cite{litinski2019magic} and magic-state cultivation \cite{gidney2024magic}. 
In this study, we employ the Reed-Muller code described in Section~\ref{sec:scenario_qec}. 
Additionally, we create PySR models for $H$, $S$, and $T$ gate counts as functions of GridSynth $\epsilon$ to obtain:
\begin{equation}
n_H \approx 30 \log_{10} \left( \frac{1}{\epsilon} \right), \quad
n_S \approx 15 \log_{10} \left( \frac{1}{\epsilon} \right), \quad
n_T \approx 30 \log_{10} \left( \frac{1}{\epsilon} \right)
\label{eqn:GridSynth_PySR_HST_counts}
\end{equation}

Using Equations~\ref{eqn:product_formula},~\ref{eqn:GridSynth_PySR_HST_counts}, and the relation $\tilde{\varepsilon}_{H} = \tilde{\varepsilon}_{S} \neq \tilde{\varepsilon}_{T}$, the study can be extended to a more realistic setting. 
For example, we may assume $\tilde{\varepsilon}_{T} = p$ and $\tilde{\varepsilon}_{H} = \tilde{\varepsilon}_{S} = 10^{-2}p$ (assuming that $T$ gate has an LER two orders of magnitude higher than that of $H$ and $S$ gates), and derive new models for $\epsilon_{\mathrm{opt}}$ using both PySR and the analytical method.

\vspace{0.5em}
\textit{\emph{Integration with Steane code}}
\vspace{0.2em}

\begin{figure}[hbt]
    \centering
    \includegraphics[clip, trim=0cm 0cm 0cm 0cm, width=0.5\textwidth]{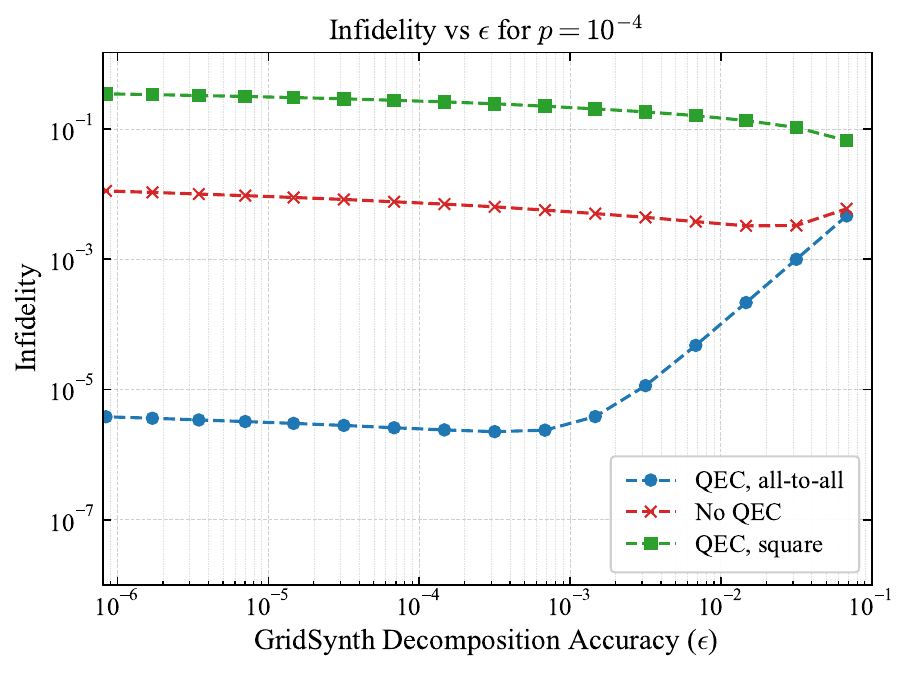}
    \caption{Infidelity ($1-F$) vs GridSynth decomposition accuracy ($\epsilon$) for physical gate depolarizing error rate $p=10^{-4}$. We used the LER models generated in Section~\ref{sec:scenario_qec} and assume $\tilde{\varepsilon}_{H} = \tilde{\varepsilon}_{S}$, $\tilde{\varepsilon}_{T} = 2 \cdot \tilde{\varepsilon}_{H}$ with clipping the maximum value to $1.0$.} 
    \label{fig:s2_infidelity_vs_decmp_accuracy_all_to_all_and_square}
\end{figure}

Next, we extend this study by incorporating LER models for the Steane code single-qubit gates $G = \{H, S, T\}$ from Section~\ref{sec:scenario_qec}. 
We assume that the PER for these gates is approximately $p$, and that the corresponding LERs are functions of $p$, i.e., $\tilde{\varepsilon}_{g_i} = f_i(p)$ for all $g_i \in G$. 
To estimate the fidelities, we use the empirical model given in Equation~\ref{eqn:fidelity_estimation}, together with the ESP formula in Equation~\ref{eqn:product_formula}. 
Figure~\ref{fig:s2_infidelity_vs_decmp_accuracy_all_to_all_and_square} shows the infidelity of QEC-encoded and unencoded circuits under all-to-all and square qubit connectivity topologies. 
As observed, for all-to-all connectivity, the encoded circuits achieve lower infidelities than their unencoded counterparts. 
In contrast, under square connectivity, the encoded circuits become impractical due to the accumulation of \textit{SWAP} gate errors.
Thus, the square connectivity would therefore have practical relevance over all-to-all connectivity in regimes with significant lower physical noise level and scalable quantum processors.

\textbf{Future work}

The analysis presented in this section can be extended to other single-qubit unitary synthesis algorithms, such as the Solovay-Kitaev~\cite{dawson2005solovay}, TraSyn~\cite{hao2026reducing}, and Morisaki-Sano-Akibue algorithm~\cite {morisaki2025optimal}. 
Furthermore, this framework can be extended to multi-qubit unitaries via decomposition techniques, including quantum Shannon decomposition~\cite{shende2005synthesis}, QSeed~\cite{weiden2023improving}, flag decomposition~\cite{kottmann2026parameter}, and can be integrated with QEC models to capture compounding effects.

\subsubsection{Ground-state energy estimation with iterative quantum phase estimation}
\label{sec:scenario_iqpe_dcmp}

\textbf{Background}

Early fault-tolerant (EFT)~\cite{arnault2024typology,katabarwa2024early} regimes occupy an increasingly relevant middle ground between noisy intermediate-scale quantum (NISQ) devices and fully fault-tolerant quantum computers (FTQC). 
NISQ devices have qubit counts potentially beyond classical simulation limits (in the range of 50-100 qubits), with physical noise levels that enable QEC codes to demonstrate below-threshold operation (typically a single logical qubit memory or basic logical gates).
Thus, full QEC encoding of logical qubits and the required operations for algorithms is not possible, limiting their applicability to low-depth unencoded circuit use cases, such as variational algorithms implemented by parametric quantum circuits and trained iteratively via a hybrid quantum-classical optimization loop.
FTQC refers to a regime in which the above issues have been addressed, so that users have a sufficient number of high-fidelity, error-corrected qubits at their disposal, with coherence maintained beyond the required circuit execution duration.
FTQC algorithms are conventional ones with well-motivated applications and theoretical complexity guarantees, such as quantum factoring, quantum phase estimation, and linear equation solvers. 
EFT bridges these two regimes via principled approaches that would enable a smooth transition.
These include incorporating low-ancilla overhead QEC codes, partial error correction, post-selection based on error detection, circuit cutting and knitting, non-uniform qubit basis calibration, error mitigation via circuit-level and pulse-level protocols, randomized algorithms, and low-ancilla iterative algorithms, among others.
Thus, even in modest problem instances, EFT stacks already combine algorithmic structure, small-scale error correction, discrete gate synthesis, and noisy backends with connectivity topologies. 
Manually reasoning about their coupled resource trade-offs rapidly becomes infeasible. 

\vspace{0.2em}
\textbf{Setup}

\addch{In this section, we consider an EFT algorithm, in particular, the iterative quantum phase estimation (iQPE)~\cite{dobvsivcek2007arbitrary,moore2021statistical}, applied to the electronic ground-state energy estimation (GSEE) of the} $\verb|N|_2 \verb|H|_4$ \addch{(Hydrazine) molecule. We import the Hamiltonian (H) expressed as a sum of Pauli strings (Equation~\ref{eqn:hamiltonian_in_pauli_basis}) from PennyLane datasets~\cite{bergholm2018pennylane}.} For a molecule not in the database, first a quantum chemical electronic structure calculation using Psi4~\cite{turney2012psi4} is run for the molecule in STO-3G basis set. Then the Psi4 executes SCF, MP2, CISD, CCSD, and FCI methods~\cite{turney2012psi4} to obtain the molecular integrals. The molecular Hamiltonian (in terms of one- and two-electron integrals) is then converted to a second-quantized Fermionic operator. Following this, the Fermionic ladder operators are mapped to qubit Pauli operators via the Jordan-Wigner transform, eventually yielding the qubit Hamiltonian expressed as a sum of Pauli strings.

We specifically focus on symbolically inferring the gate-count resource, incorporating estimates from error correction and decomposition from the previous section, for a full-stack QRE. 
We also extend this study to estimate FTQC resources by replacing Steane code with surface code.
Table \ref{tab:iqpe_stack_table} summarizes the stack, hyperparameters, corresponding ranges for Steane code (with T-teleportation from Reed-Muller code) and surface code.

%

\begin{align}
	H &= \sum_{i=1}^L c_i P_i, \quad \text{where } P_i \in \{I, X, Y, Z\}^{\otimes n},\; c_i\in\mathbb{R},\; n=28 \text{ for } \mathtt{N}_2\mathtt{H}_4.
	\label{eqn:hamiltonian_in_pauli_basis}
\end{align}

Its matrix exponential (setting $t = 1$) is the time-evolution unitary $U = e^{-i H}$.
The qubit Hamiltonian is then diagonalized to find the smallest algebraic eigenvalue corresponding to the ground-state energy and the associated eigenvector:
\begin{equation}
    H \ket{\psi_0} = E_0 \ket{\psi_0}, \qquad
    E_0 = \min_{\lambda \in \sigma(H)} \lambda
\end{equation}
The ground-state eigenvector $\ket{\psi_0}$ is prepared directly from the
classically computed amplitudes $\{c_k\}$:
\begin{equation}
    \ket{\psi_0} = \sum_{k=0}^{2^n - 1} c_k \ket{k}
\end{equation}

\begin{table}
\caption{Full-stack resource estimation table showing the layers, hyperparameters, and corresponding ranges. Hyperparameters for the Steane code (T from Reed-Muller code) and the surface code implementation are listed under their respective columns. The first two layers (Preprocessing and iQPE) use identical hyperparameters and ranges for both implementations, whereas the third (first-order Trotterization + GridSynth) and fourth (Error Correction) layers use implementation-specific hyperparameter sets.}
\centering

\begin{minipage}{1.0\textwidth}
\centering

\small

\begin{tabular}{|>{\centering\arraybackslash}m{3cm}|
                  >{\centering\arraybackslash}m{2.7cm}|
                  >{\centering\arraybackslash}m{2.7cm}|
                  >{\centering\arraybackslash}m{2.7cm}|
                  >{\centering\arraybackslash}m{2.7cm}|}

    \hline
     & 
    \multicolumn{2}{c|}{\cellcolor{lavender!30}\textbf{Steane code (T from Reed-Muller code)}} & 
    \multicolumn{2}{c|}{\cellcolor{lavender!30}\textbf{Surface code}} \\
    \hline

    \multirow{1}{*}{\cellcolor{lavender}
    \textbf{Layers}} & \cellcolor{lavender}
    \textbf{Hyperparameters} & \cellcolor{lavender}
    \textbf{Ranges} 
    & \cellcolor{lavender}
    \textbf{Hyperparameters} & \cellcolor{lavender}
    \textbf{Ranges} \\
    \hline
    
    Preprocessing 
    & Bond length ($\mathring{\mathrm{A}}$) & $[1.446]$ 
    & Bond length ($\mathring{\mathrm{A}}$) & $[1.446]$ \\ 
    \hline

    iQPE 
    & Eigenvalue precision & $[10]$ 
    & Eigenvalue precision & $[10]$ \\ 
    \hline

    \multirow{2}{*}[-0.4em]{\shortstack[c]{Trotterization\\+\\ GridSynth}}
    & Trotter steps & $[5]$ 
    & Trotter steps & $[5]$ \\
    \cline{2-5}

    & GridSynth accuracy & $[2.5 \times 10^{-4}]$ 
    & GridSynth accuracy & $[7.2\times10^{-7}]$ \\
    \cline{4-5}
    
    \hline

    \multirow{7}{*}{Error Correction}
    &  & 
    & Physical error rate & $[10^{-4}]$ \\
    \cline{4-5}
    
    &  & 
    & Code distance (data) & $[11]$ \\
    \cline{4-5}

    & Depolarizing error rate & $[10^{-4}]$  
    & No. of factories & $[5]$ \\
    \cline{4-5}

    &  & 
    & L1 distillation code distance (factory) & $[19]$ \\
    \cline{4-5}

    & & 
    & L2 distillation code distance (factory) & $[31]$ \\
    \cline{4-5}
    
    & & 
    & Cycle time ($\mu s$) & $[1]$ \\
    \hline
    
\end{tabular}

\end{minipage}%
\label{tab:iqpe_stack_table}
\end{table}

\textbf{Results}

\textit{\emph{Algorithm resources}}

\begin{figure}[htb]
\centering
\includegraphics[page=9, clip, trim=0cm 0cm 0cm 2.3cm,width=0.95\textwidth]{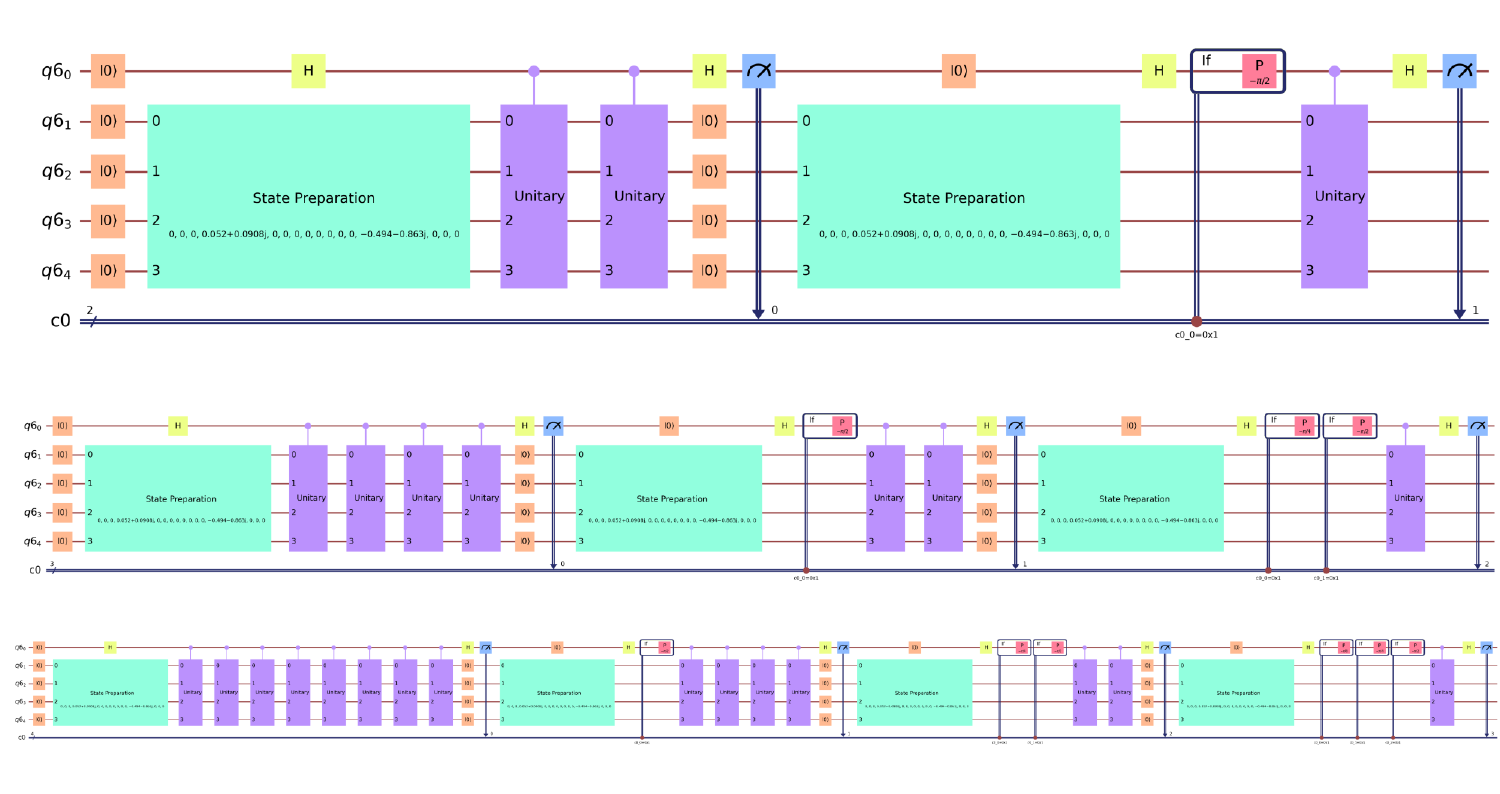}
\caption{Quantum circuits for iterative quantum phase estimation (iQPE) algorithm for precision of $2$, $3$, and $4$ bits. Subsequent circuits and gate colors highlight the scaling patterns.}
\label{fig:iqpe}
\end{figure}

In conventional QPE, an $n$-qubit control register is prepared in a superposition and entangled with a target eigenstate via controlled powers of a unitary operator, followed by an inverse quantum Fourier transform (QFT) to extract the phase, enabling parallel estimation of multiple phase bits.
The iterative quantum phase estimation (iQPE)~\cite{dobvsivcek2007arbitrary} is a resource-efficient QPE variant that replaces the multi-qubit control register with a single ancilla qubit reused across multiple rounds to estimate the phase bitwise sequentially. 
As shown in Figure~\ref{fig:iqpe}, each iteration applies a controlled-$U^{2^k}$ operation, followed by a phase correction based on previously measured bits and a Hadamard rotation prior to measurement. 
The measurement outcome probabilistically determines the $k$-th bit of the phase, and the process is repeated for increasing powers of $U$. 
This structure eliminates the need for a full QFT, making iQPE particularly suitable for EFTs constrained by qubit and circuit coherence, at the cost of higher runtime than QPE.

We are interested in modeling the gate counts of iQPE with respect to the eigenvalue precision $ep$ and the qubit size of the input unitary $qu$.
For example, from visual inspection of these circuits, qubit count scales as $qu+1$, state preparation and measurement operator scales as $ep$, Hadamard scales as $2*ep$, and reset operators scale as $(qu+1)*ep$. 
The scaling for controlled unitary and phase gates is non-trivial, with $\sum_{i = 0}^{ep-1} 2^i$ and $\sum_{i = 0}^{ep-1} i$ if $ep > 1$, respectively. 
AutoQuREO's surrogate modeling enables us to verify whether these intuitions are correct and to automate this resource analysis without requiring expertise in analytical complexity theory.
PySR-based symbolic regression infers the following symbolic resource estimates from compilation data of $ep = [1,6]$ and $qu = [1,7]$ as shown in Box~\ref{box:iqpe}.

\begin{verbbox}{Gate resources for iQPE}{box:iqpe}
\begin{verbatim}
qb                : qu + 1.0000001
reset             : qu*ep + ep
state_preparation : ep
cu_in             : 1.9999988**ep - ep/ep
h                 : ep + ep
measure           : ep
p                 : -0.500011*ep - 0.0002529222*ep*(-0.20524421) + ep/((2.0000248/ep))
\end{verbatim}
\end{verbbox}
We note that the surrogate model can replicate all the trivial equations and suggest alternate forms for the non-trivial ones within specified error bounds.
Demonstratively, the inferred equation for the controlled unitary, $2^{ep}-1$, is simpler than our analytical benchmark.  
Characteristics bloat from genetic programming (e.g., the \verb|ep/ep| instead of \verb|1|) and convergence error from simulated annealing (e.g., the \verb|1.0000001|) can be further fine-tuned using PySR's parameters (e.g., parsimony) within Optuna to refine the surrogate model.
These equations, however, are within acceptable modeling errors for our representative case study.

\vspace{0.2em}
\textit{\emph{Decomposition resources}}

The controlled unitary gates (\verb|cu_in|) in iQPE require explicit decomposition to map to the EFTQC and FTQC stack. We assume an initial $\ket{0}^{\otimes28}$ state having a non-zero overlap with the ground-state.
Since the Hamiltonian (H) is written in Pauli basis (Equation~\ref{eqn:hamiltonian_in_pauli_basis}), the time evolution operator can be implemented by exponentiating the individual Pauli terms and constructing the corresponding controlled gates. Here, we employ first-order Trotterization, as described in Equation~\ref{eqn:first_order_trotterization_step}, with $r$ Trotter steps to estimate gate resources.
Let \verb|x_i|, \verb|y_i|, and \verb|z_i| denote the number of $X-$, $Y-$, and $Z-$Pauli operators, respectively in the Pauli term $P_i$. The Pauli weight of $P_i$ is then given by, \verb|w_i| = \verb|x_i| + \verb|y_i| + \verb|z_i|, and \verb|sum_paulis()| denotes the summation over all Pauli terms in the Hamiltonian. The resulting gate counts are summarized in Box \ref{box:trotter}.

\begin{verbbox}{Gate resources for first-order Trotterization}{box:trotter}
\begin{verbatim}
cx  : sum_paulis(2 * (w_i - 1)) * r * cu_in + 2 * L * r * cu_in
h   : sum_paulis(2 * (x_i + y_i)) * r * cu_in
s   : sum_paulis(y_i) * r * cu_in
s_dg: sum_paulis(y_i) * r * cu_in
rz  : 2 * L * r * cu_in
\end{verbatim}
\end{verbbox}

Thus, the iQPE gate resources can now be expressed in terms of single- and two-qubit gates.
In situations where the Pauli basis representation is not available, we can incorporate the block-ZXZ improvement~\cite{krol2024beyond} of quantum Shannon decomposition (QSD)~\cite{shende2005synthesis,krol2022efficient} for multi-qubit unitaries synthesis.
The associated resources are $\frac{9}{16}\cdot4^n - \frac{3}{2}\cdot2^n$ for $CX$ gates and $\frac{21}{16}\cdot4^n - \frac{3}{2}\cdot2^n$ for rotation gates (with ratio of \verb|rz:ry| gates being $2:1$).
We can thereafter use the relation $R_y(\theta) = SHR_z(\theta)HS^\dagger$ to convert rotations to \verb|rz| counts.



In this case, after Trotterization, the \verb|rz| gates are decomposed using GridSynth. Using the characterization data of Steane code from Section \ref{sec:scenario_qec}, and the PySR modeled Equations \ref{eqn:rs_eps_model} and \ref{eqn:GridSynth_PySR_HST_counts} for optimal decomposition settings, the gate counts for noisy hardware (\verb|depol|) are calculated as shown in Box~\ref{box:rs}.
The function \verb|characterization_data()| corresponds to the experimental data of \ref{fig:s1_LER_PER_data_and_model_H_S_gates} for Steane code, and to the surface code logical error rate model, $p_L= 0.1 \left(\frac{p}{0.01} \right)^{\frac{d+1}{2}}$ from~\cite{Qualtran, gidney2019efficient}.

\begin{verbbox}{Gate resources for hardware-optimized GridSynth}{box:rs}
\begin{verbatim}
p = characterization_data(depol)
h	: ceiling(30*log(1/(2.28*sqrt(p) + 162.43*p))/log(10))
s	: ceiling(15*log(1/(2.28*sqrt(p) + 162.43*p))/log(10))
t	: ceiling(30*log(1/(2.28*sqrt(p) + 162.43*p))/log(10))
\end{verbatim}
\end{verbbox}
Thus, the trade-off study from the previous section is integrated into these broader application settings.
For brevity, we omit the intermediate step of replacing \verb|rz| with this cost model.


\vspace{0.2em}

\addch{\underline{\textit{Multi-layer model comparison}}}

\addch{The symbolic estimates above were obtained layer by layer and then composed manually. This becomes cumbersome motivating a single surrogate that directly maps to the final resource. The neuro-symbolic pipeline of Section~\ref{sec:autoqureo_model} is well suited to maintain inference speed and interpretability for downstream analysis in such use cases.}

\addch{To demonstrate this, we consider a multi-layer stack for the GSEE of the Hydrazine molecule ($28$ qubits). The quantum circuit composes the iQPE algorithm layer (thus, $29$ logical qubits), first-order Trotterization with $5$ steps, and GridSynth decomposition at an accuracy of $10^{-10}$. As an extrapolation test, the models are trained with compilation data for an eigenvalue precision in the range $[2, 12]$, while the test inference is done for the range $[12, 30]$ for $400$ sampled configurations.}

\begin{figure}[htb]
    \centering
    
    \begin{subfigure}[t]{0.47\linewidth}
        \centering
        \includegraphics[clip, width=\linewidth]{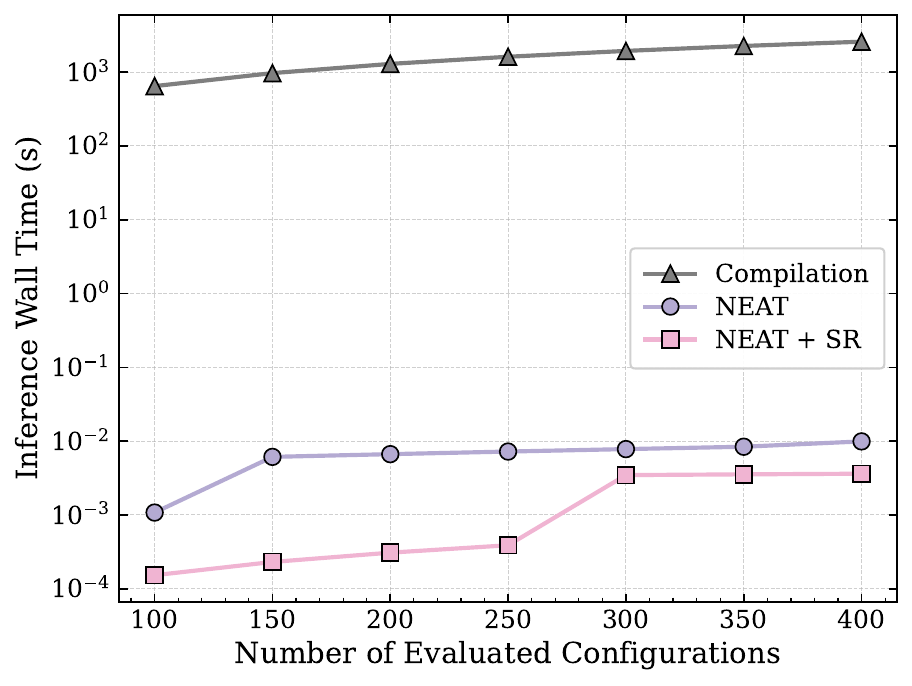}
        \caption{}
        \label{fig:s2_nesy_cost}
    \end{subfigure}
    \hfill
    \begin{subfigure}[t]{0.48\linewidth}
        \centering
        \includegraphics[clip, width=\linewidth]{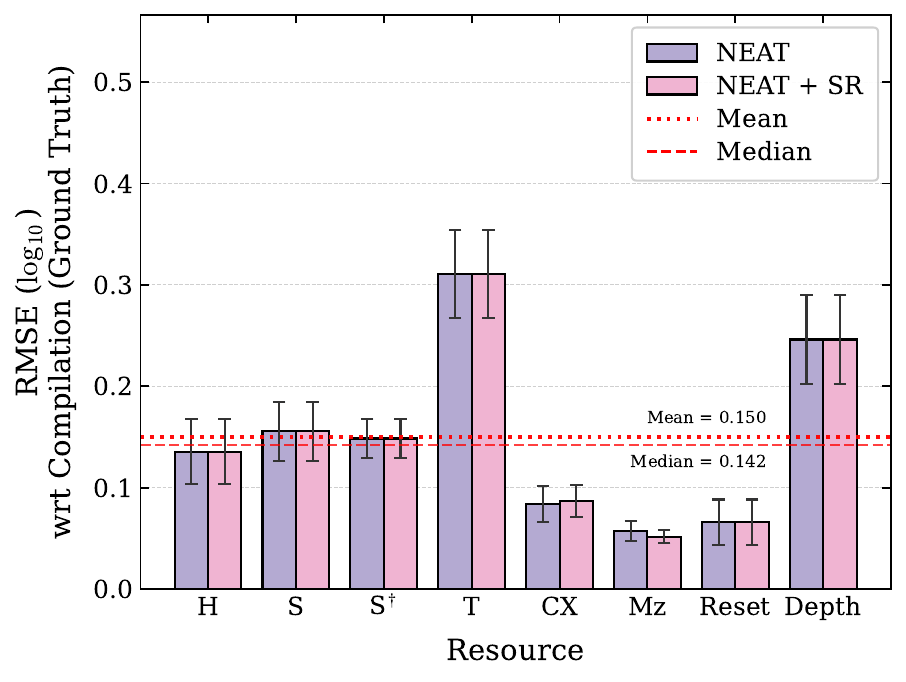}
        \caption{}
        \label{fig:s2_nesy_conf}
    \end{subfigure}
    
    \caption{Performance metric comparison of surrogate models for iterative quantum phase estimation resource prediction.
    The two models considered are, (i) neuro-evolution using augmented topologies (NEAT), and (ii) subsequent distillation of the NEAT model with PySR (modeling configurations available in the repository). 
    \textbf{(a)} Inference time on extrapolated test data.
    \textbf{(b)} RMS error on extrapolated test data.
    Note that the NEAT + SR model takes an additional $\approx$ 30 min modeling time, while being interpretable for formal analysis.}
    \label{fig:s2_nesy}
\end{figure}

\addch{The surrogate is synthesized by the two-stage neuro-symbolic pipeline. A NEAT network is first evolved on the compilation data as a sub-symbolic approximator, and PySR then distills the trained network into closed-form symbolic resource estimates. A comparison of the two models on inference cost and predictive confidence is shown in Figure~\ref{fig:s2_nesy}. As shown in Figure~\ref{fig:s2_nesy_cost}, the distilled symbolic model evaluates substantially faster than the standalone NEAT surrogate, since inference reduces to substituting hyperparameters in a formula rather than a full network forward pass. Figure~\ref{fig:s2_nesy_conf} shows that this speedup incurs no meaningful loss of accuracy, as the RMS error of the distilled model matches closely with the NEAT surrogate across the extrapolated regime. The distillation adds a one-time modeling overhead of about $30$ minutes, in exchange for interpretable equations. The benefit of the neuro-symbolic approach is therefore twofold, combining the low inference cost and transparency of a symbolic estimate with the predictive quality of the evolved network.}

\addch{These surrogate estimators follow the broader algorithmic profiling methodology described in Sections~\ref{sec:autoqureo_model} and~\ref{sec:autoqureo_surrogate}, enabling interpretable scalable extrapolation beyond directly compilable problem sizes.}

\vspace{0.2em}

\addch{\underline{\textit{Algorithm and decomposition resources under noise}}}

\begin{figure}[htb]
    \centering
    
    \begin{subfigure}[t]{0.48\linewidth}
        \centering
        \includegraphics[clip, width=\linewidth]{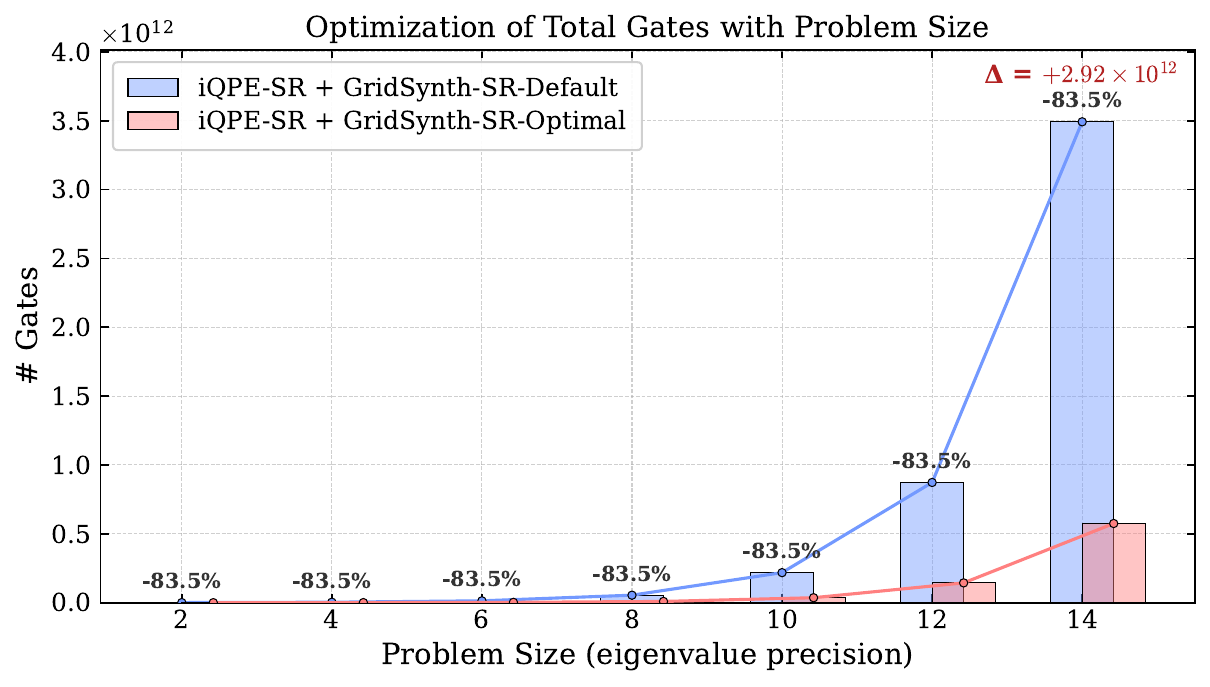}
        \caption{}
        \label{fig:s2_rsopt_joint_evp}
    \end{subfigure}
    \hfill
    \begin{subfigure}[t]{0.48\linewidth}
        \centering
        \includegraphics[clip, width=\linewidth]{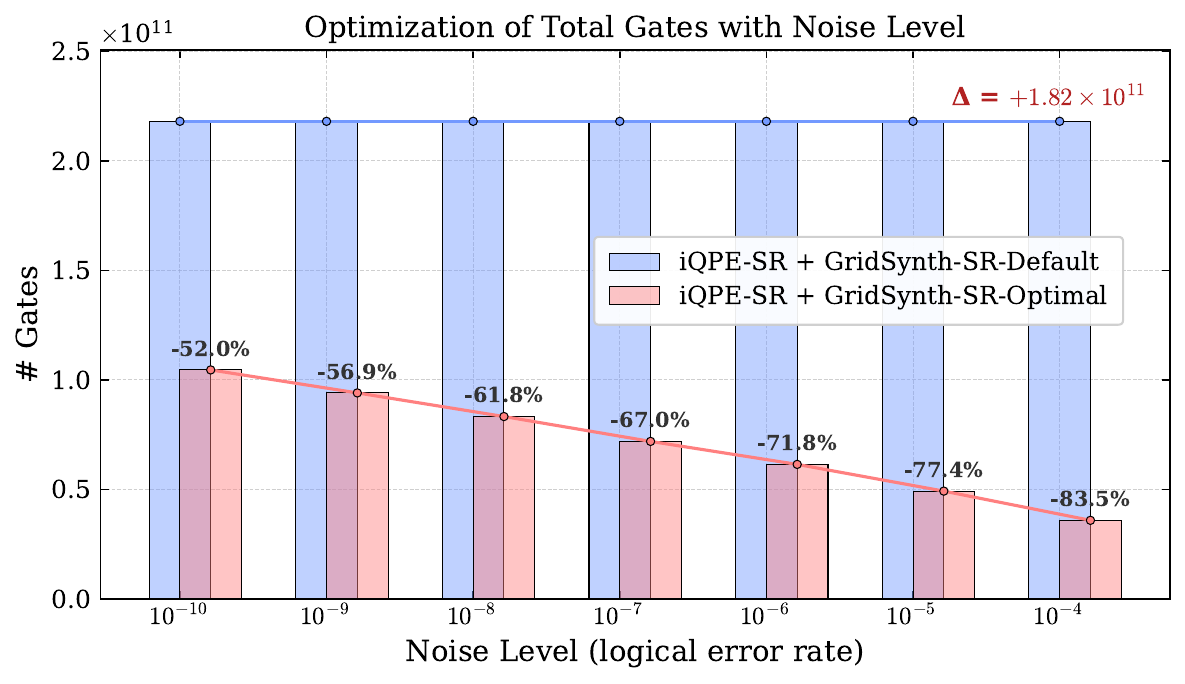}
        \caption{}
        \label{fig:s2_rsopt_joint_ler}
    \end{subfigure}
    
    \caption{Full-stack gate-count saving from the optimal GridSynth accuracy $\epsilon_{\mathrm{opt}}$ versus a fixed $\epsilon = 10^{-10}$ baseline. \textbf{(a)} Total gate count vs eigenvalue precision at a fixed LER $p = 10^{-4}$. The absolute gap widens with precision, while the relative saving stays nearly constant at ${\approx}\,83.5\%$. \textbf{(b)} Total gate count vs LER $p$ at a fixed eigenvalue precision of $10$. As $p$ increases, $\epsilon_{\mathrm{opt}}$ coarsens, so fewer $H$, $S$, and $T$ gates are synthesized per rotation and the saving grows, reaching ${\approx}\,83.5\%$ at $p = 10^{-4}$.}
    \label{fig:s2_rsopt_joint}
\end{figure}

\addch{The optimal GridSynth accuracy derived in Section~\ref{sec:scenario_qec_dcmp} was obtained for a single $R_z$ rotation, relating $\epsilon_{\mathrm{opt}}$ to the underlying noise as shown in Figure~\ref{fig:s2_infidelity_depth_vs_error_rate_and_optimal_accuracy_vs_error_rate}, and distilled into Equation~\ref{eqn:rs_eps_model}. We now propagate this single-gate result through the iQPE stack for the Hydrazine molecule. The Trotterized controlled unitaries expose the $R_z$ rotations which are thereby synthesized by GridSynth. Choosing the noise-matched $\epsilon_{\mathrm{opt}}$ rather than a fixed accuracy compounds the per-gate saving into a substantial reduction of the total gate count. In this experiment we do not commit to a particular PER-to-LER mapping induced by a specific QEC code, but instead vary the underlying logical noise $p$ directly, so that the benefit is characterized independently of the error-correction choice. Figure~\ref{fig:s2_rsopt_joint_evp} sweeps the problem size via the eigenvalue precision at a fixed noise level of $p = 10^{-4}$ (used to compute $\epsilon_{\mathrm{opt}}$). The gap between the optimized and baseline curves widens with precision, since the number of Trotterized $R_z$ rotations, and hence the number of GridSynth calls that each benefit from the coarser $\epsilon_{\mathrm{opt}}$, grows. Figure~\ref{fig:s2_rsopt_joint_ler} sweeps the underlying logical noise at a fixed eigenvalue precision of $10$. As the noise increases, $\epsilon_{\mathrm{opt}}$ coarsens as $\sqrt{p}$, so fewer $H$, $S$, and $T$ gates are needed per rotation, and the total gate count of the optimized stack drops accordingly. The single-gate decomposition trade-off scales up to a decisive full-stack resource saving, which would be impractical to establish by directly compiling the stack at every configuration.}

\subsubsection{Extension to fault-tolerant regimes}
\label{sec:scenario_ftqc}

\vspace{0.2em}

\textit{\emph{Steane resources}}

Next, we need to count the number of physical gates based on the Steane code and T-teleportation from Reed-Muller code as detailed in Section~\ref{sec:scenario_qec}.
This is shown in Box~\ref{box:qec}.
\begin{verbbox}{Gate resources for QEC (Steane code, T-teleportation from Reed-Muller code)}{box:qec}
\begin{verbatim}
phy_h         : 12*SM + 7*h + 3*reset + 12*t + 12*tdg
phy_sdg       : 7*s + 7*t
phy_s         : 7*sdg + 7*tdg
phy_cx        : 12*SM + 7*cx + 7*measure + 11*reset + 53*t + 53*tdg
phy_x         : 7*EC + 7*x
phy_z         : 7*EC + 3*t + 3*tdg + 7*z
phy_reset     : 6*SM + 7*reset + 24*t + 24*tdg
phy_t         : 8*t + 7*tdg
phy_tdg       : 7*t + 8*tdg
phy_cz        : 12*SM + 3*t + 3*tdg
phy_measure   : 6*SM + measure + 2*t + 2*tdg
\end{verbatim}
\end{verbbox}
Here, \verb|SM| refers to the syndrome measurement gadget and \verb|EC| refers to the error correction gadget.
As convention, we will consider the case where \verb|SM = EC|, and the syndrome measurement frequency is after each gate, thus the value is the sum of all logical gates.

This series of embeddings exposes four free parameters: the eigenvalue precision \verb|ep|, the number of qubits required for the Hamiltonian \verb|qu|, the Trotter steps \verb|r|, and the hardware's depolarizing noise \verb|depol|.
Interpreting eigenvalue precision as binary energy precision, the least significant bit should not be larger than the target chemical accuracy, which is taken to be $1 kcal/mol \approx 1.6*10^{-3}Ha$, requiring about $10$ precision bits.
For the 28-qubit $N_2H_4$ Hamiltonian, taking the Trotter steps of $5$ and a depolarizing error rate of $10^{-4}$ (optimistic by currently available QPUs), we obtain the estimate shown in Box~\ref{box:fs_steane}.


\begin{verbbox}{Full-stack resources for Steane code}{box:fs_steane}
\begin{center}
\begin{tabular}{@{}ll@{\hspace{2em}}ll@{}}
\texttt{num\_qubits}  & : \(\mathtt{1.1 \times 10^{3}}\) & \\
\texttt{phy\_h}       & : \(\mathtt{1.6 \times 10^{12}}\) & \quad\quad \texttt{phy\_sdg}     & : \(\mathtt{3.2 \times 10^{11}}\) \\
\texttt{phy\_cx}      & : \(\mathtt{2.6 \times 10^{12}}\) & \quad\quad \texttt{phy\_x}       & : \(\mathtt{5.7 \times 10^{11}}\) \\
\texttt{phy\_z}       & : \(\mathtt{6.6 \times 10^{11}}\) & \quad\quad \texttt{phy\_reset}   & : \(\mathtt{1.2 \times 10^{12}}\) \\
\texttt{phy\_t}       & : \(\mathtt{2.5 \times 10^{11}}\) & \quad\quad \texttt{phy\_tdg}     & : \(\mathtt{2.1 \times 10^{11}}\) \\
\texttt{phy\_cz}      & : \(\mathtt{1.1 \times 10^{12}}\) & \quad\quad \texttt{phy\_measure} & : \(\mathtt{5.5 \times 10^{11}}\) \\
\end{tabular}
\end{center}
\end{verbbox}


\vspace{0.2em}

\addch{\underline{\textit{Extension to surface code resources}}}

Next, we replace the Steane code with the surface code~\cite{gidney2019efficient} and present the corresponding resource estimates in Box~\ref{box:surface_code}. The estimates are obtained using the implementation given in \cite{Qualtran}, with the parameters listed in Table~\ref{tab:iqpe_stack_table}. Note that the parameter \verb|GridSynth accuracy| shown in the table corresponds to the optimal synthesis accuracy.

\begin{figure}[htb]
    \centering
    \includegraphics[clip, trim=0cm 0cm 0cm 0cm, width=0.5\textwidth]{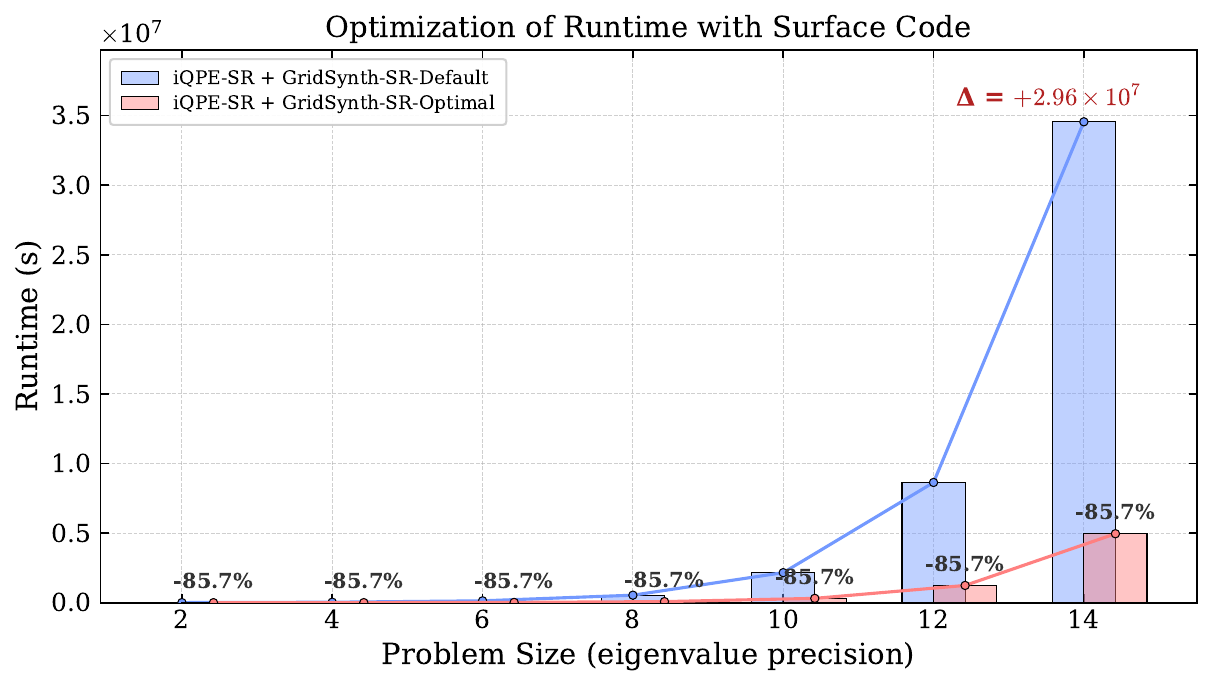}
    \caption{Surface-code runtime for the Hydrazine iQPE stack versus eigenvalue precision, comparing a fixed baseline GridSynth accuracy ($\epsilon = 10^{-10}$) against $\epsilon_{\mathrm{opt}}$. Using the surface-code parameters of Table~\ref{tab:iqpe_stack_table}, the optimal accuracy yields a nearly constant runtime reduction of ${\approx}\,85.7\%$ across all problem sizes.}
    \label{fig:rsopt_joint_qec}
\end{figure}

To assess scalability, we extend the study to a range of problem sizes. As illustrated in Figure \ref{fig:rsopt_joint_qec}, the optimal GridSynth accuracy achieves a consistent runtime reduction of approximately $85.7\%$ for \verb|ep| values ranging from $2$ to $14$.


\begin{verbbox}{Resources for surface code (with optimal GridSynth accuracy)}{box:surface_code}
\begin{center}
\begin{tabular}{@{}ll@{\hspace{2em}}ll@{}}
\texttt{runtime (s)} & : \(\mathtt{1.3 \times 10^{6}}\) & \quad\quad \texttt{num\_qubits} & : \(\mathtt{1.0 \times 10^{6}}\) \\
\end{tabular}
\end{center}
\end{verbbox}

\addch{\underline{\textit{Extension to partial FTQC}}}

\addch{The estimates above bracket two extremes of the fault-tolerance spectrum. Small codes, such as the Steane code with T-teleportation from the Reed-Muller code, are attractive in the early fault-tolerant regime for their low qubit overhead, whereas the surface code offers a scalable path to full fault tolerance at a substantially higher physical qubit cost. Partial fault tolerance on a topological patch bridges these regimes. In particular, the STAR architecture~\cite{akahoshi2024partially} error-corrects the Clifford gates while implementing rotations as space-time efficient analog operations, and its recent STAR ver.~3 refinement~\cite{toshio2026star} introduces a STAR-magic mutation protocol that realizes analog rotation gates within a single surface code patch using a Clifford+$T$+$\phi$ gate set.  
Recent application-level studies further motivate this regime, with STAR-based compilation of Trotterized time evolution enabling ground-state energy estimation of the 2D Hubbard model~\cite{akahoshi2025compilation} and chemically accurate quantum phase estimation~\cite{kanasugi2026enabling}. As a future extension, the presented logical stack model can be further composed with the resource estimates of STAR ver.~3, adding it as an alternative error-correction adapter alongside the Steane and surface code layers, so that the same iQPE and decomposition models feed directly into a partially fault-tolerant estimate.}


\textbf{Future work}

This case study can be extended to larger molecules, other EFT QPE algorithms, alternative universal QEC schemes, and decomposition strategies.
With this study, we highlight the utility of layer-specific models coalescing to form a composite digital twin of the EFT stack, enabling computationally tractable resource estimation with minimal expert involvement.

\subsection{Scenario III: Parameterized circuits and hardware constraints for optimization}
\label{sec:scenario_pqc}


\begin{tcolorbox}[colback=pink!20, colframe=pink!75, boxrule=1pt, arc=2pt,
left=5pt, right=5pt, top=5pt, bottom=5pt,
title={\vspace{0.2em}\textcolor{pink!25!black}{\sffamily{{Summary}}}},
fonttitle=\bfseries, coltitle=black]

\begin{itemize}[leftmargin=*]
    \item \textbf{Novelties used:} Full-stack flexibility ($\mathcal{N}$1), library support ($\mathcal{N}$2), and automated optimization ($\mathcal{N}$4).
    
    \item \textbf{Co-design variables:} Mixing ans\"atz, qubit connectivity and routing strategy, backend gate set, and zero-noise extrapolation (ZNE).
    
    \item \textbf{Key insight:} Reveals how sensitive performance is to ans\"atz and routing choices under noisy backend.
\end{itemize}

\end{tcolorbox}

\textbf{Background}

Parameterized quantum circuit (PQC) based variational algorithms underpin many NISQ-era algorithms due to their simplicity, ability to use shallow noisy circuits, and wide applicability across domains, such as optimization, machine learning, finance, and drug discovery \cite{bauer2020quantum, motta2022emerging, havlivcek2019supervised, bravyi2020quantum, riste2017demonstration, abbas2024challenges}. 
However, their performance is highly sensitive to parameterized ans\"atz choices, noise characteristics, and qubit connectivity. 
Hence, recent papers have explored the importance of resource estimation and appropriate design choices in variational algorithms \cite{bezdvek2025classical, patra2025resource}.

Here we study the quantum approximate optimization algorithm, or rather its generalization, the quantum alternating operator ans\"atze (QAOA) applied to MaxCut, a standard NP-complete combinatorial optimization problem. 
QAOA can be tailored to quadratic unconstrained binary optimization (QUBO) problems, with extensions to higher-order and constrained optimization problems~\cite{sunny2025extending,uotila2025higher}. 
Given its wide applicability, QAOA-based MaxCut presents a motivated benchmark for studying NISQ-era quantum resource estimation.

Most prior research implementing QAOA assumes idealized all-to-all connectivity, which is generally impractical in near-term devices. 
In this scenario, we systematically introduce constraints, such as connectivity and qubit routing, backend gates, and noisy hardware. 
We incorporate zero-noise extrapolation (ZNE) via linear and Richardson schemes and evaluate performance across noisy and mitigated settings.
AutoQuREO's generalized layer and hyperparameter abstractions, as described in Section~\ref{sec:autoqureo_layer}, enable these heterogeneous variables to coexist within the optimization framework.

By jointly exploring hyperparameter ranges, AutoQuREO reveals trade-offs among expressivity, routing overhead, and mitigation efficacy. 
These results generalize naturally to more complex QML workloads as additional data and models are incorporated.

\textbf{Setup}

The following stages define the resource estimation case study:
\begin{enumerate}
    \item \textbf{MaxCut problem}: Given a graph $G$, with $N$ nodes and $E$ edges, MaxCut aims to divide the nodes into sets $N_1$ and $N_2$ that maximize the number of edges connecting these two sets. Figure~\ref{s4_graph} shows the graph considered in this task.
    \begin{figure}[!htb]
        \centering
        \includegraphics[page=10, clip, trim=0.2cm 10.7cm 19.5cm 1.0cm,width=0.22\textwidth]{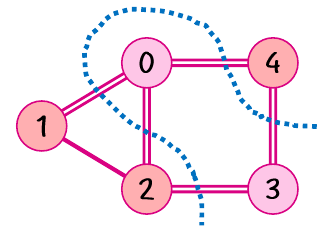}
        \caption{Input graph for MaxCut problem, with a chosen node ordering (the solution is independent of the ordering choice). For this simple example, the solution can be found by brute-force search, with the blue dotted line dividing the graph into two node sets: $\{0,3\}$ and $\{1,2,4\}$.}
        \label{s4_graph}
    \end{figure}
    
   \item \textbf{QAOA circuit} : Using $(N,E)$, we can construct a cost Hamiltonian ${H}_c$, as follows,
    \begin{equation}
        {H}_C = \sum_{(i,j)\in E} W_{ij} {Z}_i {Z}_j,
    \end{equation}
    where $W_{ij}$ is the weight of edge $(i,j)$, chosen here to be all 1. Minimization of ${H}_C$ corresponds to the solution of the MaxCut problem. 
    The QAOA is composed of alternating layers of trainable ${H}_c$ and mixer Hamiltonian $H_M$,
    \begin{equation}
        U(\boldsymbol{\alpha}, \boldsymbol{\beta}) = e^{-iH_C\alpha_1} e^{-iH_M\beta_1} ...\   e^{-iH_C\alpha_p} e^{-iH_M\beta_p},
    \end{equation}
    repeated for $p$ layers. $\boldsymbol{\alpha},\boldsymbol{\beta} \in \mathbb{R}^{p}$ correspond to vectors of angles defining the parameters for the evolution duration. Several choices of the mixer Hamiltonians have been proposed in literature\,\cite{blekos2024review}. In this work, we focus on the following:
    \begin{itemize}
        \item \textbf{Vanilla-X}: composed of the sum of $X$ operators over each qubit corresponding to the trainable mixer unitary layer $e^{-i\beta\sum_j X_j}$
        \item \textbf{Vanilla-Y}: composed of the sum of $Y$ operators over each qubit corresponding to the trainable mixer unitary layer $e^{-i\beta\sum_j Y_j}$
        \item \textbf{Multiangle-X}: similar to Vanilla-X, with independent trainable parameters for each qubit, corresponding to a trainable mixer unitary $e^{-i\sum_j \beta_j X_j}$
        \item \textbf{Grover}: inspired by the Grover Search quantum algorithm, the ans\"atz implements a Grover-like selective phase shift mixing operators combined with a trainable parameter. The exact form of the operator depends on which solutions are feasible among the search space. In our example, all solutions are feasible resulting in a mixer unitary of the form $e^{-i\beta |s\rangle\langle s|}$, where $\ket{s} = \ket{+}^{\otimes n}$.
        \item \textbf{Digitized Counter Adiabatic (DCA)}: inspired by adiabatic evolution, DCA adds further trainable counter-adiabatic driving terms, obtained through a nested commutator approach of the adiabatic gauge potential \cite{claeys2019floquet}, to the standard ans\"atze. The resulting ans\"atze is of the form,
        \begin{equation}
        U(\boldsymbol{\alpha}, \boldsymbol{\beta}, \boldsymbol{\gamma}) = e^{-iH_C\alpha_1} e^{-iH_M\beta_1} e^{-iH_{DCA}\gamma_1} ...\   e^{-iH_C\alpha_p} e^{-iH_M\beta_p}e^{-iH_{DCA}\gamma_p},
    \end{equation}
    \end{itemize}
    The specific case we consider here is with $H_M = \sum_i X_i$, and refer to the ans\"atz as DCA-X. The number of layers $p$ is also an important parameter; in general, larger layers result in higher solution accuracy. Here we choose $1$ and $3$ layers for our analysis.
    
    \item \textbf{Routing}: In hardware backends, the quantum circuit needs to be routed depending on the connectivity and allowed backend gates. Here we focus on two types of hardware whose connectivity is shown in Figure~\ref{s4_backend},
    \begin{itemize}
        \item \textbf{Trapped-ion backend}: Allows entangling gates between all pairs of qubits, and very high fidelity single-qubit gates, albeit with slower gate-speed (of the order of $\mu s$), as listed in Table~\ref{backend_table}.
        \item \textbf{Superconducting backend}: Allows entangling gates only between neighboring qubits, with much higher gate-speed (of the order of $n s$), as listed in Table~\ref{backend_table}.
    \end{itemize}
    For both backends, we use the SABRE routing method provided by Qiskit~\cite{qiskit2024}.
    
    \begin{figure}[!tbh]
        \centering
        \begin{subfigure}{0.48\linewidth}
        \centering
        \includegraphics[page=10, clip, trim=0cm 0cm 18cm 9cm,width=0.5\linewidth]{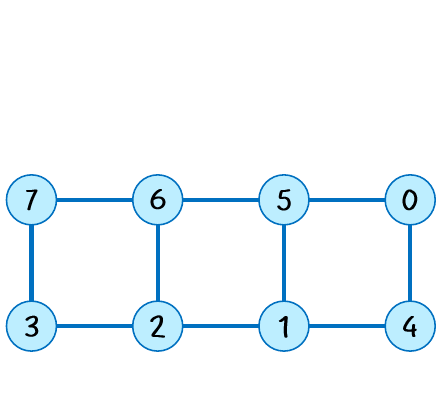}
        \caption{Superconducting backend}
        \end{subfigure}
        \begin{subfigure}{0.48\linewidth}
        \centering
        \includegraphics[page=10, clip, trim=20cm 0cm 0cm 9cm,width=0.4\linewidth]{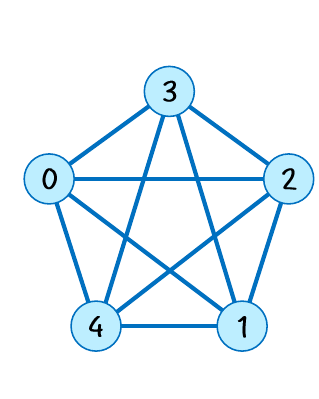}
        \caption{Trapped-ion backend}
        \end{subfigure}
        \caption{Backend connectivities. \textbf{(a)} Superconducting backend with grid connectivity among $4\times2$ qubits, and \textbf{(b)} Trapped-ion backend with all-to-all connectivity among $5$ qubits are considered in this scenario. Note: mapping is agnostic to the arbitrary physical qubit IDs. The gate-times, fidelity values, measurement error and $T_1,T_2$ values are mentioned in Table~\ref{backend_table}. We use the same time and fidelity values for all qubits in a backend, and the $T_1,T_2$ values are randomly sampled for each qubit from the range of experimentally obtainable values.}
        \label{s4_backend}
    \end{figure}
    
\begin{table}[!tbh]
\centering
\caption{Runtime and fidelities of quantum gates, measurements and $T_1,T_2$ for superconducting and trapped-ion backend obtained from recent literature.}
\begin{tabular}{|c|cc|cc|}
\hline
\multirow{2}{*}{}     & \multicolumn{2}{c|}{\textbf{Superconducting}}          & \multicolumn{2}{c|}{\textbf{Trapped-ion}}             \\ \cline{2-5} 
                      & \multicolumn{1}{c|}{\textbf{Gate time}} & \textbf{Fidelity} & \multicolumn{1}{c|}{\textbf{Gate time}} & \textbf{Fidelity} \\ \hline
$1$-qubit rotation & \multicolumn{1}{c|}{$8 ns$}        & $99.997\%$\cite{rower2024suppressing}        & \multicolumn{1}{c|}{$13 \mu s$}    & $99.99998\%$ \cite{smith2025single}     \\ \hline
$CX$ gate             & \multicolumn{1}{c|}{$60 ns$}       & $99.96\%$ \cite{lin202524}        & \multicolumn{1}{c|}{$60 \mu s$}    & $99.97\%$ \cite{loschnauer2025scalable}        \\ \hline
Measurement           & \multicolumn{1}{c|}{-}             & $99.9\%$\cite{wang202499}        & \multicolumn{1}{c|}{-}             & $99.9993\%$ \cite{sotirova2024high}       \\ \hline
$T_1$                 & \multicolumn{1}{c|}{$50-300\mu s$} & -                 & \multicolumn{1}{c|}{$1-10s$}       & -                 \\ \hline
$T_2$                 & \multicolumn{1}{c|}{$50-200\mu s$} & -                 & \multicolumn{1}{c|}{$1-10s$}       & -                 \\ \hline
\end{tabular}
\label{backend_table}
\end{table}

    \item \textbf{Circuit optimization}: The routed quantum circuits are further simplified by using Qiskit's built-in circuit optimization routine. The level of optimization ranges from $0$ to $3$, with level $0$ performing no optimization and just the minimal amount of work to make the circuit runnable on the selected backend, and level $3$ spending the most amount of effort (and typically runtime) to try to optimize the circuit by combining multiple gates into more efficient operations. Here, the optimization level is set to $3$.

    \item \textbf{Error mitigation}: In NISQ devices, error-mitigation is an important part of obtaining useful information from quantum measurements. Here, we use the ZNE technique~\cite{giurgica2020digital}, which runs the circuit at various noise levels by increasing the circuit depth and extrapolates to the zero-noise measurement outcome. We compare the bare noisy circuit against ZNE using linear extrapolation with global circuit folding, and Richardson extrapolation with random gate folding. We use the ZNE modules from the Mitiq library~\cite{mitiq}.

    \item \textbf{Training preparation}: To train the routed and optimized parametrized circuit, we need a classical optimizer. Here we choose the gradient-free COBYLA optimizer~\cite{bonet2023performance} with maximum iterations 1000 and tolerance $10^{-6}$. For each evaluation, we set the number of measurement shots to be 10000.

    \item \textbf{Training execution}: This layer simply runs the final training pipeline based on all the configurations set in the above layers.

    \item \textbf{MaxCut evaluation}: The solution is evaluated by measuring the approximation ratio defined as,
    \begin{equation}
        \text{Approximation Ratio (AR)} = \frac{\langle\psi(\theta^*)| \hat{H}_C|\psi(\theta^*)\rangle}{\text{min\_eigval}(H_C)}
    \end{equation}
    where $|\psi(\theta^*)\rangle$ is the optimized ans\"atze and $\text{min\_eigval}(H_C)$ is the minimum eigenvalue of $H_C$. The correct solution corresponds to $\text{AR} = 1$, with higher values corresponding to better accuracy of the solution.
\end{enumerate}

Given these layers and corresponding hyperparameter choices, our aim is to identify how these choices affect the final approximation ratio and the resources required. 
The hyperparameter choices and resources are summarized in Table~\ref{fig:s4_overview}.

\begin{table}[!tbh]
\caption{QRE layers and associated hyperparameter choices considered in this scenario. The optimization is done to minimize runtime and maximize the Approximation Ratio.}
\label{fig:s4_overview}
\centering
\small
\begin{minipage}{0.73\textwidth}
\centering
\begin{tabular}{|>{\centering\arraybackslash}m{3cm}|
                  >{\centering\arraybackslash}m{3cm}|
                  >{\centering\arraybackslash}m{3.5cm}|}
\hline
\rowcolor{lavender}
\textbf{Layers} & \textbf{Hyperparameters} & \textbf{Ranges}\\
\hline
MaxCut problem & Graph & \\
\hline
\multirow{2}{*}{\makecell{QAOA circuit}} & QAOA mixer & [Vanilla-X, Multiangle-X, DCA-X, Vanilla-Y, Grover] \\
\cline{2-3} & Layers & [1, 3] \\
\hline
\multirow{2}{*}{\makecell{Routing}} & Backend & [Trapped Ion all-to-all, Superconducting grid] \\
\cline{2-3} & Method & [SABRE] \\
\hline
Circuit optimization & Optimization level & [3]\\
\hline
Error mitigation & \small Strategy  & [None, ZNE Linear, ZNE Richardson] \\
\hline
\multirow{4}{*}{\makecell{Training preparation}} & Tolerance & [1e-6] \\
\cline{2-3} & Max iterations & [1000] \\
\cline{2-3} & Optimizer & [COBYLA] \\
\cline{2-3} & Shots & [10000] \\
\hline
Train execution & - &\\
\hline
MaxCut evaluation & -  &\\
\hline
\end{tabular}
\end{minipage}%
\begin{minipage}{0.27\textwidth}
\centering
\begin{tabular}{|>{\centering\arraybackslash}m{3.5cm}|}
\hline
\rowcolor{pink}
\textbf{Resources} \\
\hline
Number of qubits  \\
\hline
Approximation ratio  \\
\hline
Runtime  \\
\hline
\end{tabular}
\end{minipage}
\end{table}

\textbf{Results}

\begin{figure}[!tbh]
    \includegraphics[width=1\linewidth]{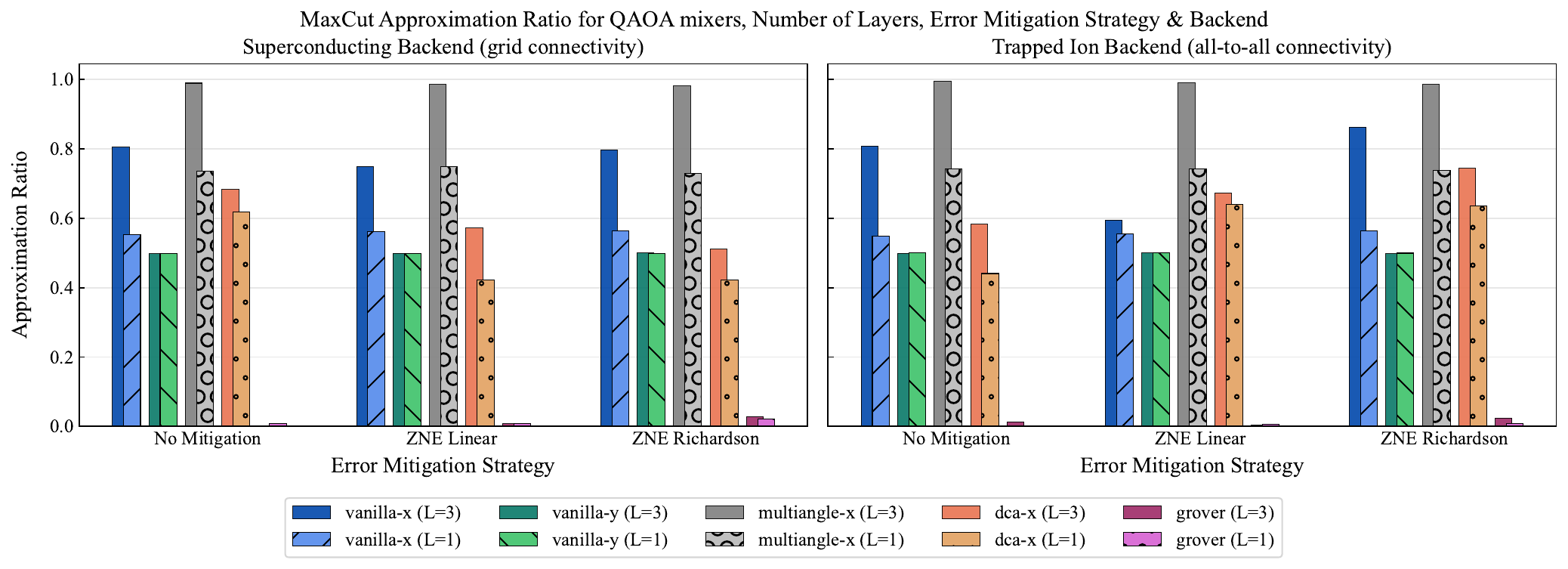}
    \caption{Comparison of Approximation Ratio for the MaxCut problem using different ans\"atze choices (marked in different colors), number of layers (marked in solid and patterns), backend and error-mitigation strategies in the QRE stack.}
    \label{fig:s4_ar}
\end{figure}

\begin{figure}[!tbh]
    \includegraphics[width=1\linewidth]{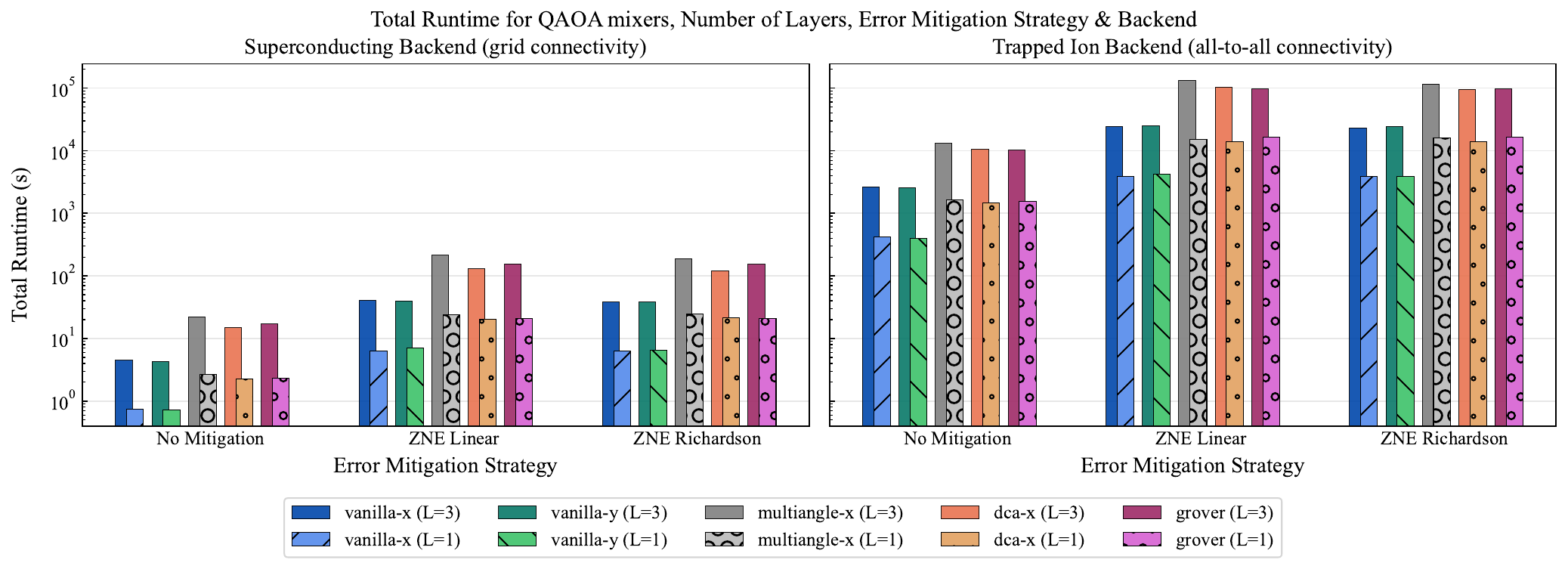}
    \caption{Comparison of total runtime for the MaxCut problem using different ans\"atze choices (marked in different colors), number of layers (marked in solid and patterns), backend, and error-mitigation strategies in the QRE stack.}
    \label{fig:s4_runtime}
\end{figure}

For each backend, we plot the AR and runtime for various error-mitigation strategies, ans\"atze choice and number of layers in Figures~\ref{fig:s4_ar} and \ref{fig:s4_runtime}. 
To remove biases introduced by parameter initialization, we run $5$ independent trials for each configuration and choose the best AR as the outcome for that configuration. 
Each ans\"atze choice is marked in separate colors, with solid color for $3$ layers and patterned for 1 layer, and the error-mitigation strategies shown on the x-axis. 
Using Figure~\ref{fig:s4_ar}, a few insightful trends can be identified:
\begin{itemize}
    \item Most prominently, multiangle-X ans\"atze consistently outperforms all other ans\"atze choices which can be attributed to the additional trainable parameters. On the other hand, Grover ans\"atze shows very low AR for all configurations.
    \item Increasing the number of layers has improved AR for all ans\"atze choices except Vanilla-Y, suggesting that its maximum possible accuracy might be limited for this problem.
    \item The impact of error mitigation on the outcome is dependent on other factors. The change is most noticeable for Vanilla-X and DCA-X. Vanilla-X with a superconducting backend shows similar performance with and without error mitigation, with a slight dip in AR for the ZNE linear scheme. On the other hand, for the trapped-ion backend, the ZNE linear scheme shows a significant dip, whereas the ZNE Richardson scheme slightly improves the AR. The trend for DCA-X is more clearly visible. For superconducting hardware, performance degrades as we move from bare circuits to the ZNE Richardson scheme, whereas the opposite is true for the trapped-ion backend. The circuits in DCA-X are deeper than the other ans\"atze choices except Grover, which explains the degradation in performance when using error mitigation on superconducting hardware with short coherence timescales ($\sim \mu s$) compared to increasing AR on trapped ion, which can support much longer coherence times ($\sim s$).
    
\end{itemize}

A similar comparison of their total runtime shows that multiangle-X, DCA-X, and Grover require much larger total runtime, compared to other ans\"atze choices. 
Since multiangle-X has many more trainable parameters, it requires many more function evaluations in the classical optimizer COBYLA, resulting in a larger runtime despite having similar depth to vanilla-X and vanilla-Y. 
Grover and DCA-X both require deeper circuits by design, and DCA-X also contains additional trainable parameters. 
Due to much slower gate-times, the trapped ion platform requires $\sim 10^3-10^4$ times longer. 
Additionally, the error-mitigation strategies add an order of magnitude of overhead to the total runtime in all cases.

To quantify the impact of various hyperparameter choices on the AR, we use the $\eta^2$ effect size, which measures the contribution of each hyperparameter to the AR. 
Unlike $p$-value, which gives a binary decision if the correlation between two variables is statistically significant, $\eta^2$ measures the size of the correlation. 
The largest impact is due to the QAOA type reflected in its $\eta^2$ value of $0.9073$. 
The next significant contributing factor is the number of layers with an $\eta^2$ value of $0.0357$, as more layers have generally demonstrated improved AR values. 
The contribution from error-mitigation and backend, on the other hand, has been mixed and dependent on specific configuration choices.
Thus, for larger instances, where conducting extensive DSE becomes intractable, the user can focus their efforts on tuning the design choices with higher sensitivity.

\begin{figure}[htb]
    \centering
    \includegraphics[width=0.45\linewidth]{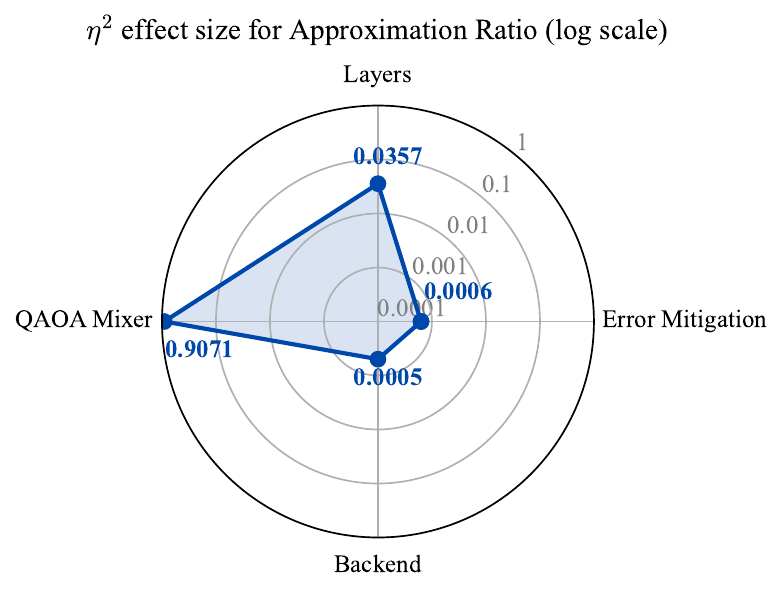}
    \caption{$\eta^2$ effect size quantifying the impact of each hyperparameter on Approximation Ratio. $\eta^2 \sim 0.01$ corresponds to a small effect, $\eta^2 \sim 0.06$ corresponds to a medium effect and $\eta^2 \sim 0.14$ higher values correspond to a large effect.}
    \label{fig:s4_eta2}
\end{figure}

Using AutoQuREO, we can analyze the impact of various hyperparameters and backends on AR and runtime within a single framework, and identify the best configurations that maximize AR while minimizing runtime. 
For heuristic algorithms like QAOA, such analysis is especially helpful for understanding which hyperparameter choices to make as the number of qubits increases to obtain reasonable answers on noisy hardware. 
Due to its flexibility, AutoQuREO can be used to study a much broader range of variational algorithms~\cite{ni2025adaptive,kundu2024kanqas,wang2019accelerated} and to analyze useful trends~\cite{sanz2026efficiently,wadhwa2026model} to predict optimal choices under realistic hardware constraints.
This transition from idealized simulation assumptions to realistic hardware-aware digital twins is precisely the intended use case of AutoQuREO workflow introduced in Section~\ref{sec:autoqureo3}.

\addtocontents{toc}{\vspace{-1.0em}}
\section{Discussion}
\label{sec:conclusion}

This research presents the AutoQuREO quantum resource estimation (QRE) framework.
AutoQuREO is designed with a design philosophy to democratize full-stack QRE across heterogeneous layers of the quantum computing stack, spanning algorithms, compilation, error correction, and hardware backends. 
Rather than constraining users to a fixed abstraction (e.g., FTQC-centric pipelines), the framework exposes layer-wise modularity, allowing researchers, for example, to focus on a target layer while leveraging standardized tooling for the remaining stack for co-design and optimization. 
This is enabled through a reusable library of codes and surrogate models, reducing the overhead of cross-domain expertise. 
AutoQuREO bridges estimation and optimization by embedding multi-objective design space exploration (DSE) directly into the workflow, enabling not only prediction of resource metrics but also selection of optimal configurations for hardware deployment. 
The framework enables flexibility in analysis, allowing users to arbitrarily visualize trade-offs across hyperparameters and resources. 
A key distinguishing feature is the integration of neuro-symbolic (NeSy) explainable AI (XAI) resource surrogate modeling.
This integrates the advantages from expressive learning (e.g., neural networks, neuro-evolution) with interpretable symbolic regression, enabling scalability across regimes while remaining amenable to analytical reasoning and formalization.

Through the representative case studies, we demonstrate AutoQuREO's ability to uncover non-trivial insights in multi-layer quantum solutions. 
The scenario-driven methodology enables users to formalize questions of interest, such as optimizing a target resource while constraining others, and systematically explore the resulting trade-offs towards realistic deployment. 

Looking forward, several backend and systems-level extensions can further enhance the applicability of AutoQuREO.
Integration with emerging intermediate representations such as Bloqs, QREF, QIR~\cite{QIRspec}, and MLIR~\cite{hopf2026integrating,stade2024towards,mccaskey2021mlir,ittah2022qiro}, can provide interoperability across quantum software ecosystems. 
Additionally, GPU-accelerated QRE and model training can significantly accelerate surrogate synthesis and large-scale DSE. 
Another promising direction of interest is compiler phase ordering~\cite{quetschlich2023compiler,xu2025optimizing}, treating compilation passes as tunable parameters that influence overall resource allocation. 
Such extensions would further blur the boundary between quantum compilation in deployment pipelines and QRE, enabling holistic optimization of quantum middleware.

AutoQuREO can benefit from advanced automation methods, such as active learning~\cite{balaprakash2013active}, to achieve surrogate modeling with less data. 
Static analysis techniques can complement compiler-driven QRE by providing formal constraints and invariants. 
Enhancements to symbolic regression, including the incorporation of reusable gadgets~\cite{grayeli2024symbolic}, can improve compositionality and interpretability of learned models. 
Furthermore, large language models (LLMs) offer opportunities for autoformalization~\cite{szegedy2020promising,wu2022autoformalization,ren2026merlean} of code-to-model pipelines, while the adapter abstraction can be extended as a model context protocol (MCP)~\cite{hou2025model}, enabling seamless interoperability between heterogeneous modeling agents.

In the broader context of AutoQC~\cite{sarkar2024automated}, AutoQuREO contributes a critical component toward quantifying quantum advantage in an automated quantum solution workflow. 
A natural extension is the integration of AutoQuREO with quantum algorithm design automation (QADA), in which resource estimation and algorithm synthesis interact in a closed loop.

Finally, we envision AutoQuREO as a community-driven platform for advancing full-stack quantum computing research.
The extensible library of layers, models, adapters, and scenarios is designed to facilitate contributions from researchers across domains, fostering a shared ecosystem of reusable components and benchmarks. 

\section*{Acknowledgements}

The authors thank Kushagra Garg for insightful discussions regarding error composability; and Muhilan Murugesan and Tania Sidana for discussions on correction schemes for modeling. We thank Yasuhiro Endo, Hirotaka Oshima, Shinji Kikuchi, Shintaro Sato, and Vivek Mahajan for valuable project guidance and comments on the article.

The authors used AI-assisted editing tools to improve grammar and clarity during manuscript preparation. 
The manuscript has been carefully reviewed by the authors, and they take full responsibility for the final content.

\section*{Author contributions}

A.S. and H.O. conducted the requirement analysis to identify gaps in current methods, conceptualized the distinguishing novelties, designed the software workflow and led the AutoQuREO implementation.
S.N.A. integrated PySR with NEAT and Optuna for surrogate synthesis and conducted the LER-PER results presented in \S\;\ref{sec:scenario_qec}.
A.P. conducted the co-design scenario in \S\;\ref{sec:scenario_hsim}, H.O. and A.S. conducted the co-design scenario in \S\;\ref{sec:scenario_dcmp}, and R.B. conducted the co-design scenario in \S\;\ref{sec:scenario_pqc}.
P.P.K. conducted the model benchmarking studies in \S\;\ref{sec:scenario_iqpe_dcmp}. 
K.S. was instrumental in initiating the project, securing necessary resources and overseeing project road map. 
All authors participated in the review and editing of the manuscript.


\bibliographystyle{unsrt}
\bibliography{references}

@article{harrigan2024expressing,
  title={Expressing and analyzing quantum algorithms with qualtran},
  author={Harrigan, Matthew P and Khattar, Tanuj and Yuan, Charles and Peduri, Anurudh and Yosri, Noureldin and Malone, Fionn D and Babbush, Ryan and Rubin, Nicholas C},
  journal={arXiv preprint arXiv:2409.04643},
  year={2024}
}

@article{campbell2026resource,
  title={Resource Estimation via Efficient Compilation of Key Quantum Primitives},
  author={Campbell, Colin and Rines, Rich and Omole, Victory and Oberoi, Tina and Goiporia, Palash and Roy, Rayat and Cline, R Peyton and Jones, Eric B and Tomesh, Teague},
  journal={arXiv preprint arXiv:2604.01376},
  year={2026}
}

@misc{PsiQ, 
  title={PSIQ/Bartiq: Bartiq}, 
  url={https://github.com/PsiQ/bartiq}, 
  journal={GitHub}, 
  author={PsiQ},
  note= {Accessed: 2026-02-23}
}

@article{mishra2026eqisa,
  title={{EQISA}: Energy-efficient Quantum Instruction Set Architecture using Sparse Dictionary Learning},
  author={Mishra, Sibasish and Sarkar, Aritra and Feld, Sebastian},
  journal={arXiv preprint arXiv:2603.20646},
  year={2026}
}

@article{suzuki2022quantum,
  title={Quantum error mitigation as a universal error reduction technique: Applications from the NISQ to the fault-tolerant quantum computing eras},
  author={Suzuki, Yasunari and Endo, Suguru and Fujii, Keisuke and Tokunaga, Yuuki},
  journal={PRX quantum},
  volume={3},
  number={1},
  pages={010345},
  year={2022},
  publisher={APS}
}

@inproceedings{hopf2026integrating,
  title={Integrating Quantum Software Tools with (in) MLIR},
  author={Hopf, Patrick and Ochoa, Erick and Stade, Yannick and Rovara, Damian and Quetschlich, Nils and Florea, Ioan Albert and Izaac, Josh and Wille, Robert and Burgholzer, Lukas},
  booktitle={Proceedings of the Supercomputing Asia and International Conference on High Performance Computing in Asia Pacific Region},
  pages={42--54},
  year={2026}
}

@inproceedings{wahl2023zero,
  title={Zero noise extrapolation on logical qubits by scaling the error correction code distance},
  author={Wahl, Misty A and Mari, Andrea and Shammah, Nathan and Zeng, William J and Ravi, Gokul Subramanian},
  booktitle={2023 IEEE International Conference on Quantum Computing and Engineering (QCE)},
  volume={1},
  pages={888--897},
  year={2023},
  organization={IEEE}
}

@article{moore2021statistical,
  title={Statistical approach to quantum phase estimation},
  author={Moore, Alexandria J and Wang, Yuchen and Hu, Zixuan and Kais, Sabre and Weiner, Andrew M},
  journal={New Journal of Physics},
  volume={23},
  number={11},
  pages={113027},
  year={2021},
  publisher={IOP Publishing}
}

@article{ni2025adaptive,
  title={An Adaptive Mixer Allocation Strategy for the Quantum Alternating Operator Ansatz},
  author={Ni, Xiao-Hui and Wu, Yu-Sen and Cai, Bin-Bin and Li, Wen-Min and Qin, Su-Juan and Gao, Fei},
  journal={Advanced Quantum Technologies},
  pages={e00487},
  year={2025},
  publisher={Wiley Online Library}
}

@article{breuckmann2021quantum,
  title={Quantum low-density parity-check codes},
  author={Breuckmann, Nikolas P and Eberhardt, Jens Niklas},
  journal={PRX quantum},
  volume={2},
  number={4},
  pages={040101},
  year={2021},
  publisher={APS}
}

@article{tham2025optimized,
  title={Optimized Clifford Noise Reduction: Theory, Simulations and Experiments},
  author={Tham, Edwin and Delfosse, Nicolas},
  journal={Quantum},
  volume={9},
  pages={1829},
  year={2025},
  publisher={Verein zur F{\"o}rderung des Open Access Publizierens in den Quantenwissenschaften}
}

@article{wadhwa2026model,
  title={Model selection in hybrid quantum neural networks with applications to quantum transformer architectures},
  author={Wadhwa, Harsh and Bhowmick, Rahul and Raj, Naipunnya and Sangle, Rajiv and Bhat, Ruchira V and Sabapathy, Krishnakumar},
  journal={arXiv preprint arXiv:2603.21749},
  year={2026}
}

@article{sanz2026efficiently,
  title={Efficiently architecting VQAs: Expressibility--Trainability--Resources Pareto-Optimality},
  author={Sanz, Rodrigo M and Angles-Castillo, Andreu and Alarcon, Eduard and Almudever, Carmen G},
  journal={arXiv preprint arXiv:2603.22142},
  year={2026}
}

@article{gratsea2025achieving,
  title={Achieving Utility-Scale Applications through Full Stack Co-Design of Fault Tolerant Quantum Computers},
  author={Gratsea, Katerina and Otten, Matthew},
  journal={arXiv preprint arXiv:2510.26547},
  year={2025}
}

@article{purohit2024building,
  title={Building a quantum-ready ecosystem},
  author={Purohit, Abhishek and Kaur, Maninder and Seskir, Zeki Can and Posner, Matthew T and Venegas-Gomez, Araceli},
  journal={IET Quantum Communication},
  volume={5},
  number={1},
  pages={1--18},
  year={2024},
  publisher={Wiley Online Library}
}

@article{bowen2025design,
  title={Design and efficiency in graph-state computation},
  author={Bowen, Greg and Caesura, Athena and Devitt, Simon and Vijayan, Madhav Krishnan},
  journal={arXiv preprint arXiv:2502.18985},
  year={2025}
}

@article{fan2024lego_hqec,
  title={LEGO\_HQEC: Automating the Analysis, Construction, and Decoding of Holographic Quantum Codes},
  author={Fan, Junyu and Steinberg, Matthew and Jahn, Alexander and Cao, Chunjun and Sarkar, Aritra and Feld, Sebastian},
  journal={arXiv preprint arXiv:2410.22861},
  year={2024}
}

@article{hoefler2023disentangling,
  title={Disentangling hype from practicality: On realistically achieving quantum advantage},
  author={Hoefler, Torsten and H{\"a}ner, Thomas and Troyer, Matthias},
  journal={Communications of the ACM},
  volume={66},
  number={5},
  pages={82--87},
  year={2023},
  publisher={ACM New York, NY, USA}
}

@article{bultrini2023battle,
  title={The battle of clean and dirty qubits in the era of partial error correction},
  author={Bultrini, Daniel and Wang, Samson and Czarnik, Piotr and Gordon, Max Hunter and Cerezo, Marco and Coles, Patrick J and Cincio, Lukasz},
  journal={Quantum},
  volume={7},
  pages={1060},
  year={2023},
  publisher={Verein zur F{\"o}rderung des Open Access Publizierens in den Quantenwissenschaften}
}

@article{wang2019accelerated,
  title={Accelerated variational quantum eigensolver},
  author={Wang, Daochen and Higgott, Oscar and Brierley, Stephen},
  journal={Physical review letters},
  volume={122},
  number={14},
  pages={140504},
  year={2019},
  publisher={APS}
}

@article{litinski2019game,
  title={A game of surface codes: Large-scale quantum computing with lattice surgery},
  author={Litinski, Daniel},
  journal={Quantum},
  volume={3},
  pages={128},
  year={2019},
  publisher={Verein zur F{\"o}rderung des Open Access Publizierens in den Quantenwissenschaften}
}

@article{kundu2024kanqas,
  title={Kanqas: Kolmogorov-arnold network for quantum architecture search},
  author={Kundu, Akash and Sarkar, Aritra and Sadhu, Abhishek},
  journal={EPJ Quantum Technology},
  volume={11},
  number={1},
  pages={76},
  year={2024},
  publisher={Springer}
}

@article{mohseni2024build,
  title={How to build a quantum supercomputer: Scaling from hundreds to millions of qubits},
  author={Mohseni, Masoud and Scherer, Artur and Johnson, K Grace and Wertheim, Oded and Otten, Matthew and Aadit, Navid Anjum and Alexeev, Yuri and Bresniker, Kirk M and Camsari, Kerem Y and Chapman, Barbara and others},
  journal={arXiv preprint arXiv:2411.10406},
  year={2024}
}

@inproceedings{shende2005synthesis,
  title={Synthesis of quantum logic circuits},
  author={Shende, Vivek V and Bullock, Stephen S and Markov, Igor L},
  booktitle={Proceedings of the 2005 Asia and South Pacific Design Automation Conference},
  pages={272--275},
  year={2005}
}

@article{krol2024beyond,
  title={Beyond quantum Shannon decomposition: Circuit construction for n-qubit gates based on block-ZXZ decomposition},
  author={Krol, Anna M and Al-Ars, Zaid},
  journal={Physical Review Applied},
  volume={22},
  number={3},
  pages={034019},
  year={2024},
  publisher={APS}
}

@article{kremer2024practical,
  title={Practical and efficient quantum circuit synthesis and transpiling with reinforcement learning},
  author={Kremer, David and Villar, Victor and Paik, Hanhee and Duran, Ivan and Faro, Ismael and Cruz-Benito, Juan},
  journal={arXiv preprint arXiv:2405.13196},
  year={2024}
}

@article{muthyala2025symantic,
  title={Symantic: An efficient symbolic regression method for interpretable and parsimonious model discovery in science and beyond},
  author={Muthyala, Madhav R and Sorourifar, Farshud and Peng, You and Paulson, Joel A},
  journal={Industrial \& Engineering Chemistry Research},
  volume={64},
  number={6},
  pages={3354--3369},
  year={2025},
  publisher={ACS Publications}
}

@article{bonet2023performance,
  title={Performance comparison of optimization methods on variational quantum algorithms},
  author={Bonet-Monroig, Xavier and Wang, Hao and Vermetten, Diederick and Senjean, Bruno and Moussa, Charles and B{\"a}ck, Thomas and Dunjko, Vedran and O'Brien, Thomas E},
  journal={Physical Review A},
  volume={107},
  number={3},
  pages={032407},
  year={2023},
  publisher={APS}
}

@inproceedings{balaprakash2013active,
  title={Active-learning-based surrogate models for empirical performance tuning},
  author={Balaprakash, Prasanna and Gramacy, Robert B and Wild, Stefan M},
  booktitle={2013 IEEE International Conference on Cluster Computing (CLUSTER)},
  pages={1--8},
  year={2013},
  organization={IEEE}
}

@article{muthyala2024torchsisso,
  title={Torchsisso: a pytorch-based implementation of the sure independence screening and sparsifying operator for efficient and interpretable model discovery},
  author={Muthyala, Madhav and Sorourifar, Farshud and Paulson, Joel A},
  journal={Digital Chemical Engineering},
  volume={13},
  pages={100198},
  year={2024},
  publisher={Elsevier}
}

@article{ring1997child,
  title={Child: A first step towards continual learning},
  author={Ring, Mark B},
  journal={Machine Learning},
  volume={28},
  number={1},
  pages={77--104},
  year={1997},
  publisher={Springer}
}

@article{beverland2022assessing,
  title={Assessing requirements to scale to practical quantum advantage},
  author={Beverland, Michael E and Murali, Prakash and Troyer, Matthias and Svore, Krysta M and Hoefler, Torsten and Kliuchnikov, Vadym and Low, Guang Hao and Soeken, Mathias and Sundaram, Aarthi and Vaschillo, Alexander},
  journal={arXiv preprint arXiv:2211.07629},
  year={2022}
}

@article{sarkar2024automated,
  title={Automated quantum software engineering},
  author={Sarkar, Aritra},
  journal={Automated Software Engineering},
  volume={31},
  number={1},
  pages={36},
  year={2024},
  publisher={Springer}
}

@article{daguerre2025code,
  title={Code switching revisited: Low-overhead magic state preparation using color codes},
  author={Daguerre, Lucas and Kim, Isaac H},
  journal={Physical Review Research},
  volume={7},
  number={2},
  pages={023080},
  year={2025},
  publisher={APS}
}

@article{meurer2017sympy,
  title={SymPy: symbolic computing in Python},
  author={Meurer, Aaron and Smith, Christopher P and Paprocki, Mateusz and {\v{C}}ert{\'\i}k, Ond{\v{r}}ej and Kirpichev, Sergey B and Rocklin, Matthew and Kumar, AMiT and Ivanov, Sergiu and Moore, Jason K and Singh, Sartaj and others},
  journal={PeerJ Computer Science},
  volume={3},
  pages={e103},
  year={2017},
  publisher={PeerJ Inc.}
}

@inproceedings{suchara2013qure,
  title={Qure: The quantum resource estimator toolbox},
  author={Suchara, Martin and Kubiatowicz, John and Faruque, Arvin and Chong, Frederic T and Lai, Ching-Yi and Paz, Gerardo},
  booktitle={2013 IEEE 31st international conference on computer design (ICCD)},
  pages={419--426},
  year={2013},
  organization={IEEE}
}

@incollection{mcgeoch2002using,
  title={Using finite experiments to study asymptotic performance},
  author={McGeoch, Catherine and Sanders, Peter and Fleischer, Rudolf and Cohen, Paul R and Precup, Doina},
  booktitle={Experimental algorithmics: from algorithm design to robust and efficient software},
  pages={93--126},
  year={2002},
  publisher={Springer}
}

@article{graham1982gprof,
  title={Gprof: A call graph execution profiler},
  author={Graham, Susan L and Kessler, Peter B and McKusick, Marshall K},
  journal={ACM Sigplan Notices},
  volume={17},
  number={6},
  pages={120--126},
  year={1982},
  publisher={ACM New York, NY, USA}
}

@article{pham2025integrating,
  title={Integrating Resource Analyses via Resource Decomposition},
  author={Pham, Long and Niu, Yue and Glover, Nathan and Saad, Feras and Hoffmann, Jan},
  journal={Proceedings of the ACM on Programming Languages},
  volume={9},
  number={OOPSLA2},
  pages={3811--3840},
  year={2025},
  publisher={ACM New York, NY, USA}
}

@article{sivarajah2021t,
  title={t| ket>: a retargetable compiler for NISQ devices},
  author={Sivarajah, Seyon and Dilkes, Silas and Cowtan, Alexander and Simmons, Will and Edgington, Alec and Duncan, Ross},
  journal={Quantum Science \& Technology},
  volume={6},
  number={1},
  pages={014003},
  year={2021},
  publisher={IOP Publishing}
}

@article{colledan2025flexible,
  title={Flexible type-based resource estimation in quantum circuit description languages},
  author={Colledan, Andrea and Dal Lago, Ugo},
  journal={Proceedings of the ACM on Programming Languages},
  volume={9},
  number={POPL},
  pages={1386--1416},
  year={2025},
  publisher={ACM New York, NY, USA}
}

@article{turney2012psi4,
  title={Psi4: an open-source ab initio electronic structure program},
  author={Turney, Justin M and Simmonett, Andrew C and Parrish, Robert M and Hohenstein, Edward G and Evangelista, Francesco A and Fermann, Justin T and Mintz, Benjamin J and Burns, Lori A and Wilke, Jeremiah J and Abrams, Micah L and others},
  journal={Wiley Interdisciplinary Reviews: Computational Molecular Science},
  volume={2},
  number={4},
  pages={556--565},
  year={2012},
  publisher={Wiley Online Library}
}

@inproceedings{li2019tackling,
  title={Tackling the qubit mapping problem for NISQ-era quantum devices},
  author={Li, Gushu and Ding, Yufei and Xie, Yuan},
  booktitle={Proceedings of the twenty-fourth international conference on architectural support for programming languages and operating systems},
  pages={1001--1014},
  year={2019}
}

@book{gamma1995design,
  title={Design patterns: elements of reusable object-oriented software},
  author={Gamma, Erich},
  year={1995},
  publisher={Pearson Education India}
}

@article{pham2024robust,
  title={Robust resource bounds with static analysis and Bayesian inference},
  author={Pham, Long and Saad, Feras A and Hoffmann, Jan},
  journal={Proceedings of the ACM on Programming Languages},
  volume={8},
  number={PLDI},
  pages={76--101},
  year={2024},
  publisher={ACM New York, NY, USA}
}

@article{deutsch1989quantum,
  title={Quantum computational networks},
  author={Deutsch, David Elieser},
  journal={Proceedings of the royal society of London. A. mathematical and physical sciences},
  volume={425},
  number={1868},
  pages={73--90},
  year={1989},
  publisher={The Royal Society London}
}

@misc{BenchQ,
    author = {{Zapata AI}}, 
    title = {benchq: {R}esource estimation for fault-tolerant quantum computation.},
    year = {2023},
    url = {https://github.com/zapatacomputing/benchq}
}

@misc{Qualtran,
    author = {Google},
    title = {Qualtran documentation},
    year = {2023},
    url = {https://qualtran.readthedocs.io/en/latest/}
}

@article{gidney2025factor,
  title={How to factor 2048 bit RSA integers with less than a million noisy qubits},
  author={Gidney, Craig},
  journal={arXiv preprint arXiv:2505.15917},
  year={2025}
}

@article{bagherimehrab2022nearly,
  title={Nearly optimal quantum algorithm for generating the ground state of a free quantum field theory},
  author={Bagherimehrab, Mohsen and Sanders, Yuval R and Berry, Dominic W and Brennen, Gavin K and Sanders, Barry C},
  journal={PRX Quantum},
  volume={3},
  number={2},
  pages={020364},
  year={2022},
  publisher={APS}
}

@article{sanders2020compilation,
  title={Compilation of fault-tolerant quantum heuristics for combinatorial optimization},
  author={Sanders, Yuval R and Berry, Dominic W and Costa, Pedro CS and Tessler, Louis W and Wiebe, Nathan and Gidney, Craig and Neven, Hartmut and Babbush, Ryan},
  journal={PRX quantum},
  volume={1},
  number={2},
  pages={020312},
  year={2020},
  publisher={APS}
}

@article{javadiabhari2015scaffcc,
  title={ScaffCC: Scalable compilation and analysis of quantum programs},
  author={JavadiAbhari, Ali and Patil, Shruti and Kudrow, Daniel and Heckey, Jeff and Lvov, Alexey and Chong, Frederic T and Martonosi, Margaret},
  journal={Parallel Computing},
  volume={45},
  pages={2--17},
  year={2015},
  publisher={Elsevier}
}

@misc{pyLIQTR,
    author = {Kevin Obenland and Justin Elenewski and Arthur Kurlej and Joe Belarge and John Blue and Robert Rood},
    title = {pyLIQTR (LIncoln Laboratory Quantum algorithm Test and Research)},
    year = {2023},
    url = {https://github.com/isi-usc-edu/pyLIQTR}
}

@misc{QIRspec,
    author = {{QIR Alliance}}, 
    title = {{Q}uantum {I}ntermediate {R}epresentation ({QIR}) specification},
    url = {https://github.com/qir-alliance/qir-spec/blob/main/specification/README.md},
    year={2022}
}

@article{stade2024towards,
  title={Towards Supporting QIR: Thoughts on Adopting the Quantum Intermediate Representation},
  author={Stade, Yannick and Burgholzer, Lukas and Wille, Robert},
  journal={arXiv preprint arXiv:2411.18682},
  year={2024}
}

@inproceedings{mccaskey2021mlir,
  title={A MLIR dialect for quantum assembly languages},
  author={McCaskey, Alexander and Nguyen, Thien},
  booktitle={2021 IEEE International Conference on Quantum Computing and Engineering (QCE)},
  pages={255--264},
  year={2021},
  organization={IEEE}
}

@article{ittah2022qiro,
  title={QIRO: A static single assignment-based quantum program representation for optimization},
  author={Ittah, David and H{\"a}ner, Thomas and Kliuchnikov, Vadym and Hoefler, Torsten},
  journal={ACM Transactions on Quantum Computing},
  volume={3},
  number={3},
  pages={1--32},
  year={2022},
  publisher={ACM New York, NY}
}

@article{cross2022openqasm,
  title={OpenQASM 3: A broader and deeper quantum assembly language},
  author={Cross, Andrew and Javadi-Abhari, Ali and Alexander, Thomas and De Beaudrap, Niel and Bishop, Lev S and Heidel, Steven and Ryan, Colm A and Sivarajah, Prasahnt and Smolin, John and Gambetta, Jay M and others},
  journal={ACM Transactions on Quantum Computing},
  volume={3},
  number={3},
  pages={1--50},
  year={2022},
  publisher={ACM New York, NY}
}

@article{kakuko2025practical,
  title={A Practical Open-Source Software Stack for a Cloud-Based Quantum Computing System},
  author={Kakuko, Norihiro and Gokita, Shun and Masumoto, Naoyuki and Matsumoto, Keita and Miyaji, Kosuke and Miyanaga, Takafumi and Mori, Toshio and Nakayama, Haruki and Sasada, Keita and Takamiya, Yasuhito and others},
  journal={arXiv preprint arXiv:2507.23165},
  year={2025}
}

@inproceedings{zaparanuks2012algorithmic,
  title={Algorithmic profiling},
  author={Zaparanuks, Dmitrijs and Hauswirth, Matthias},
  booktitle={Proceedings of the 33rd ACM SIGPLAN conference on Programming Language Design and Implementation},
  pages={67--76},
  year={2012}
}

@article{suau2022qprof,
  title={Qprof: A gprof-inspired quantum profiler},
  author={Suau, Adrien and Staffelbach, Gabriel and Todri-Sanial, Aida},
  journal={ACM Transactions on Quantum Computing},
  volume={4},
  number={1},
  pages={1--28},
  year={2022},
  publisher={ACM New York, NY}
}

@article{zilk2025breaking,
  title={Breaking Down Quantum Compilation: Profiling and Identifying Costly Passes},
  author={Zilk, Felix and Tundo, Alessandro and De Maio, Vincenzo and Brandic, Ivona},
  journal={arXiv preprint arXiv:2504.15141},
  year={2025}
}

@article{stanley2002evolving,
  title={Evolving neural networks through augmenting topologies},
  author={Stanley, Kenneth O and Miikkulainen, Risto},
  journal={Evolutionary computation},
  volume={10},
  number={2},
  pages={99--127},
  year={2002},
  publisher={MIT Press}
}

@article{cranmer2023interpretable,
  title={Interpretable machine learning for science with PySR and SymbolicRegression. jl},
  author={Cranmer, Miles},
  journal={arXiv preprint arXiv:2305.01582},
  year={2023}
}

@inproceedings{akiba2019optuna,
  title={Optuna: A next-generation hyperparameter optimization framework},
  author={Akiba, Takuya and Sano, Shotaro and Yanase, Toshihiko and Ohta, Takeru and Koyama, Masanori},
  booktitle={Proceedings of the 25th ACM SIGKDD international conference on knowledge discovery \& data mining},
  pages={2623--2631},
  year={2019}
}

@article{dobvsivcek2007arbitrary,
  title={Arbitrary accuracy iterative quantum phase estimation algorithm using a single ancillary qubit: A two-qubit benchmark},
  author={Dob{\v{s}}{\'\i}{\v{c}}ek, Miroslav and Johansson, G{\"o}ran and Shumeiko, Vitaly and Wendin, G{\"o}ran},
  journal={Physical Review A—Atomic, Molecular, and Optical Physics},
  volume={76},
  number={3},
  pages={030306},
  year={2007},
  publisher={APS}
}

@article{katabarwa2024early,
  title={Early fault-tolerant quantum computing},
  author={Katabarwa, Amara and Gratsea, Katerina and Caesura, Athena and Johnson, Peter D},
  journal={PRX quantum},
  volume={5},
  number={2},
  pages={020101},
  year={2024},
  publisher={APS}
}

@inproceedings{magalhaes2018optimization,
  title={An optimization approach to generate accurate and efficient lookup tables for engineering applications},
  author={Magalh{\~a}es, H and Marques, F and Liu, B and Pombo, Jo{\~a}o and Flores, P and Ambr{\'o}sio, Jorge and Bruni, S},
  booktitle={International Conference on Engineering Optimization},
  pages={1446--1457},
  year={2018},
  organization={Springer}
}

@article{dawson2005solovay,
  title={The solovay-kitaev algorithm},
  author={Dawson, Christopher M and Nielsen, Michael A},
  journal={arXiv preprint quant-ph/0505030},
  year={2005}
}

@inproceedings{hao2026reducing,
  title={Reducing T gates with unitary synthesis},
  author={Hao, Tianyi and Xu, Amanda and Tannu, Swamit},
  booktitle={Proceedings of the 31st ACM International Conference on Architectural Support for Programming Languages and Operating Systems, Volume 2},
  pages={1589--1604},
  year={2026}
}

@article{ross2016optimal,
  title={Optimal ancilla-free Clifford+ T approximation of z-rotations.},
  author={Ross, Neil J and Selinger, Peter},
  journal={Quantum Inf. Comput.},
  volume={16},
  number={11\&12},
  pages={901--953},
  year={2016}
}

@incollection{hatano2005finding,
  title={Finding exponential product formulas of higher orders},
  author={Hatano, Naomichi and Suzuki, Masuo},
  booktitle={Quantum annealing and other optimization methods},
  pages={37--68},
  year={2005},
  publisher={Springer}
}

@article{rice1953classes,
  title={Classes of recursively enumerable sets and their decision problems},
  author={Rice, Henry Gordon},
  journal={Transactions of the American Mathematical society},
  volume={74},
  number={2},
  pages={358--366},
  year={1953},
  publisher={JSTOR}
}

@article{lindblad1976generators,
  title={On the generators of quantum dynamical semigroups},
  author={Lindblad, Goran},
  journal={Communications in mathematical physics},
  volume={48},
  number={2},
  pages={119--130},
  year={1976},
  publisher={Springer}
}

@article{daley2022practical,
  title={Practical quantum advantage in quantum simulation},
  author={Daley, Andrew J and Bloch, Immanuel and Kokail, Christian and Flannigan, Stuart and Pearson, Natalie and Troyer, Matthias and Zoller, Peter},
  journal={Nature},
  volume={607},
  number={7920},
  pages={667--676},
  year={2022},
  publisher={Nature Publishing Group UK London}
}

@article{di2024efficient,
  title={Efficient quantum algorithm to simulate open systems through a single environmental qubit},
  author={Di Bartolomeo, Giovanni and Vischi, Michele and Feri, Tommaso and Bassi, Angelo and Donadi, Sandro},
  journal={Physical Review Research},
  volume={6},
  number={4},
  pages={043321},
  year={2024},
  publisher={APS}
}

@article{rendon2024improved,
  title={Improved accuracy for Trotter simulations using Chebyshev interpolation},
  author={Rendon, Gumaro and Watkins, Jacob and Wiebe, Nathan},
  journal={Quantum},
  volume={8},
  pages={1266},
  year={2024},
  publisher={Verein zur F{\"o}rderung des Open Access Publizierens in den Quantenwissenschaften}
}

@article{mohammadipour2025reducing,
  title={Reducing circuit depth in lindblad simulation via step-size extrapolation},
  author={Mohammadipour, Pegah and Li, Xiantao},
  journal={arXiv preprint arXiv:2507.22341},
  year={2025}
}

@article{su2026resource,
  title={Resource Estimation for Fault-Tolerant Quantum Programs},
  author={Su, Bonan and Feng, Yuan and Zhou, Li and Ying, Mingsheng},
  journal={arXiv preprint arXiv:2608.04573},
  year={2026}
}

@inproceedings{giortamis2025qos,
  title={$\{$QOS$\}$: Quantum operating system},
  author={Giortamis, Emmanouil and Rom{\~a}o, Francisco and Tornow, Nathaniel and Bhatotia, Pramod},
  booktitle={19th USENIX Symposium on Operating Systems Design and Implementation (OSDI 25)},
  pages={429--447},
  year={2025}
}

@article{johansson2012qutip,
  title={QuTiP: An open-source Python framework for the dynamics of open quantum systems},
  author={Johansson, J Robert and Nation, Paul D and Nori, Franco},
  journal={Computer physics communications},
  volume={183},
  number={8},
  pages={1760--1772},
  year={2012},
  publisher={Elsevier}
}

@article{meuli2020enabling,
  title={Enabling accuracy-aware quantum compilers using symbolic resource estimation},
  author={Meuli, Giulia and Soeken, Mathias and Roetteler, Martin and H{\"a}ner, Thomas},
  journal={Proceedings of the ACM on Programming Languages},
  volume={4},
  number={OOPSLA},
  pages={1--26},
  year={2020},
  publisher={ACM New York, NY, USA}
}

@inproceedings{szegedy2020promising,
  title={A promising path towards autoformalization and general artificial intelligence},
  author={Szegedy, Christian},
  booktitle={International Conference on Intelligent Computer Mathematics},
  pages={3--20},
  year={2020},
  organization={Springer}
}

@article{wu2022autoformalization,
  title={Autoformalization with large language models},
  author={Wu, Yuhuai and Jiang, Albert Qiaochu and Li, Wenda and Rabe, Markus and Staats, Charles and Jamnik, Mateja and Szegedy, Christian},
  journal={Advances in neural information processing systems},
  volume={35},
  pages={32353--32368},
  year={2022}
}

@article{hou2025model,
  title={Model context protocol (mcp): Landscape, security threats, and future research directions},
  author={Hou, Xinyi and Zhao, Yanjie and Wang, Shenao and Wang, Haoyu},
  journal={ACM Transactions on Software Engineering and Methodology},
  year={2025},
  publisher={ACM New York, NY}
}

@article{martinez2026pauli,
  title={Pauli Correlation Encoding for Budget-Contraint Optimization},
  author={Mart{\'\i}nez, Jacobo Pad{\'\i}n and Soloviev, Vicente P and Rentero, Alejandro Borrallo and Otero, Ant{\'o}n Rodr{\'\i}guez and Rodr{\'\i}guez, Raquel Alfonso and Krompiec, Michal},
  journal={arXiv preprint arXiv:2602.17479},
  year={2026}
}

@inproceedings{quetschlich2023compiler,
  title={Compiler optimization for quantum computing using reinforcement learning},
  author={Quetschlich, Nils and Burgholzer, Lukas and Wille, Robert},
  booktitle={2023 60th ACM/IEEE Design Automation Conference (DAC)},
  pages={1--6},
  year={2023},
  organization={IEEE}
}

@inproceedings{xu2025optimizing,
  title={Optimizing quantum circuits, fast and slow},
  author={Xu, Amanda and Molavi, Abtin and Tannu, Swamit and Albarghouthi, Aws},
  booktitle={Proceedings of the 30th ACM International Conference on Architectural Support for Programming Languages and Operating Systems, Volume 1},
  pages={777--793},
  year={2025}
}

@article{toshio2025practical,
  title={Practical quantum advantage on partially fault-tolerant quantum computer},
  author={Toshio, Riki and Akahoshi, Yutaro and Fujisaki, Jun and Oshima, Hirotaka and Sato, Shintaro and Fujii, Keisuke},
  journal={Physical Review X},
  volume={15},
  number={2},
  pages={021057},
  year={2025},
  publisher={APS}
}

@article{kanasugi2026enabling,
  title={Enabling Chemically Accurate Quantum Phase Estimation in the Early Fault-Tolerant Regime},
  author={Kanasugi, Shota and Toshio, Riki and Maruyama, Kazunori and Oshima, Hirotaka},
  journal={arXiv preprint arXiv:2603.22778},
  year={2026}
}

@article{saadatmand2024superconducting,
  title={Superconducting qubits at the utility scale: The potential and limitations of modularity},
  author={Saadatmand, SN and Wilson, Tyler L and Hodson, Mark J and Field, Mark and Devitt, Simon J and Vijayan, Madhav Krishnan and Robertson, Alan and Le, Thinh P and Ruh, Jannis and Paler, Alexandru and others},
  journal={arXiv preprint arXiv:2406.06015},
  year={2024}
}

@article{krzyszkowski2025analysis,
  title={Analysis of Surface Code Algorithms on Quantum Hardware Using the Qrisp Framework},
  author={Krzyszkowski, Jan and Niemiec, Marcin},
  journal={Electronics},
  volume={14},
  number={23},
  pages={4707},
  year={2025},
  publisher={MDPI}
}

@article{bergholm2018pennylane,
  title={Pennylane: Automatic differentiation of hybrid quantum-classical computations},
  author={Bergholm, Ville and Izaac, Josh and Schuld, Maria and Gogolin, Christian and Ahmed, Shahnawaz and Ajith, Vishnu and Alam, M Sohaib and Alonso-Linaje, Guillermo and AkashNarayanan, Bharath and Asadi, Ali and others},
  journal={arXiv preprint arXiv:1811.04968},
  year={2018}
}

@article{grayeli2024symbolic,
  title={Symbolic regression with a learned concept library},
  author={Grayeli, Arya and Sehgal, Atharva and Costilla Reyes, Omar and Cranmer, Miles and Chaudhuri, Swarat},
  journal={Advances in Neural Information Processing Systems},
  volume={37},
  pages={44678--44709},
  year={2024}
}

@article{akahoshi2024partially,
  title={Partially fault-tolerant quantum computing architecture with error-corrected clifford gates and space-time efficient analog rotations},
  author={Akahoshi, Yutaro and Maruyama, Kazunori and Oshima, Hirotaka and Sato, Shintaro and Fujii, Keisuke},
  journal={PRX quantum},
  volume={5},
  number={1},
  pages={010337},
  year={2024},
  publisher={APS}
}

@article{self2024protecting,
  title={Protecting expressive circuits with a quantum error detection code},
  author={Self, Chris N and Benedetti, Marcello and Amaro, David},
  journal={Nature Physics},
  volume={20},
  number={2},
  pages={219--224},
  year={2024},
  publisher={Nature Publishing Group UK London}
}

@article{steane1996multiple,
  title={Multiple-particle interference and quantum error correction},
  author={Steane, Andrew},
  journal={Proceedings of the Royal Society of London. Series A: Mathematical, Physical and Engineering Sciences},
  volume={452},
  number={1954},
  pages={2551--2577},
  year={1996},
  publisher={The Royal Society London}
}

@article{arnault2024typology,
  title={A typology of quantum algorithms},
  author={Arnault, Pablo and Arrighi, Pablo and Herbert, Steven and Kasnetsi, Evi and Li, Tianyi},
  journal={arXiv preprint arXiv:2407.05178},
  year={2024}
}

@article{saadatmand2024fault,
  title={Fault-tolerant resource estimation using graph-state compilation on a modular superconducting architecture},
  author={Saadatmand, SN and Wilson, Tyler L and Field, Mark and Vijayan, Madhav Krishnan and Le, Thinh P and Ruh, Jannis and Maan, Arshpreet Singh and Moflic, Ioana and Caesura, Athena and Paler, Alexandru and others},
  journal={arXiv preprint arXiv:2406.06015},
  year={2024}
}

@article{bertels2020quantum,
  title={Quantum computer architecture toward full-stack quantum accelerators},
  author={Bertels, Koen and Sarkar, Aritra and Hubregtsen, Thomas and Serrao, M and Mouedenne, Abid A and Yadav, Amitabh and Krol, A and Ashraf, Imran and Almudever, C Garcia},
  journal={IEEE Transactions on Quantum Engineering},
  volume={1},
  pages={1--17},
  year={2020},
  publisher={IEEE}
}

@article{shi2020resource,
  title={Resource-efficient quantum computing by breaking abstractions},
  author={Shi, Yunong and Gokhale, Pranav and Murali, Prakash and Baker, Jonathan M and Duckering, Casey and Ding, Yongshan and Brown, Natalie C and Chamberland, Christopher and Javadi-Abhari, Ali and Cross, Andrew W and others},
  journal={Proceedings of the IEEE},
  volume={108},
  number={8},
  pages={1353--1370},
  year={2020},
  publisher={IEEE}
}

@article{arute2019quantum,
  title={Quantum supremacy using a programmable superconducting processor},
  author={Arute, Frank and Arya, Kunal and Babbush, Ryan and Bacon, Dave and Bardin, Joseph C and Barends, Rami and Biswas, Rupak and Boixo, Sergio and Brandao, Fernando GSL and Buell, David A and others},
  journal={Nature},
  volume={574},
  number={7779},
  pages={505--510},
  year={2019},
  publisher={Nature Publishing Group UK London}
}

@article{quetschlich2023mqt,
  title={MQT Bench: Benchmarking software and design automation tools for quantum computing},
  author={Quetschlich, Nils and Burgholzer, Lukas and Wille, Robert},
  journal={Quantum},
  volume={7},
  pages={1062},
  year={2023},
  publisher={Verein zur F{\"o}rderung des Open Access Publizierens in den Quantenwissenschaften}
}

@inproceedings{van2023using,
  title={Using azure quantum resource estimator for assessing performance of fault tolerant quantum computation},
  author={van Dam, Wim and Mykhailova, Mariia and Soeken, Mathias},
  booktitle={Proceedings of the SC'23 Workshops of the International Conference on High Performance Computing, Network, Storage, and Analysis},
  pages={1414--1419},
  year={2023}
}

@article{babbush2025grand,
  title={The grand challenge of quantum applications},
  author={Babbush, Ryan and King, Robbie and Boixo, Sergio and Huggins, William and Khattar, Tanuj and Low, Guang Hao and McClean, Jarrod R and O'Brien, Thomas and Rubin, Nicholas C},
  journal={arXiv preprint arXiv:2511.09124},
  year={2025}
}

@inproceedings{quetschlich2024utilizing,
  title={Utilizing resource estimation for the development of quantum computing applications},
  author={Quetschlich, Nils and Soeken, Mathias and Murali, Prakash and Wille, Robert},
  booktitle={2024 IEEE International Conference on Quantum Computing and Engineering (QCE)},
  volume={1},
  pages={232--238},
  year={2024},
  organization={IEEE}
}

@article{google2025quantum,
  title={Quantum error correction below the surface code threshold},
  author= {Google Quantum AI and Collaborators},
  journal={Nature},
  volume={638},
  number={8052},
  pages={920--926},
  year={2025},
  publisher={Nature Publishing Group UK London}
}

@inproceedings{bernstein1993quantum,
  title={Quantum complexity theory},
  author={Bernstein, Ethan and Vazirani, Umesh},
  booktitle={Proceedings of the twenty-fifth annual ACM symposium on Theory of computing},
  pages={11--20},
  year={1993}
}

@inproceedings{campbell2023superstaq,
  title={Superstaq: Deep optimization of quantum programs},
  author={Campbell, Colin and Chong, Frederic T and Dahl, Denny and Frederick, Paige and Goiporia, Palash and Gokhale, Pranav and Hall, Benjamin and Issa, Salahedeen and Jones, Eric and Lee, Stephanie and others},
  booktitle={2023 IEEE International Conference on Quantum Computing and Engineering (QCE)},
  volume={1},
  pages={1020--1032},
  year={2023},
  organization={IEEE}
}

@article{blekos2024review,
  title={A review on quantum approximate optimization algorithm and its variants},
  author={Blekos, Kostas and Brand, Dean and Ceschini, Andrea and Chou, Chiao-Hui and Li, Rui-Hao and Pandya, Komal and Summer, Alessandro},
  journal={Physics Reports},
  volume={1068},
  pages={1--66},
  year={2024},
  publisher={Elsevier}
}

@article{bauer2020quantum,
  title={Quantum algorithms for quantum chemistry and quantum materials science},
  author={Bauer, Bela and Bravyi, Sergey and Motta, Mario and Chan, Garnet Kin-Lic},
  journal={Chemical Reviews},
  volume={120},
  number={22},
  pages={12685--12717},
  year={2020},
  publisher={ACS Publications}
}

@article{motta2022emerging,
  title={Emerging quantum computing algorithms for quantum chemistry},
  author={Motta, Mario and Rice, Julia E},
  journal={Wiley Interdisciplinary Reviews: Computational Molecular Science},
  volume={12},
  number={3},
  pages={e1580},
  year={2022},
  publisher={Wiley Online Library}
}

@article{riste2017demonstration,
  title={Demonstration of quantum advantage in machine learning},
  author={Rist{\`e}, Diego and Da Silva, Marcus P and Ryan, Colm A and Cross, Andrew W and C{\'o}rcoles, Antonio D and Smolin, John A and Gambetta, Jay M and Chow, Jerry M and Johnson, Blake R},
  journal={npj Quantum Information},
  volume={3},
  number={1},
  pages={16},
  year={2017},
  publisher={Nature Publishing Group UK London}
}

@article{bravyi2020quantum,
  title={Quantum advantage with noisy shallow circuits},
  author={Bravyi, Sergey and Gosset, David and K{\"o}nig, Robert and Tomamichel, Marco},
  journal={Nature Physics},
  volume={16},
  number={10},
  pages={1040--1045},
  year={2020},
  publisher={Nature Publishing Group UK London}
}

@article{havlivcek2019supervised,
  title={Supervised learning with quantum-enhanced feature spaces},
  author={Havl{\'\i}{\v{c}}ek, Vojt{\v{e}}ch and C{\'o}rcoles, Antonio D and Temme, Kristan and Harrow, Aram W and Kandala, Abhinav and Chow, Jerry M and Gambetta, Jay M},
  journal={Nature},
  volume={567},
  number={7747},
  pages={209--212},
  year={2019},
  publisher={Nature Publishing Group}
}

@article{abbas2024challenges,
  title={Challenges and opportunities in quantum optimization},
  author={Abbas, Amira and Ambainis, Andris and Augustino, Brandon and B{\"a}rtschi, Andreas and Buhrman, Harry and Coffrin, Carleton and Cortiana, Giorgio and Dunjko, Vedran and Egger, Daniel J and Elmegreen, Bruce G and others},
  journal={Nature Reviews Physics},
  pages={1--18},
  year={2024},
  publisher={Nature Publishing Group UK London}
}

@article{sunny2025extending,
  title={Extending QAOA-GPT to Higher-Order Quantum Optimization Problems},
  author={Sunny, Leanto and Rijal, Abhinav and Siopsis, George},
  journal={arXiv preprint arXiv:2511.07391},
  year={2025}
}

@inproceedings{uotila2025higher,
  title={Higher-order portfolio optimization with quantum approximate optimization algorithm},
  author={Uotila, Valter and Ripatti, Julia and Zhao, Bo},
  booktitle={2025 IEEE International Conference on Quantum Computing and Engineering (QCE)},
  volume={1},
  pages={01--12},
  year={2025},
  organization={IEEE}
}

@inproceedings{giurgica2020digital,
  title={Digital zero noise extrapolation for quantum error mitigation},
  author={Giurgica-Tiron, Tudor and Hindy, Yousef and LaRose, Ryan and Mari, Andrea and Zeng, William J},
  booktitle={2020 IEEE international conference on quantum computing and engineering (QCE)},
  pages={306--316},
  year={2020},
  organization={IEEE}
}

@article{mitiq,
  title     = {Mitiq: A software package for error mitigation on noisy quantum computers},
  author    = {Ryan LaRose and Andrea Mari and Sarah Kaiser and Peter J. Karalekas and Andre A. Alves and Piotr Czarnik and Mohamed El Mandouh and Max H. Gordon and Yousef Hindy and Aaron Robertson and Purva Thakre and Misty Wahl and Danny Samuel and Rahul Mistri and Maxime Tremblay and Nick Gardner and Nathaniel T. Stemen and Nathan Shammah and William J. Zeng},
  journal   = {Quantum},
  year      = {2022},
  month     = {Aug},
  doi       = {10.22331/q-2022-08-11-774},
  url       = {https://doi.org/10.22331/q-2022-08-11-774},
  publisher = {Verein zur Forderung des Open Access Publizierens in den Quantenwissenschaften},
  volume    = {6},
  pages     = {774},
}

@article{daguerre2025experimental,
  title={Experimental demonstration of high-fidelity logical magic states from code switching},
  author={Daguerre, Lucas and Blume-Kohout, Robin and Brown, Natalie C and Hayes, David and Kim, Isaac H},
  journal={Physical Review X},
  volume={15},
  number={4},
  pages={041008},
  year={2025},
  publisher={APS}
}

@article{morisaki2025optimal,
  title={Optimal ancilla-free Clifford+ T synthesis for general single-qubit unitaries},
  author={Morisaki, Hayata and Sano, Kaoru and Akibue, Seiseki},
  journal={arXiv preprint arXiv:2510.05816},
  year={2025}
}

@article{escofet2025accurate,
  title={An accurate and efficient analytic model of fidelity under depolarizing noise oriented to large scale quantum system design},
  author={Escofet, Pau and Rodrigo, Santiago and Garcia-S{\'a}ez, Artur and Alarc{\'o}n, Eduard and Abadal, Sergi and Almud{\'e}ver, Carmen G},
  journal={Quantum Science and Technology},
  volume={10},
  number={3},
  pages={035061},
  year={2025},
  publisher={IOP Publishing}
}

@article{litinski2019magic,
  title={Magic state distillation: Not as costly as you think},
  author={Litinski, Daniel},
  journal={Quantum},
  volume={3},
  pages={205},
  year={2019},
  publisher={Verein zur F{\"o}rderung des Open Access Publizierens in den Quantenwissenschaften}
}

@article{gidney2024magic,
  title={Magic state cultivation: growing T states as cheap as CNOT gates},
  author={Gidney, Craig and Shutty, Noah and Jones, Cody},
  journal={arXiv preprint arXiv:2409.17595},
  year={2024}
}

@article{avtandilyan2024optimal,
  title={Optimal-order Trotter--Suzuki decomposition for quantum simulation on noisy quantum computers: AA Avtandilyan, WV Pogosov},
  author={Avtandilyan, AA and Pogosov, WV},
  journal={Quantum Information Processing},
  volume={24},
  number={1},
  pages={8},
  year={2024},
  publisher={Springer}
}

@article{xu2025exponentially,
  title={Exponentially Decaying Quantum Simulation Error with Noisy Devices},
  author={Xu, Jue and Zhao, Chu and Fan, Junyu and Zhao, Qi},
  journal={arXiv preprint arXiv:2504.10247},
  year={2025}
}

@article{chatterjee2025comprehensive,
  title={A comprehensive cross-model framework for benchmarking the performance of quantum hamiltonian simulations},
  author={Chatterjee, Avimita and Rappaport, Sonny and Giri, Anish and Johri, Sonika and Proctor, Timothy and Neira, David E Bernal and Sathe, Pratik and Lubinski, Thomas},
  journal={IEEE Transactions on Quantum Engineering},
  year={2025},
  publisher={IEEE}
}

@article{siekierski2025software,
  title={Software for Creating Scalable Benchmarks from Quantum Algorithms},
  author={Siekierski, Noah and Seritan, Stefan and Patel, Neer and Niu, Siyuan and Lubinski, Thomas and Proctor, Timothy},
  journal={arXiv preprint arXiv:2511.02134},
  year={2025}
}

@misc{qiskit2024,
      title={Quantum computing with {Q}iskit},
      author={Javadi-Abhari, Ali and Treinish, Matthew and Krsulich, Kevin and Wood, Christopher J. and Lishman, Jake and Gacon, Julien and Martiel, Simon and Nation, Paul D. and Bishop, Lev S. and Cross, Andrew W. and Johnson, Blake R. and Gambetta, Jay M.},
      year={2024},
      doi={10.48550/arXiv.2405.08810},
      eprint={2405.08810},
      archivePrefix={arXiv},
      primaryClass={quant-ph}
}

@article{campbell2018random,
  title={A random compiler for fast Hamiltonian simulation},
  author={Campbell, Earl},
  journal={arXiv preprint arXiv:1811.08017},
  year={2018}
}

@article{daley2014quantum,
  title={Quantum trajectories and open many-body quantum systems},
  author={Daley, Andrew J},
  journal={Advances in Physics},
  volume={63},
  number={2},
  pages={77--149},
  year={2014},
  publisher={Taylor \& Francis}
}

@article{bezdvek2025classical,
  title={Classical Optimization Strategies for Variational Quantum Algorithms: A Systematic Study of Noise Effects and Parameter Efficiency},
  author={Bezd{\v{e}}k, Tom{\'a}{\v{s}} and Yuan, Haomu and Nov{\'a}k, Vojt{\v{e}}ch and Ill{\'e}sov{\'a}, Silvie and Beseda, Martin},
  journal={arXiv preprint arXiv:2511.09314},
  year={2025}
}

@article{patra2025resource,
  title={Resource Estimation for VQE on Small Molecules: Impact of Fermion Mappings and Hamiltonian Reductions},
  author={Patra, Ashish Kumar and Ghevade, Vikas Dattatraya and Bhat, Ruchika and Maitra, Rahul and others},
  journal={arXiv preprint arXiv:2512.01605},
  year={2025}
}

@article{rower2024suppressing,
  title={Suppressing counter-rotating errors for fast single-qubit gates with fluxonium},
  author={Rower, David A and Ding, Leon and Zhang, Helin and Hays, Max and An, Junyoung and Harrington, Patrick M and Rosen, Ilan T and Gertler, Jeffrey M and Hazard, Thomas M and Niedzielski, Bethany M and others},
  journal={PRX Quantum},
  volume={5},
  number={4},
  pages={040342},
  year={2024},
  publisher={APS}
}

@article{lin202524,
  title={24 days-stable CNOT gate on fluxonium qubits with over 99.9\% fidelity},
  author={Lin, Wei-Ju and Cho, Hyunheung and Chen, Yinqi and Vavilov, Maxim G and Wang, Chen and Manucharyan, Vladimir E},
  journal={PRX Quantum},
  volume={6},
  number={1},
  pages={010349},
  year={2025},
  publisher={APS}
}

@article{wang202499,
  title={99.9\%-fidelity in measuring a superconducting qubit},
  author={Wang, Can and Liu, Feng-Ming and Chen, He and Du, Yi-Fei and Ying, Chong and Wang, Jian-Wen and Huo, Yong-Heng and Peng, Cheng-Zhi and Zhu, Xiaobo and Chen, Ming-Cheng and others},
  journal={arXiv preprint arXiv:2412.13849},
  year={2024}
}

@article{smith2025single,
  title={Single-qubit gates with errors at the 10-7 level},
  author={Smith, Molly C and Leu, Aaron D and Miyanishi, Koichiro and Gely, Mario F and Lucas, David M},
  journal={Physical Review Letters},
  volume={134},
  number={23},
  pages={230601},
  year={2025},
  publisher={APS}
}

@article{loschnauer2025scalable,
  title={Scalable, high-fidelity all-electronic control of trapped-ion qubits},
  author={L{\"o}schnauer, CM and Mosca Toba, J and Hughes, AC and King, SA and Weber, MA and Srinivas, R and Matt, R and Nourshargh, R and Allcock, DTC and Ballance, CJ and others},
  journal={PRX Quantum},
  volume={6},
  number={4},
  pages={040313},
  year={2025},
  publisher={APS}
}

@article{sotirova2024high,
  title={High-fidelity heralded quantum state preparation and measurement},
  author={Sotirova, AS and Leppard, JD and Vazquez-Brennan, A and Decoppet, SM and Pokorny, F and Malinowski, M and Ballance, CJ},
  journal={arXiv preprint arXiv:2409.05805},
  year={2024}
}

@inproceedings{weiden2023improving,
  title={Improving quantum circuit synthesis with machine learning},
  author={Weiden, Mathias and Younis, Ed and Kalloor, Justin and Kubiatowicz, John and Iancu, Costin},
  booktitle={2023 IEEE International Conference on Quantum Computing and Engineering (QCE)},
  volume={1},
  pages={1--11},
  year={2023},
  organization={IEEE}
}

@article{kottmann2026parameter,
  title={Parameter-optimal unitary synthesis with flag decompositions},
  author={Kottmann, Korbinian and Wierichs, David and Alonso-Linaje, Guillermo and Killoran, Nathan},
  journal={arXiv preprint arXiv:2603.20376},
  year={2026}
}

@article{zou2024lightsabre,
  title={LightSABRE: A lightweight and enhanced SABRE algorithm},
  author={Zou, Henry and Treinish, Matthew and Hartman, Kevin and Ivrii, Alexander and Lishman, Jake},
  journal={arXiv preprint arXiv:2409.08368},
  year={2024}
}

@article{claeys2019floquet,
  title={Floquet-engineering counterdiabatic protocols in quantum many-body systems},
  author={Claeys, Pieter W and Pandey, Mohit and Sels, Dries and Polkovnikov, Anatoli},
  journal={Physical review letters},
  volume={123},
  number={9},
  pages={090602},
  year={2019},
  publisher={APS}
}

@article{gidney2019efficient,
  title={Efficient magic state factories with a catalyzed {$| CCZ\rangle $} to {$2| T\rangle $} transformation},
  author={Gidney, Craig and Fowler, Austin G},
  journal={Quantum},
  volume={3},
  pages={135},
  year={2019},
  publisher={Verein zur F{\"o}rderung des Open Access Publizierens in den Quantenwissenschaften}
}

@article{toshio2026star,
  title={STAR-magic mutation: Even more efficient analog rotation gates for early fault-tolerant quantum computer},
  author={Toshio, Riki and Kanasugi, Shota and Fujisaki, Jun and Oshima, Hirotaka and Sato, Shintaro and Fujii, Keisuke},
  journal={arXiv preprint arXiv:2603.22891},
  year={2026}
}

@article{akahoshi2025compilation,
  title={Compilation of trotter-based time evolution for partially fault-tolerant quantum computing architecture},
  author={Akahoshi, Yutaro and Toshio, Riki and Fujisaki, Jun and Oshima, Hirotaka and Sato, Shintaro and Fujii, Keisuke},
  journal={PRX Quantum},
  volume={6},
  number={4},
  pages={040319},
  year={2025},
  publisher={APS}
}

@article{ren2026merlean,
  title={Merlean: An agentic framework for autoformalization in quantum computation},
  author={Ren, Yuanjie and Li, Jinzheng and Qi, Yidi},
  journal={arXiv preprint arXiv:2602.16554},
  year={2026}
}

@article{krol2022efficient,
  title={Efficient decomposition of unitary matrices in quantum circuit compilers},
  author={Krol, Anna M and Sarkar, Aritra and Ashraf, Imran and Al-Ars, Zaid and Bertels, Koen},
  journal={Applied Sciences},
  volume={12},
  number={2},
  pages={759},
  year={2022},
  publisher={MDPI}
}

\ifreviseshow
    \input{xtra_review_frj}
\fi

\end{document}